\documentclass[12pt]{iopart}
\pdfoutput=1 

\usepackage[pdftex]{graphicx,color}

\usepackage{textcomp}
\usepackage{cite}
\usepackage{float}
\usepackage{bm} 
\expandafter\let\csname equation*\endcsname\relax
\expandafter\let\csname endequation*\endcsname\relax
\usepackage{amsmath,amssymb,amsthm,eucal,graphicx,float,epstopdf,ulem}

 \newcommand{\ctwotwo}{\raisebox{-2mm} {\includegraphics[width=6mm]{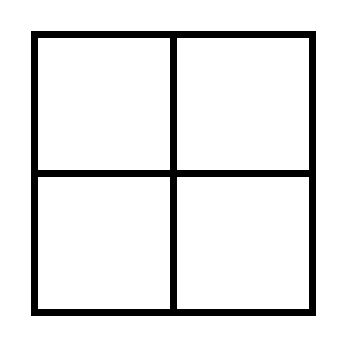}}}
 \newcommand{\cplus}{\raisebox{-2mm} {\includegraphics[width=6mm]{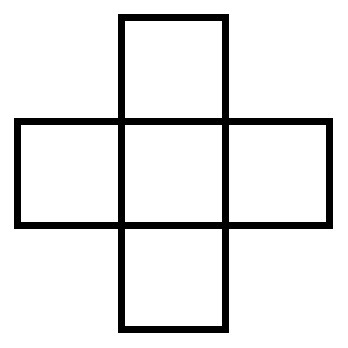}}}
 \newcommand{\conetwo}{\raisebox{-2mm} {\includegraphics[width=3mm]{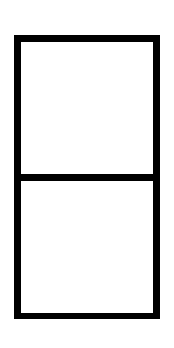}}}
 \newcommand{\ctwoone}{\raisebox{-0.5mm} {\includegraphics[width=6mm]{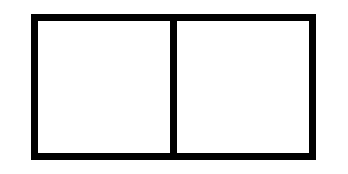}}}
 \newcommand{\hard}{\raisebox{-0.2mm} {\includegraphics[width=9.5mm]{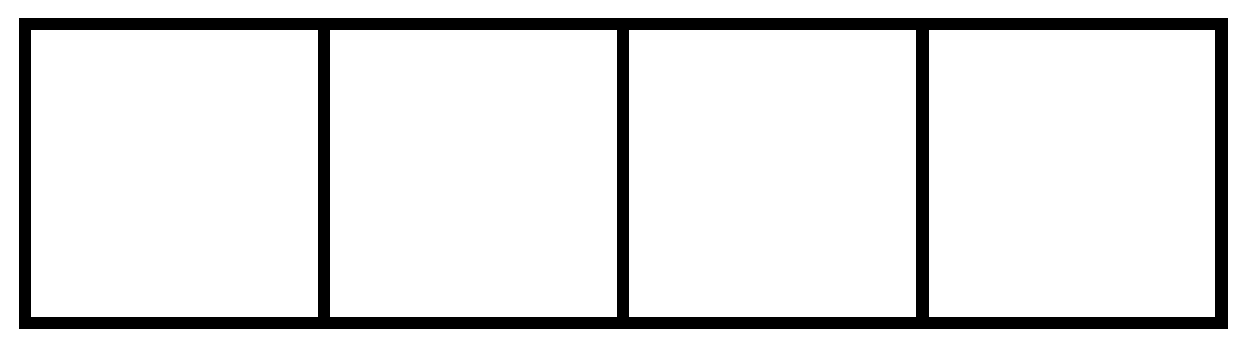}}}
 \newcommand{\hardv}{\raisebox{-4.5mm} {\includegraphics[width=2.8mm]{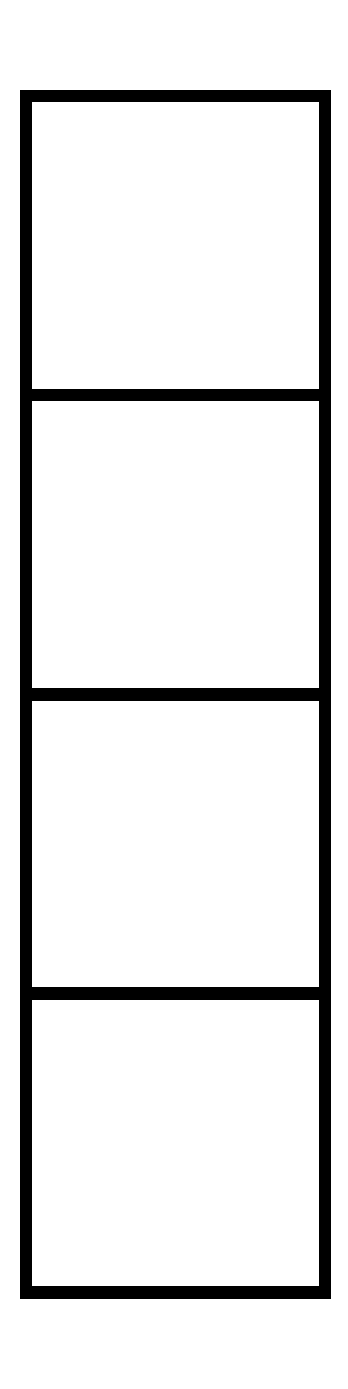}}}

\begin{document}

\title{Bulk diffusion in a kinetically constrained lattice gas}

\author{Chikashi Arita$^1$, P L Krapivsky$^{2,3}$, Kirone Mallick$^3$} 

\address{$^1$ Theoretische Physik, Universit\"at des Saarlandes, 66041 Saarbr\"ucken, Germany} 
\address{$^2$ Department of Physics, Boston University, Boston, MA 02215, USA}
\address{$^3$ Institut de Physique Th\'eorique, Universit\'e Paris-Saclay, CEA and CNRS,
91191 Gif-sur-Yvette, France} 
 
\begin{abstract}
In the hydrodynamic regime, the evolution of a stochastic lattice gas with symmetric hopping rules is described by a diffusion equation with density-dependent diffusion coefficient encapsulating all microscopic details of the dynamics. This diffusion coefficient is, in principle, determined by a Green-Kubo formula. In practice, even when the equilibrium properties of a lattice gas are analytically known, the diffusion coefficient cannot be computed except when a lattice gas additionally satisfies the gradient condition. We develop a procedure to systematically obtain analytical approximations for the diffusion coefficient for non-gradient lattice gases with known equilibrium. The method relies on a variational formula found by Varadhan and Spohn which is a version of the Green-Kubo formula particularly suitable for diffusive lattice gases. Restricting the variational formula to finite-dimensional sub-spaces allows one to perform the minimization and gives upper bounds for the diffusion coefficient. We apply this approach to a kinetically constrained non-gradient lattice gas in two dimensions, viz. to the Kob-Andersen model on the square lattice. 
\end{abstract}

\section{Introduction} 
 
Deriving a hydrodynamic limit is an important challenge in non-equilibrium statistical physics going back to Maxwell and Boltzmann. The derivations of this limit for molecular gases with deterministic dynamics, which originally triggered the development of kinetic theory, remain heuristic and incomplete \cite{Balescu,RL77}. The situation is much better for stochastic variants of molecular gas dynamics, see e.g. \cite{bib:Spohn1991,Anna,Yau,bib:KL,Var}. For stochastic lattice gases with symmetric hopping, the hydrodynamic description is particularly simple: the coarse-grained density $\rho(\boldsymbol{r} , t)$ satisfies the (non-linear) diffusion equation 
\begin{equation}
 \partial_t \rho = - \nabla \cdot \boldsymbol{J} ,
 \qquad \boldsymbol{J} = - {\bf D}(\rho) \nabla \rho\, . 
\end{equation} 
 In $d$ dimensions, ${\bf D}(\rho)$ is generally a symmetric $d\times d$ invertible diffusion matrix.
 In the simplest models ${\bf D}(\rho)=D(\rho) \boldsymbol{1}$ where $\boldsymbol{1}$ is the unit matrix, so to probe the relaxation on the hydrodynamic level one must know a single diffusion coefficient $D(\rho)$ encapsulating the microscopic hopping rules. 

The calculation of $D(\rho)$ is a challenging problem. First of all, even for very simple interacting lattice gases the equilibrium is unknown\footnote{For instance, the equilibrium behavior of the lattice gas with infinitely strong repulsion between particles occupying neighboring sites and zero interaction otherwise depends only on the density (the temperature is irrelevant), yet a phase transition between a low density disordered state and a high density ordered state is not fully understood even on the square lattice \cite{Fisher65,Runnels,Baxter,Arenzon_etal}.} in $d\geq 2$ dimensions. Further, even if the equilibrium properties of the lattice gas are known (this may occur in one dimension, or in higher dimensions for lattice gases with trivial equilibrium), the calculation of $D(\rho)$ is feasible only when a stochastic lattice gas satisfies the gradient condition\cite{bib:Spohn1991,bib:KL}. This special property states that the microscopic current is the gradient of a local function, {\it i.e.}, loosely speaking, the Fick law $\boldsymbol{J} = - {\bf D}(\rho) \nabla \rho$ is already valid at the discrete microscopic level. The simplest lattice gas obeying the gradient property is a collection of non-interacting random walkers, whereas the simplest interacting gradient lattice gas is the symmetric simple exclusion process \cite{bib:Spohn1991,bib:KL,bib:Spohn}; in these two models the diffusion coefficient does not depend on the density. Usually in a gradient lattice gas the diffusion coefficient depends on the density; some examples of such gradient lattice gases are the Katz-Lebowitz-Spohn model with symmetric hopping \cite{bib:KLS,bib:HKPS}, repulsion processes \cite{bib:Krapivsky}, a lattice gas of leap-frogging particles \cite{bib:CCGS,bib:GK}, and an exclusion process with avalanches \cite{bib:EPA}. However, generic interacting lattice gases do {\it not} satisfy the gradient condition.

The goal of this paper is to develop a procedure allowing one to probe the density dependence of the diffusion coefficient in non-gradient lattice gases. The basic tool which we use is the variational formula for the diffusion coefficient derived by Varadhan and Spohn, see \cite{bib:Spohn1991} and also \cite{bib:KL,bib:KLO,bib:KLO2}. This variational formula is a version of the Green-Kubo formula which is particularly suitable for diffusive lattice gases and is generally valid regardless of the presence of phase transitions and of the gradient condition. The Varadhan-Spohn formula requires one to minimize a functional over an infinite-dimensional function space. We recently demonstrated \cite{bib:AKM} that the Varadhan-Spohn variational formula can be used as a tool to derive explicit (albeit approximate) formulas for the diffusion coefficient. Essentially, we employed the Ritz method, namely we performed the minimization over finite-dimensional sub-spaces. The resulting minima give upper bounds for the diffusion coefficient. A similar approximation scheme has been used in \cite{bib:GG} for the computation of the thermal conductivity in stochastic energy exchange models \cite{bib:Sasada}. 

The complexity of the calculations increases with the dimension of the space of test functions 
 and with the spatial dimension of the model. Therefore in \cite{bib:AKM} we studied a one-dimensional lattice gas, namely we considered a generalized exclusion process with maximal occupancy equal to 2. Generalized exclusion processes \cite{bib:KLO,bib:KLO2,Timo,spider1d,BNCPV,AKM} are parametrized by hopping rates depending on the number of particles in the departure site and the (neighboring) target site and generically these lattice gases are non-gradient, although they contain a sub-class of gradient lattice gases (the misanthrope process \cite{bib:Cocozza-Thivent}). Increasing the dimensionality of the sub-space of the test functions we obtained more and more accurate results. The simplest `mean-field' prediction is already very accurate and after a few iterations we obtained a precision of the order of one part in a million. For the misanthrope process, which is gradient, the simplest approximation yields the exact result for the diffusion coefficient.

In this article we extend the method of Ref.~\cite{bib:AKM} to non-gradient lattice gases in higher dimensions. The dynamical properties of a lattice gas obviously cannot be understood if its equilibrium properties are unknown. An interesting class of non-gradient lattice gases which by construction have trivial equilibrium states are kinetically constrained lattice gases. These lattice gases were proposed \cite{bib:KA,JK} as toy models of the dynamics of structural glasses and, indeed,
 some of their properties, such as non-exponential relaxation and aging, do resemble those of glasses (see \cite{bib:RS,bib:GST} for a review). An accurate computation of the diffusion coefficient for a kinetically constrained lattice gas may therefore be useful from the point of view of applications to the dynamics of structural glasses. The calculations are quite laborious, so we limit ourselves to a specific kinetically constrained lattice gas in two dimensions, the Kob-Andersen (KA) model \cite{bib:KA} on the square lattice. 

The remainder of this work is organized as follows. In the next section, we give the precise definition of the KA model on the square lattice, and review some of its basic properties. In section \ref{sec:variational}, we present the approximation scheme of computing the upper bounds for the diffusion coefficient. In section \ref{sec:results} we describe upper bounds and in section \ref{sec:N-results} we compare these bounds with numerical results extracted from simulating of the steady state in a system with open boundaries. We investigate the high-density limit in section \ref{sec:high-density}. Simulations are performed in an open system, namely on a cylinder connected to reservoirs with fixed densities; the set-up is explained in section \ref{sec:simulations}. We conclude with a discussion (section \ref{sec:conclusions}). The details of the calculations are relegated to the Appendices.

\section{The model} 
\label{sec:model}

Kinetically constrained models are lattice models without static interactions other than hard core exclusion. These models have been originally proposed to mimic the dynamics of structural glasses. By design, the equilibrium state in these modes is trivial making them relatively tractable. The dynamics of these models are interesting e.g. they exhibit non-exponential relaxation, aging and other dynamical properties of glasses. One of the first kinetically constrained lattice gases, the Kob-Andersen (KA) model \cite{bib:KA}, can be defined on any hyper-cubic lattice $\mathbb{Z}^d$ as well as on other lattices, e.g., on the triangular lattice \cite{JK}. The KA model is an exclusion process, that is, each site is occupied by at most one particle. Hopping to nearest-neighbor sites is assumed to be stochastic and symmetric. We set the hopping rate to any of the $2d$ neighbors to unity, so that the total hopping rate is $2d$. The jump cannot occur when the destination site is occupied (exclusion). The key feature of the KA model is that the jump is allowed only if before and after the jump the particle has at least $m$ empty neighbors. The allowed range of the parameter $m$ is $1\leq m\leq 2d$. The case $m=1$ is the symmetric simple exclusion process for which the diffusion coefficient is identical to the hopping rate. When $m>d$, the dynamics is too constrained, e.g. the hypercube can never be broken up. Thus the interesting range is $2\leq m\leq d$.
On the square lattice the only interesting possibility is therefore $m=2$. 

We limit ourselves to the KA model with $m=2$ on the square lattice and call it, for brevity, the KA model. The process occurs on the infinite square lattice $\mathbb{Z}^2$; in simulations we treat finite lattices. The element $ \tau_{ i,j } $ of a configuration $ \tau $ represents the state of site $ (i,j) $ of the lattice; it is either empty $( \tau_{i,j} =0 )$ or occupied by a particle $ (\tau_{ i,j} =1 ) $. The relaxation in the KA model is very slow in the $\rho\to 1$ limit and earlier simulations \cite{bib:KA,Kurchan,Parisi} suggested the break of ergodicity at a certain $\rho_c<1$; another seemingly pathological feature of the KA model is 
that a fully occupied double column which spans the lattice can never be destroyed. It was proved \cite{TBF,BT17}, however, that the KA model on the infinite lattice is ergodic and exhibits a hydrodynamic behavior for any $\rho<1$.

\begin{figure}
\begin{center}
\includegraphics[width=33mm]{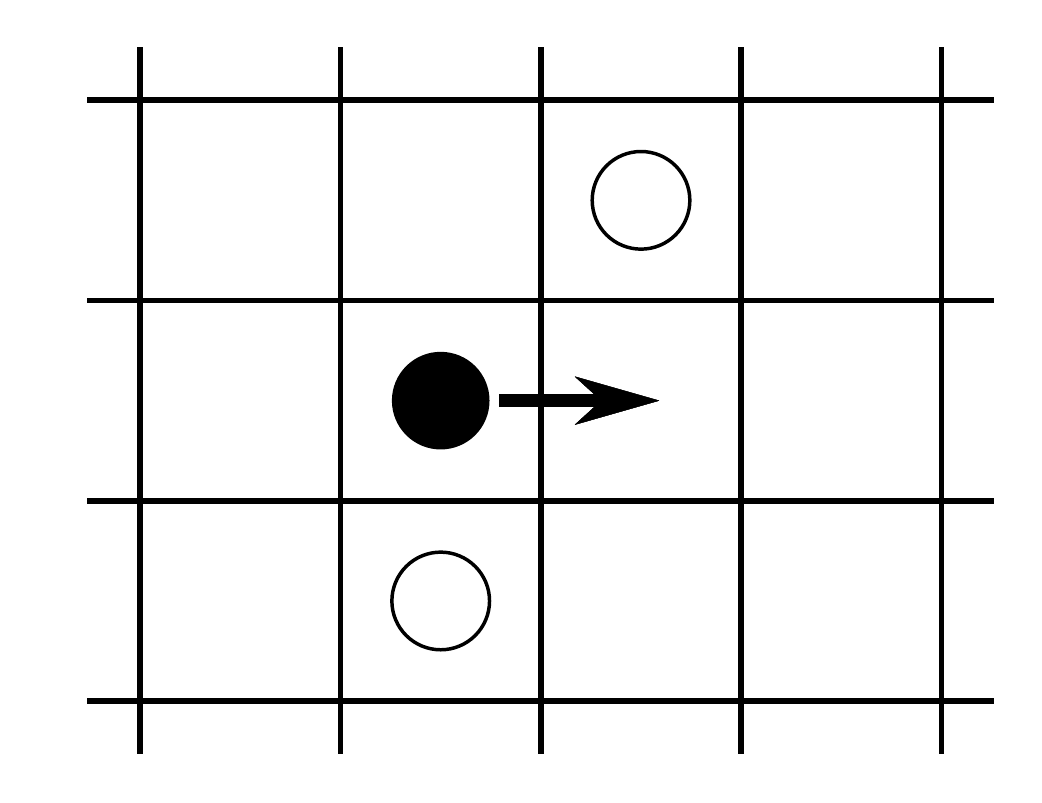}
\ \includegraphics[width=33mm]{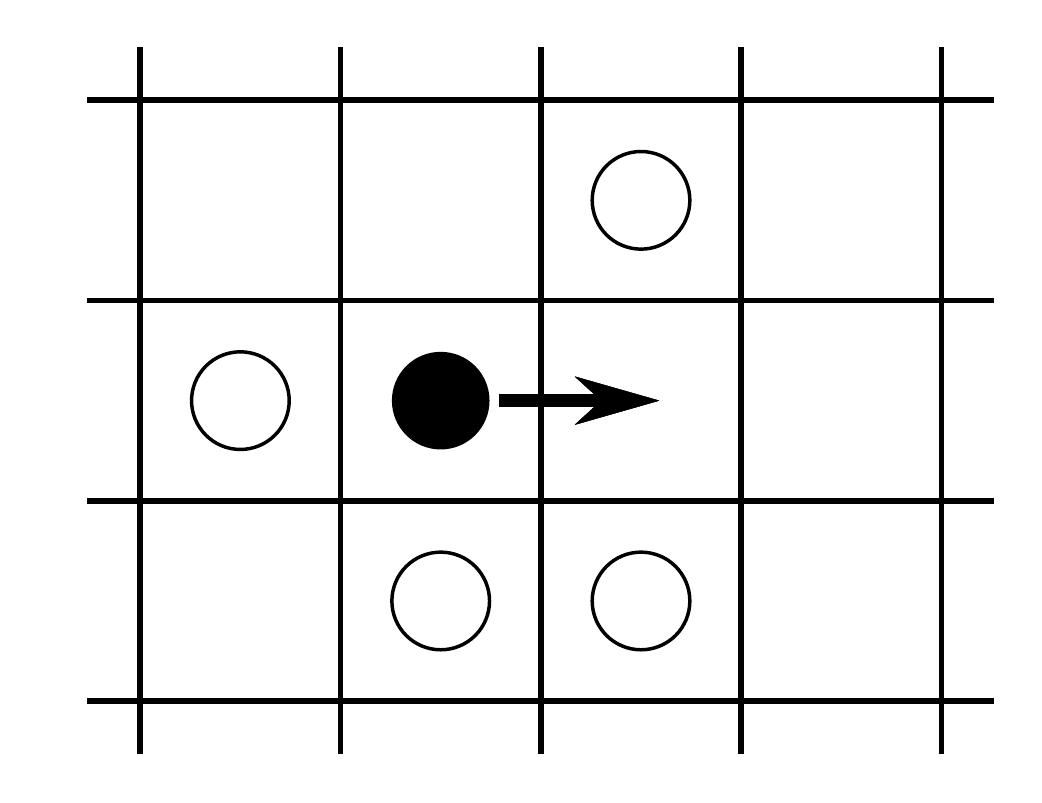}
\ \includegraphics[width=33mm]{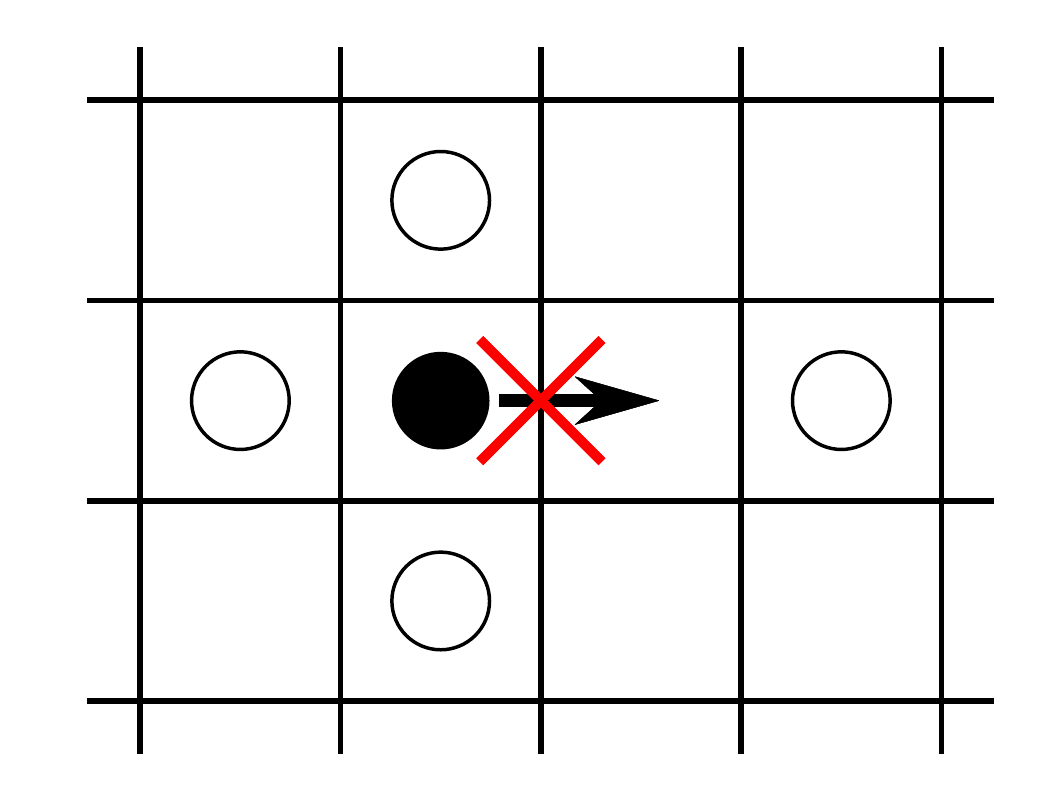}
\ \includegraphics[width=33mm]{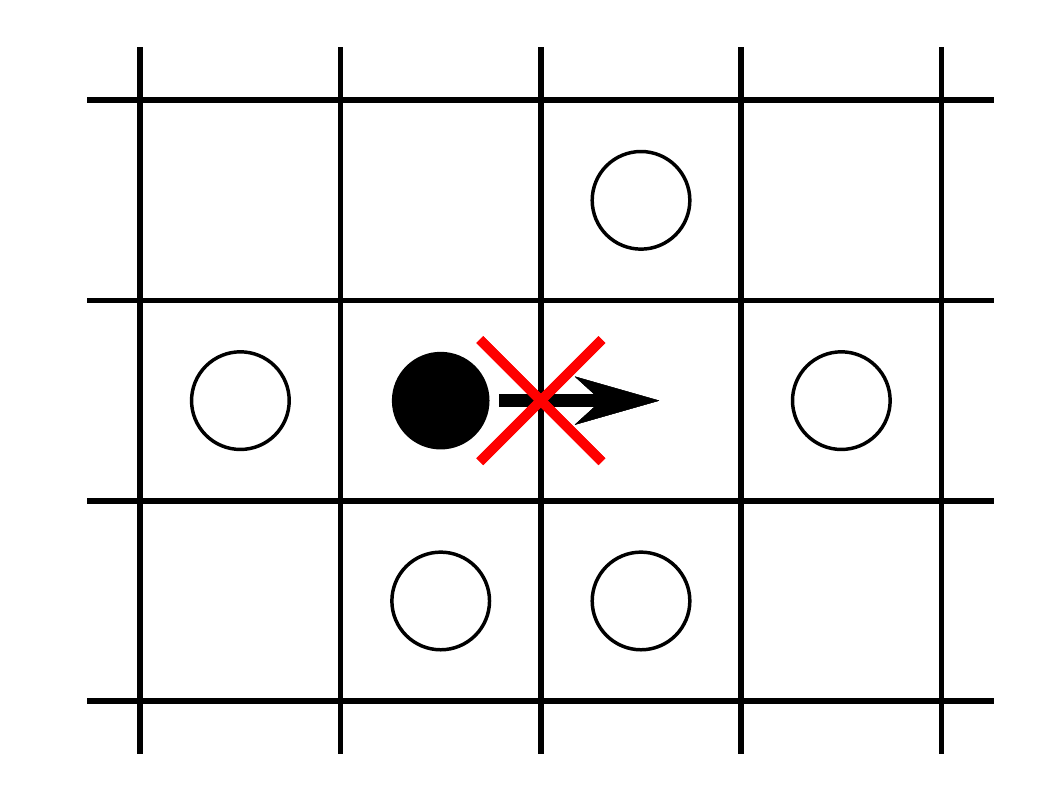}
\caption{Illustration of the hopping rules of the KA model on the square lattice. Particles hop with unit rate to neighboring empty sites and a hop is possible if before and after the hop the particle has at least two empty neighbors. In the presented examples the particle (shown as $\bullet$) attempts to jump rightward, other occupied sites are shown by $ \circ $. The local environment is too congested in the two examples on the right, so the attempted rightward jump is forbidden. }
\label{fig:rule}
\end{center}
\end{figure}

In the standard exclusion process, a particle at site $ (i,j)$ jumps to neighboring ``target'' sites $ (i\pm 1 ,j) $, $ (i ,j \pm 1 ) $ with unit rate if the target site is empty, otherwise the rate is zero. The hopping of the particle is not affected by the states of other sites. In the KA model, the hopping is kinetically constrained. The rate is still set to unity whenever the hopping is allowed, i.e., if at least 2 neighbors of the departure site are empty and at least 2 neighbors of the target site are empty. In figure~\ref{fig:rule}, we show some examples. 
In the two cases on the left, 
the hopping is allowed; in the other cases, three neighboring sites are occupied before or after the hopping. 

The instantaneous rightward current $ P^{ ( 1, 0 ) }_{i,j} ( \tau ) $ can be written as 
\begin{equation}
\label{eq:instant10}
P^{ ( 1, 0 ) }_{i,j} (\tau) = \tau_{ i,j } ( 1 - \tau_{ i+1,j } ) H_{i,j} (\tau ) . 
\end{equation}
The terms $\tau_{ i,j } ( 1 - \tau_{ i+1,j } )$ are the same as in the simple exclusion process, while the factor 
\begin{equation}
 H_{i,j} (\tau ) = ( 1 - \tau_{ i , j+1 }\tau_{ i-1 , j } \tau_{ i , j-1 })
 ( 1- \tau_{ i+1, j+1 } \tau_{ i+2 , j } \tau_{ i+1, j-1 } ) 
 \label{eq:H=}
\end{equation}
is either 1 or 0 depending on whether the hopping is allowed or not. 
The instantaneous leftward current $ P^{ ( -1, 0 ) }_{i,j} ( \tau ) $ from site $ (i+1,j) $ to $ (i ,j) $ is given by 
\begin{equation}
 \label{eq:instant-10}
 P_{i,j}^{ (-1, 0 ) } (\tau) = \tau_{ i+1,j } ( 1- \tau_{ i,j } ) H_{i,j} (\tau ) , 
\end{equation}
with the same factor $ H_{i,j} $ as it follows from the definition of the KA model. 
Similarly the instantaneous upward and downward currents are given by
\begin{subequations}
\begin{align}
\label{V-up}
P^{ ( 0,1) }_{i,j} (\tau) &= \tau_{ i,j } ( 1 - \tau_{ i,j + 1 } ) V_{i,j} (\tau ) \\
P_{i,j}^{ (0, -1 ) } (\tau) & = \tau_{ i,j +1 } ( 1- \tau_{ i,j } ) V_{i,j} (\tau )
\label{V-down}
\end{align}
\end{subequations}
with factor
\begin{equation}
V_{i,j} (\tau ) =
 ( 1 - \tau_{ i+1 , j }\tau_{ i , j-1 } \tau_{ i-1 , j } ) 
 ( 1- \tau_{ i +1, j+1 } \tau_{ i , j+2 } \tau_{ i-1, j+1 } ) 
\label{eq:V=}
\end{equation}
playing the same role as $ H_{i,j} (\tau ) $ in \eqref{eq:instant10} and \eqref{eq:instant-10}. 
The KA model satisfies the detailed balance condition due to the crucial requirement that the minimal number $m$ of empty neighbors of the departure site and of the target site (after the jump) is the same. The detailed balance condition implies that the KA model (and other kinetically constrained lattice gases) has simple equilibrium described by the product measure. Therefore for a given average density $ \rho $, any correlation function factorizes at equilibrium:
\begin{equation}
 \langle \tau_{ i_1 ,j_1 } \cdots \tau_{ i_n ,j_n } \rangle = \rho^n . 
\end{equation}
Here we assume that $ (i_\alpha ,j_\alpha) \neq (i_\beta ,j_\beta ) $ for $ \alpha\neq \beta$; the factorization formula is not true if $ (i_\alpha ,j_\alpha) = (i_\beta ,j_\beta ) $ for some $ \alpha\neq \beta $, for instance $\langle \tau_{ i ,j }^2 \rangle = \langle \tau_{ i ,j }^3 \rangle = \cdots = \rho $. Thanks to the product measure, the compressibility reads
\begin{equation}
\label{eq:chi}
 \chi = \langle \tau_{ i,j }^2 \rangle - \langle \tau_{ i,j } \rangle^2 = \rho (1-\rho) . 
\end{equation}

Our goal is to quantitatively probe hydrodynamic characteristics of the KA model. Specifically, we focus on the diffusion coefficient. We now estimate it, using mean-field arguments. One may regard the KA model as a symmetric simple exclusion process with effective hopping rates $ H_{i,j} $ and $ V_{i,j} $ given by equations~\eqref{eq:H=} and \eqref{eq:V=}. When the system is in a hydrodynamic regime, i.e., the spatial density profile varies slowly in space and time, one replaces $ \tau_{i,j+1} $ by $ \rho $, etc. Then the effective rates become $ (1-\rho^3)^2 $. Recalling that the diffusion coefficient of the symmetric simple exclusion process is given by its hopping rate, one would expect that 
\begin{equation}
\label{MF}
D = (1-\rho^3)^2\,.
\end{equation}

This is a mean-field type approximation \cite{TS17}. In section \ref{sec:results} and \ref{Ap:11} we show that \eqref{MF} follows from the variational formula if one performs the minimization over low-dimensional subspaces. Therefore $(1-\rho^3)^2$ is actually an upper bound for the diffusion coefficient. 

\section{Variational formula}
\label{sec:variational}

For stochastic lattice gases with symmetric dynamics, the diffusion coefficient can be expressed as an integral of a current-current correlation function \cite{bib:Spohn1991}. It is convenient to recast the computation of the integral into solving a variational problem; this was first realized by Varadhan and is presented in the book of Spohn \cite{bib:Spohn1991} (see also \cite{bib:KL,bib:KLO2}). 

Before proceeding, we emphasize that for lattice gases with symmetric hopping, the hydrodynamic behavior is \textit{believed} to be diffusive. For various lattice gases with symmetric hopping, the diffusive behavior has been rigorously established (see e.g. \cite{bib:Spohn1991,Anna,Yau,bib:KL,Var}), but even for the simplest models it was \textit{not} proved that the diffusive behavior is isotropic on the macroscopic scale. Thus a diffusion matrix underlies the hydrodynamic behavior. This matrix is symmetric and hence in $d$ dimensions it contains $d(d+1)/2$ independent elements. For the hyper-cubic lattice $\mathbb{Z}^d$, the diagonal elements are equal and all non-diagonal elements are identical, thus there are just two independent elements\footnote{We tacitly assume that each particle occupies a single lattice site; if a particle occupies a few lattice sites and its shape is not symmetric with respect to the symmetries of the $\mathbb{Z}^d$ lattice, the diffusion matrix may have more independent elements.}. Hence, for the KA model on the square lattice
\begin{align}
\label{D-matrix}
{\bf D}(\rho) = 
\begin{bmatrix}
D(\rho) & N(\rho)\\
N(\rho) & D(\rho)
\end{bmatrix}
\end{align}

In the following we shall discuss only the diagonal component $D(\rho)$ which we call the diffusion coefficient. More precisely, we use the Varadhan-Spohn variational formula to compute $D_{11}(\rho)$. This variational formula can be written in a neat form \cite{bib:Spohn1991}
\begin{equation}
\label{eq:D=inf/2chi}
 D = \frac{1}{2\chi} \inf_{ \varphi } \langle Q (\varphi) \rangle\,.
\end{equation}
Here $ \chi $ is the compressibility, equation~\eqref{eq:chi}, which is usually known only if we understand the equilibrium properties of the lattice gas. The infimum is taken on the space of {\it local} functions $\varphi(\tau)$ that depend only on a finite number of sites of the configuration $\tau$.

In our case of the lattice gas on the square lattice,
 the functional $Q(\varphi)$ appearing in \eqref{eq:D=inf/2chi} can be expressed as a sum of four functionals
\begin{equation}
Q(\varphi) = Q^{ (1,0) }(\varphi) + Q^{ (-1,0) }(\varphi) + Q^{ (0,1) }(\varphi) + Q^{ (0,-1) }(\varphi)\,. 
 \label{eq:Q=Q+Q+Q+Q}
\end{equation}
Each $ Q^{ ( \alpha, \beta ) } $ is a quadratic functional on the space of functions $\varphi(\tau)$. 
The expectation value $ \langle \cdot \rangle $ in equation~(\ref{eq:D=inf/2chi}) 
is taken with respect to the equilibrium measure on the configuration space. The functionals $ Q^{( \alpha, \beta )} $ depend on the microscopic dynamical rules \cite{bib:Spohn1991}. For the KA model (and generally for exclusion processes) each site can be in two states, $\tau_{i,j} = 0$ or $\tau_{i,j} = 1$, and the functionals $ Q^{( \alpha, \beta )} $ read
\begin{equation}
\label{eq:Qalphabeta}
 Q^{ ( \alpha, \beta ) } = P^{ ( \alpha, \beta ) }_{ 0,0 } ( \tau ) 
 \Bigg[ \alpha - \sum_{ ( u,v ) \in \mathbb Z^2 } \big\{ \varphi \big( A_{ u,v } \, \tau^{ ( \alpha, \beta ) }\big) 
 - \varphi ( A_{ u,v } \, \tau) \big\} \Bigg]^2 .
\end{equation}
Here $ \tau^{ ( \alpha, \beta ) }$ describes the change of the configuration $\tau$ after the single particle hop in the $ ( \alpha,\beta ) $ direction. More precisely
\begin{equation}
 \big( \tau^{ (\pm1 ,0 ) } \big)_{i,j}
 = \begin{cases}
 \tau_{ 0,0 } \mp 1 & (i,j ) = ( 0,0) , \\
 \tau_{ 1,0 } \pm 1 & (i,j ) = ( 1,0) , \\
 \tau_{ i,j }& \text{otherwise,} \\ 
 \end{cases}
\end{equation}
and
\begin{equation}
 \big( \tau^{ (0,\pm1 ) } \big)_{ i,j}
 = \begin{cases}
 \tau_{ 0,0 } \mp 1 & (i,j ) = ( 0,0) , \\
 \tau_{ 0,1 } \pm 1 & (i,j ) = ( 0,1) , \\
 \tau_{ i,j } & \text{otherwise}. \\ 
 \end{cases} 
\end{equation}
The operator $ A_{ u,v } $ shifts the configuration 
\begin{equation}
\big( A_{ u,v } \, \tau \big)_{ i,j } 
 = \tau_{ i-u, j-v } \, . 
\label{eq:Atau}
\end{equation}

The Varadhan-Spohn formula \eqref{eq:D=inf/2chi} is compact, but it is generally impossible to find the true minimum and hence to establish the true diffusion coefficient. The explicit computations are possible for lattice gases satisfying the gradient property. For such lattice gases the minimum is reached on a low-dimensional sub-space of test functions and explicit results for the diffusion coefficient are possible.

The idea of \cite{bib:AKM,bib:GG} is to use the variational formula on restricted finite-dimensional subspaces of test functions. In \cite{bib:AKM} we applied this method to a class of one-dimensional generalized exclusion processes where each site can accommodate at most two particles. A similar systematic approximation scheme has been used 
to compute the thermal conductivity of stochastic energy exchange models in \cite{bib:GG}.

In this work, we follow the logic of Refs.~\cite{bib:AKM,bib:GG} and consider the restricted version of the variational problem \eqref{eq:D=inf/2chi}, namely we perform the minimization over a restricted set of functions. Specifically we seek the minimum
\begin{align} 
\label{eq:DS=min/2chi}
 q [ S ] = \min_{ \varphi \in \mathcal{S} } \langle Q ( \varphi ) \rangle, \quad 
 D [ S ] = \frac{1}{2\chi } q [ S ] , 
\end{align}
where $ \mathcal{S} $ denotes a class of functions specified by the finite subset $S\subset \mathbb Z^2$. By evaluating \eqref{eq:DS=min/2chi}, we derive an upper bound on the diffusion coefficient of the KA model. Note that we only restricted the function space, but equations~\eqref{eq:Q=Q+Q+Q+Q}--\eqref{eq:Atau} are unchanged. 

The definition \eqref{eq:DS=min/2chi} leads to the inequalities $ D[ S ] \le D[S'] $ for $ S \supset S' $,
meaning that the larger subset $S$, the more precise upper bound we get. 
These upper bounds are explicit and provide good quantitative approximations as long as the density is not too large. Practical calculations remain limited to small sets $S$, because the number of configurations in $ \langle Q \rangle $ grows roughly as $2^{4|S|}$ (see section \ref{sec:high-density} for a more precise discussion). 

\section{Upper bounds for the diffusion coefficient}
\label{sec:results}

The simplest example is the empty set $S= \emptyset$. The minimization is not needed in this case as the corresponding functions are constant. The functionals are $ Q^{ ( \pm 1 , 0 ) } = P^{ (\pm 1 , 0 ) } $ and $ Q^{ ( 0,\pm 1 ) } = 0 $. One finds
\begin{equation}
\langle Q \rangle = \big\langle P^{ (1 , 0 ) } \big\rangle + \big\langle P^{ ( - 1 , 0 ) } \big\rangle 
 = 2 \rho (1-\rho ) (1-\rho^3)^2\, . 
\end{equation} 
Using $D [\emptyset ] = (2\chi)^{-1}\langle Q \rangle$ and the expression \eqref{eq:chi} for the compressibility we recover the mean-field prediction \eqref{MF}. For the simplest non-empty set $S= \square$ ({\it i.e.}, $S$ is a single site)
 the computations are more involved (see \ref{Ap:11}), but the outcome still coincides with the mean-field prediction:
\begin{equation}
\label{D11}
 D[ \emptyset ] = D[ \square ] = (1-\rho^3)^2\,. 
\end{equation}

In principle, one can consider arbitrary finite subsets $S\subset \mathbb Z^2$. Among sets of the same size $|S|$, connected sets provide better approximations. It is also useful to choose sets invariant under rotations by 90$^\circ$. To appreciate this we recall that equations~\eqref{eq:Q=Q+Q+Q+Q}--\eqref{eq:Atau} determine the $D_{11}$ element of the diffusion matrix ${\bf D}(\rho)$; the variational formula for $D_{22}$ is essentially the same in the general case when the minimization is performed over all functions. The restricted minimization which is performed in \eqref{eq:DS=min/2chi} yields the bound on $D_{11}$, but if the set $S$ is \textit{not} invariant under rotations by 90$^\circ$, the bound on $D_{22}$ may differ. Let us consider hard rods. There are two types on the square lattice, horizontal and vertical, schematically $\ctwoone$ and $\conetwo$ in the case of hard rods of length two. Using \eqref{eq:DS=min/2chi} we obtain the bounds on $D_{11}$. It is clear that the bound on $D_{11}$ for one set gives the bound on $D_{22}$ for another set. The detailed calculations are quite involved even in the case of hard rods of length two, so we just describe the outcome. For hard rods of length up to three, the bounds coincide and are both equal to $(1-\rho^3)^2$.

\begin{figure}
\begin{center}
 \includegraphics[width=80mm]{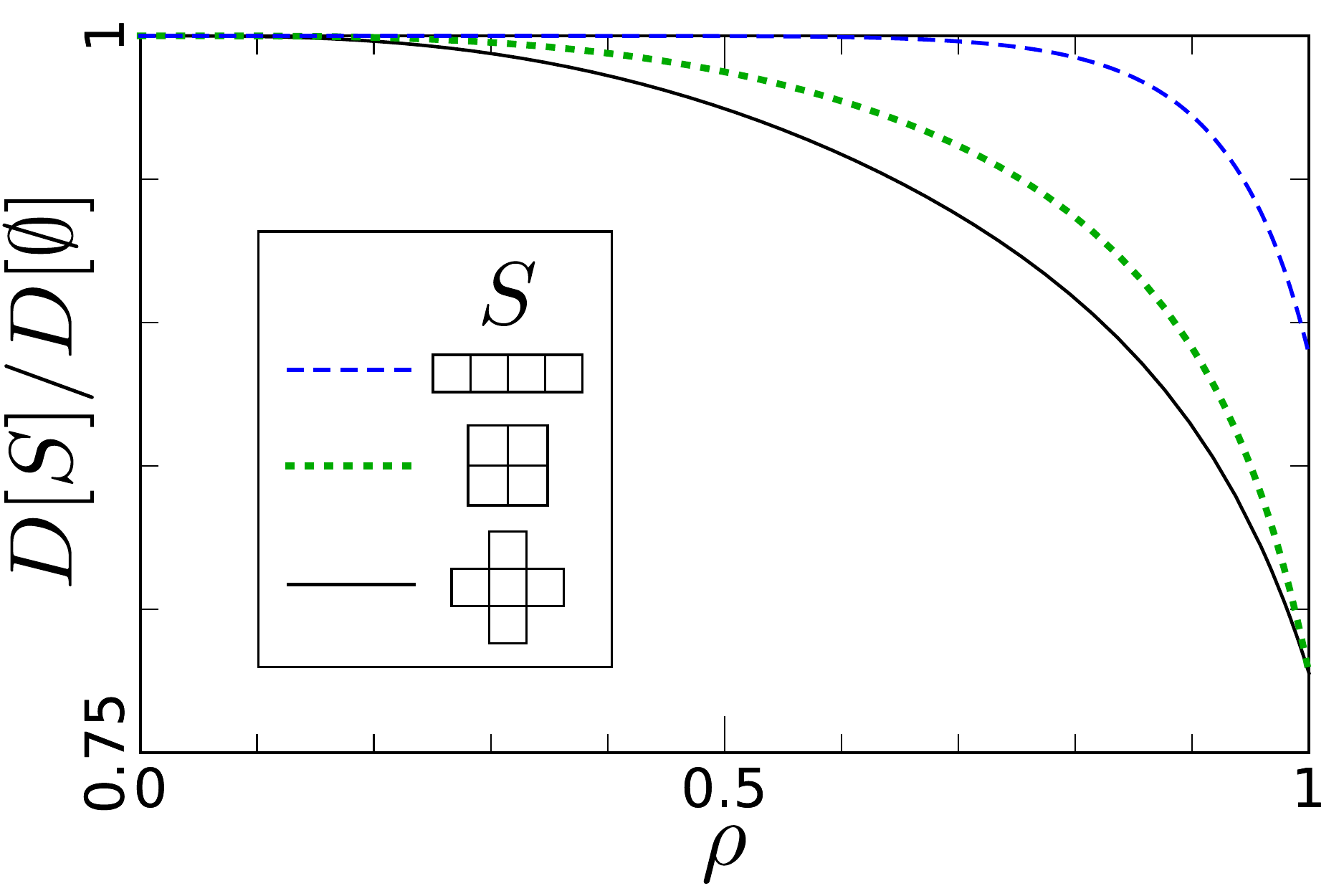}
 \caption{The ratio of the upper bounds given by \eqref{D14}--\eqref{D5} to the mean-field bound \eqref{D11} as a function of density (from top to bottom).}
\label{fig:DDD}
\end{center}
\end{figure}

 For the hard rods of length four, 
 we obtain an improved bound
\begin{equation}
\label{D14}
D [\hard] =D_{11}[\hard] =
(1-\rho^3)^2 - \frac{2(1-\rho)^2 \rho^{10}}{11 - 10 \rho^2 - 2 \rho^3 + 2\rho^4 + 2\rho^5 - \rho^6}, 
\end{equation}
 see figure~\ref{fig:DDD}. 
The other diagonal element is still the same $D_{22}[\hard] =(1-\rho^3)^2$, 
 or equivalently we have 
\begin{equation}
D \bigg[ \,\hardv \,\bigg] = D_{11} \bigg[ \,\hardv\, \bigg] = (1-\rho^3)^2. 
\end{equation}

The computations quickly become very cumbersome as the size $|S|$ increases, so it is crucial to choose sets of such shapes that the bounds are really improving when the size increases. We have found that it is profitable to choose sets which are invariant under rotations by 90$^\circ$, and also under vertical or horizontal reflections. Furthermore, we restrict our consideration to convex shapes. The symmetric convex sets are numerous. We show some examples in figure~\ref{fig:DDD}, where the sets in the $n^\text{th}$ column have the span (i.e. the maximal horizontal and vertical size) $n$. One generic property of symmetric convex sets is that $|S|\equiv 0 (\text{mod} 4)$ when $n$ is even and $|S|\equiv 1 (\text{mod} 4)$ when $n$ is odd.

The most obvious infinite family of symmetric convex sets is the family of squares; in figure~\ref{fig:DDD} we show $n\times n$ squares with $n\leq 6$. We already know the bound \eqref{D11} implied by the $1\times 1$ square. For the $2\times 2$ square, calculations are already very involved (see \ref{Ap:22} for details). The outcome
\begin{equation}
\label{D22}
 D \Big[ \ctwotwo \Big] = (1-\rho^3)^2 - \frac{2 (1-\rho)^2 \rho^4 }{ 4-\rho-2\rho^3 }
\end{equation}
is an improvement over \eqref{D11}, and it is also notably better (see figure~\ref{fig:DDD}) than the bound \eqref{D14} for the rod with the same number of elements as the $2\times 2$ square. 

\begin{figure}
\begin{center}
\includegraphics[width=80mm]{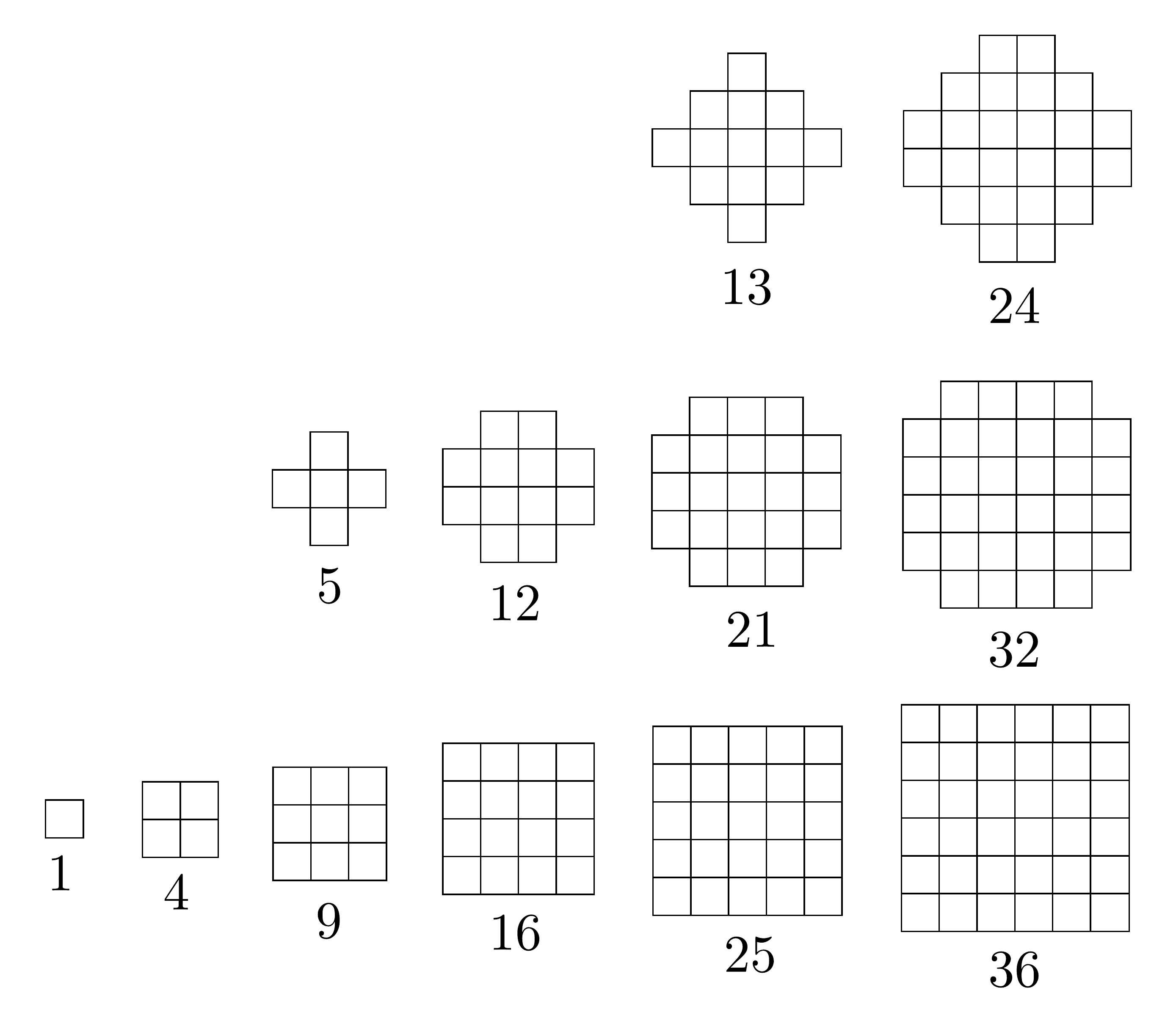}
\caption{Examples of symmetric convex shapes $S$. For each $S$, the number of sites $ | S | $ is indicated. 
We show the shapes with \textit{span} $n$ in the $n^\text{th}$ column. 
The set in the bottom are squares, the set on the left in each row is a rhombus. }
\label{fig:shapes}
\end{center}
\end{figure}

Another simple infinite family of symmetric convex sets consists of ``rhombi''. The sizes of rhombi are $|S|= \frac{n^2+1}{2}$ with $n=1,3,5,7,\ldots$; in figure~\ref{fig:shapes} we show rhombi with $n=1,3,5$. The rhombus of unit size is $S= \square$, so the bound is given by \eqref{D11}. The next rhombus has size $|S|=5$ and the corresponding upper bound is (see \ref{Ap:5} for the derivation)
\begin{equation}
\label{D5}
D \Big[\cplus \Big] =
 (1-\rho^3 )^2 - \frac{2}{3}\, \rho^4 (1 - \rho)^2 - \frac{2}{3} \rho^3 (1 - \rho)^2\, \frac{U}{V}
\end{equation}
with 
\begin{subequations}
\begin{align}
\label{U:def}
U&= 141 + 117 \rho + 36 \rho^2 - 94 \rho^3 - 70 \rho^4 +
 44 \rho^5 - 36 \rho^6 - 6 \rho^7 - 2 \rho^8 + 6 \rho^9 , \\
\label{V:def}
V &= 423 + 384 \rho - 368 \rho^2 - 578 \rho ^3 - 296 \rho ^4 + 273 \rho ^5 \nonumber\\
&\quad + 216 \rho ^6 + 98 \rho ^7 - 42 \rho ^8 - 30 \rho ^9- 12 \rho ^{10} . 
\end{align}
\end{subequations}
The bound \eqref{D5} is better than the bounds \eqref{D11}, \eqref{D14} and \eqref{D22}, see figure~\ref{fig:DDD}.

\section{Simulation results}
\label{sec:N-results}

\begin{figure}
\begin{center}
 \includegraphics[width=80mm]{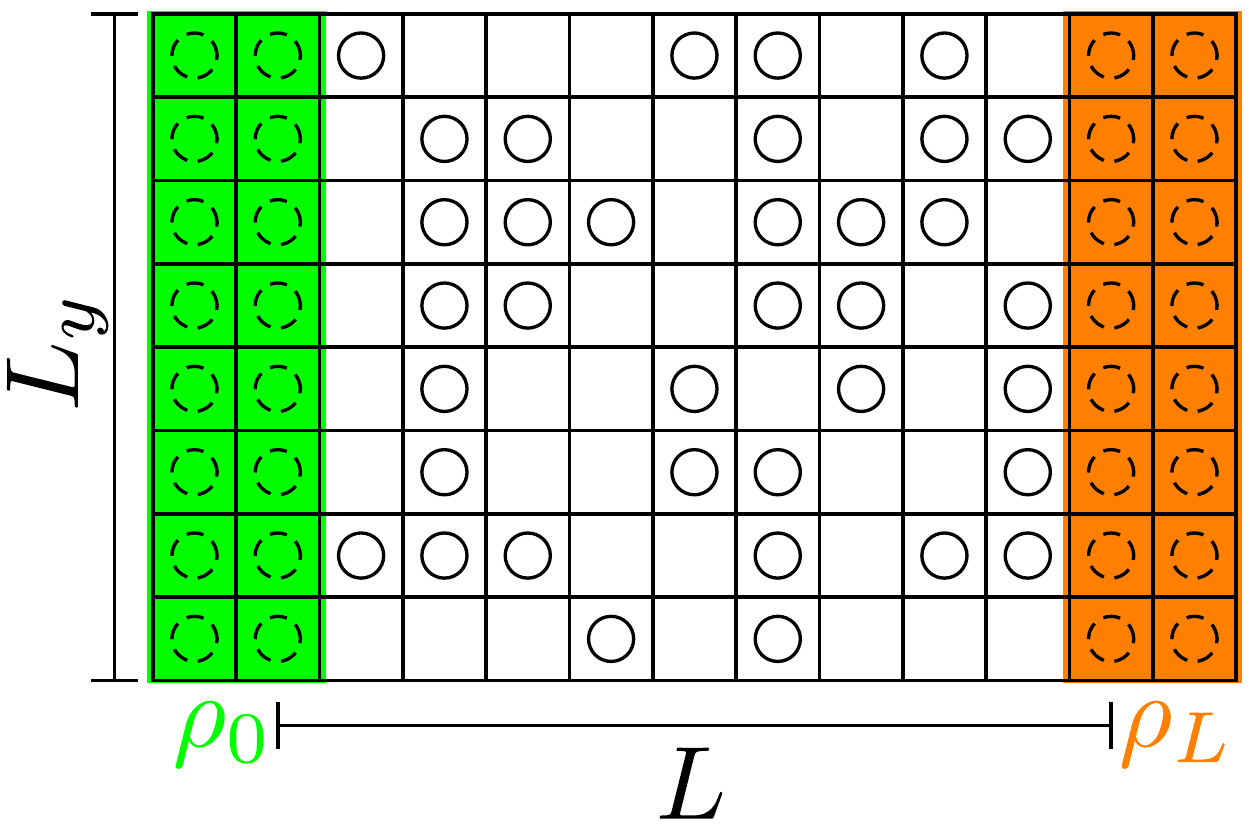}
 \caption{Illustration of the boundary conditions used in simulations. The left and right reservoirs have densities $ \rho_0 $ and $ \rho_L $. Our two-dimensional system is periodic in the vertical direction, i.e., the top sites are neighbors of the corresponding bottom sites.}
\label{fig:cylinder}
\end{center}
\end{figure}

We perform Monte Carlo simulations on a finite square lattice with $ 1 \le i \le L - 1 $ and $1 \le j \le L_y $. 
This two-dimensional lattice is connected to reservoirs with constant densities at the left and right ends,
 and periodic in the vertical direction (a cylinder), see figure~\ref{fig:cylinder}. 
 The `virtual' sites $ (0,j) $, $ (-1,j) $, $ (L,j) $ and $ (L+1,j) $ are regarded as particle reservoirs. In order to realize the boundary densities $ \rho_0 $ and $ \rho_L $, we impose injection and extraction of particles at sites $ ( 1, j ) $ and $ ( L-1,j ) $ ($ 1\le j \le L_y $), see section \ref{sec:simulations} for details. 

We simulate the system long enough, so that a steady state is reached. Essentially nothing is known about this steady state, e.g. in contrast to the equilibrium the correlation functions do not factorize, otherwise the mean-field expression \eqref{MF} for the diffusion coefficient would be exact.

The steady state helps to appreciate why the diffusion is isotropic.
By the construction we have the density gradient only in the horizontal direction.
According to the Fick law, the current in the vertical direction is given as $N(\rho)\frac{d\rho}{dx}$ 
($x=i/L$). One expects this current to be equal to zero, which implies that $ N(\rho) \equiv 0$.
Thus the diffusion matrix \eqref{D-matrix} is expected to be diagonal for the KA model,
and actually for any exclusion process on the square lattice with hopping rules compatible
with the symmetries of the square lattice.

\begin{figure}
\begin{center}
 \includegraphics[width=80mm]{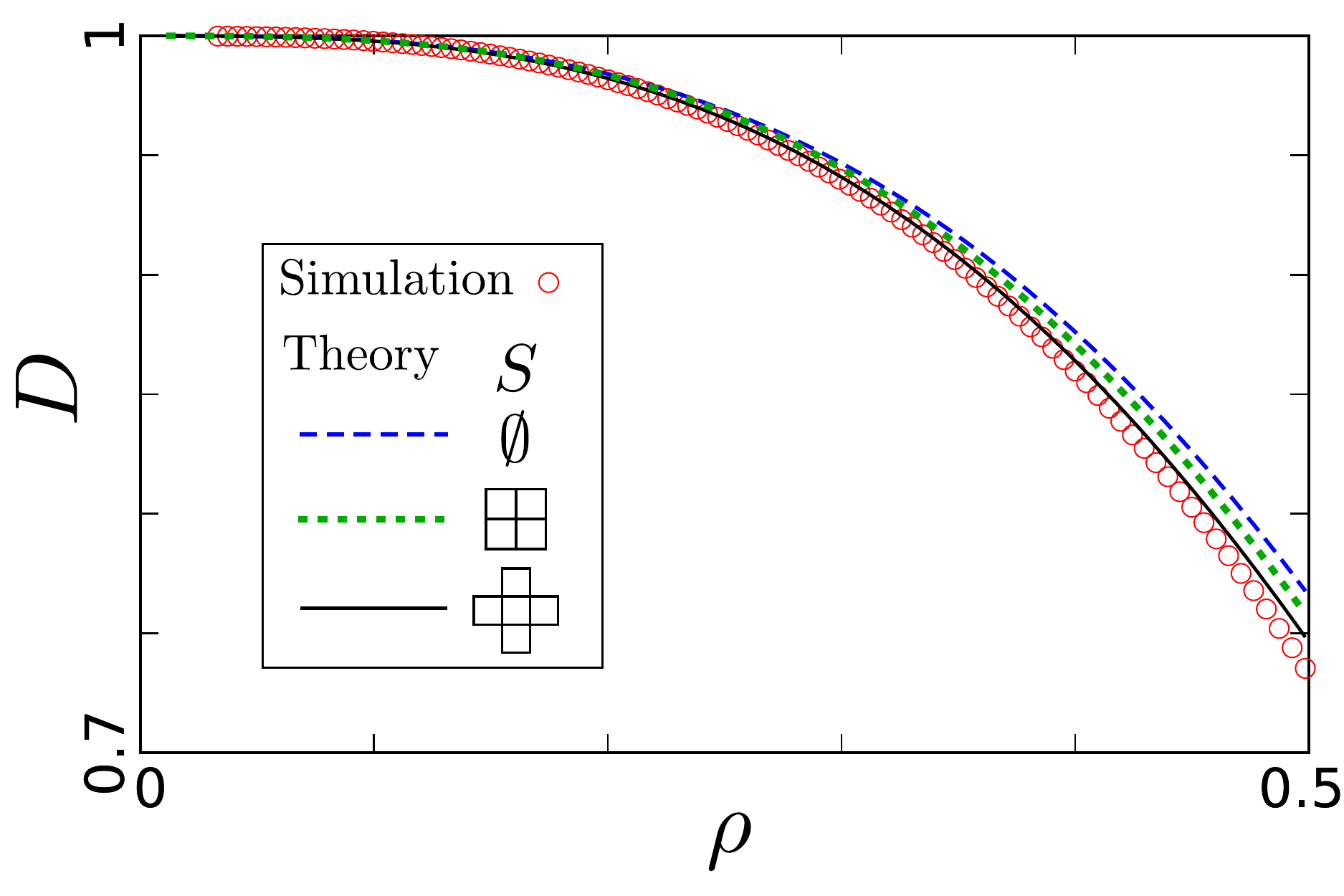}
 \caption{The diffusion coefficient versus $\rho$ extracted from simulations with boundary densities 
$ \rho_0 =0.6 $ and $ \rho_L=0 $ and system size $ ( L,L_y ) = ( 256, 200) $. The lines (dashed, dotted and solid) are theoretical upper bounds corresponding respectively to equations~\eqref{D11}, \eqref{D22} and \eqref{D5}.
 }\label{fig:diffusivity}
\end{center}
\end{figure} 

\begin{figure}
\begin{center}
 \includegraphics[width=80mm]{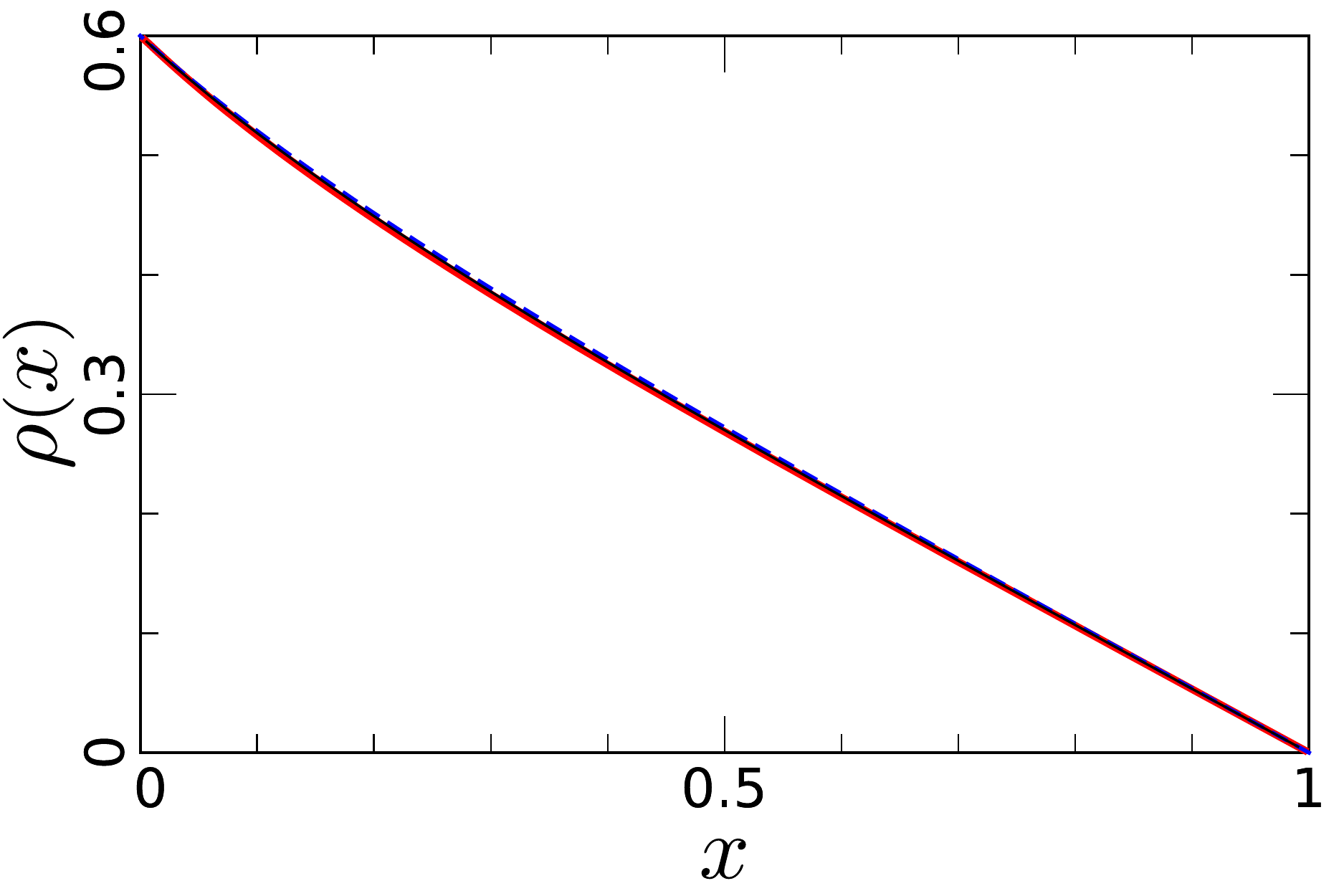}\\ 
 \includegraphics[width=80mm]{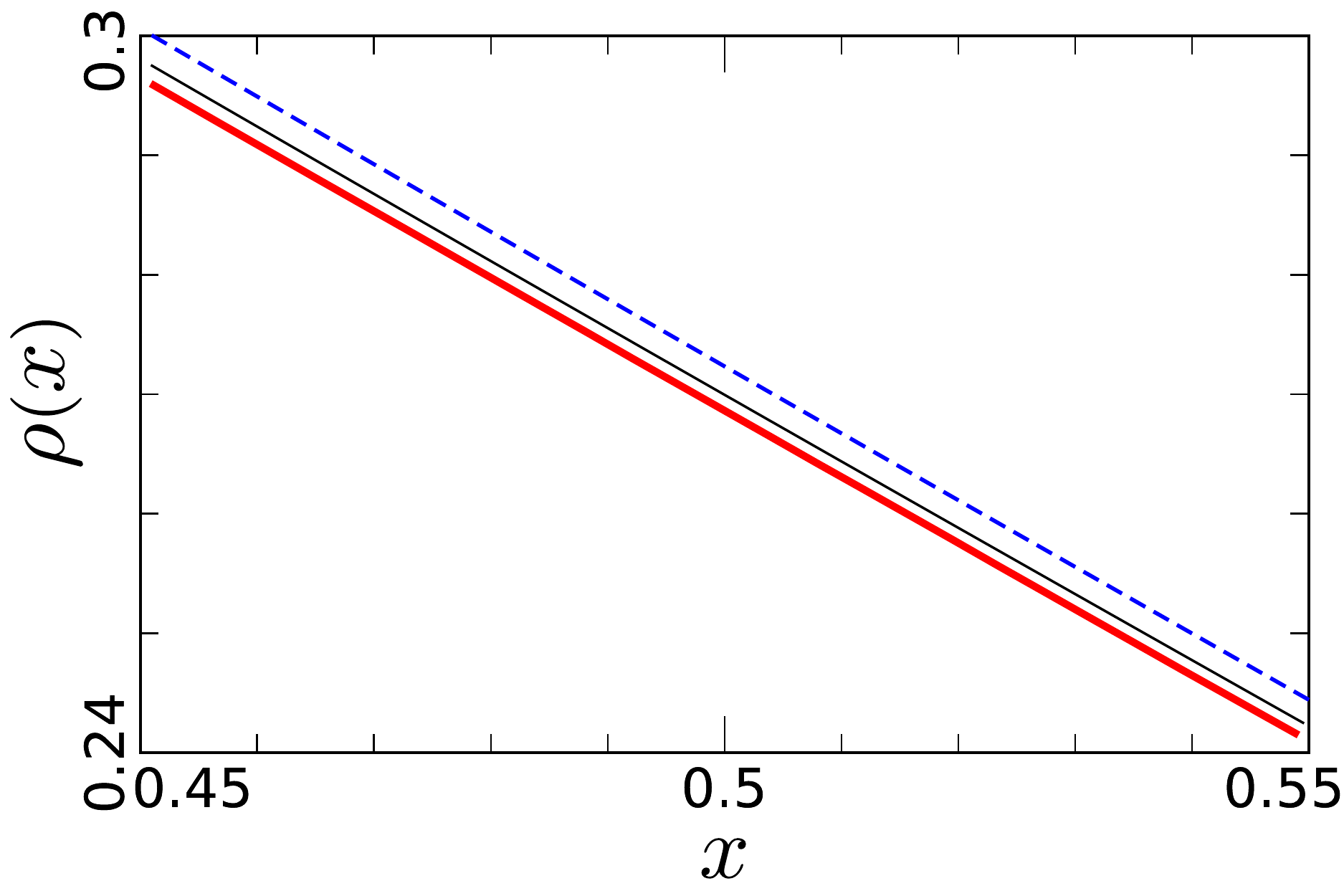}
 \caption{The density profile $\rho(x) $ in the open system 
 with boundary densities $ \rho(0) = 0.6 $ and $ \rho (1) = 0 $. 
 Simulation results for the open system with the size $ (L,L_y)=(256,200 ) $
 are shown by the thick line
 for the entire range $ 0<x<1 $ (top) and near the middle $ 0.45<x<0.55 $ (bottom). 
 The theoretical predictions are derived using the integral formula \eqref{eq:int=xint}. The dashed line corresponds to the choice of equation~\eqref{D11} for $ D (\rho ) $, and the thin line corresponds to the choice of equation~\eqref{D5}. 
 }\label{fig:density-profile}
\end{center}\end{figure}

Figure \ref{fig:diffusivity} shows simulation results for the diffusion coefficient for sufficiently low densities where the agreement with theoretical bounds is very good. The explicit approximate forms of the diffusion coefficients are useful to predict the stationary density profile. Recall that the diffusion coefficient characterizes the relationship between the density and the current in the horizontal direction,
\begin{align}
\label{eq:J=}
 J = - \frac{1}{L} D ( \rho ) \frac{d \rho }{ d x } \, , 
\end{align}
Hereinafter $ x = i / L $ is the scaled spatial coordinate. In the stationary state, the current $ J $ is independent of the position $x$ because of the conservation of particles in the bulk. Thus we have 
\begin{align}
 \frac{d }{ d x } \bigg[ D ( \rho ) \frac{d \rho }{ d x } \bigg] = 0 . 
\end{align}
The density profile $ \rho (x) $ should be the solution to this equation with the boundary conditions 
$ \rho(0) = \rho_0 $ and $ \rho (1) = \rho_L $. The implicit form of the solution reads 
\begin{align} 
\label{eq:int=xint}
 \int_{ \rho_0 }^{ \rho(x) } D( \rho ) d\rho = 
 x \int_{ \rho_0 }^{ \rho_L } D( \rho ) d\rho . 
\end{align}
Replacing $ D $ by approximation formulae we numerically plot $ \rho(x) $ versus the horizontal coordinate $x $. 
For the theoretical lines in figure~\ref{fig:density-profile}, we used $ D[ \emptyset ] = (1-\rho^3)^2 $ and $ D\Big[ \cplus\Big] $ as given by \eqref{D5}. 

The measurement of the density profile from simulations is straightforward. We observe $ \tau_{i,j} $ and take the average over $ j $ and over a time window. Comparing the theory with simulation result, 
one finds that the simplest mean-field approximation already gives a visually nice curve (see the top panel). 
However there is a systematic deviation from simulations (see the bottom panel). We find that $ D\Big[ \cplus\Big] $ indeed better agrees with the density profile in the middle of the system.

\begin{figure}
\begin{center}
\includegraphics[width=80mm]{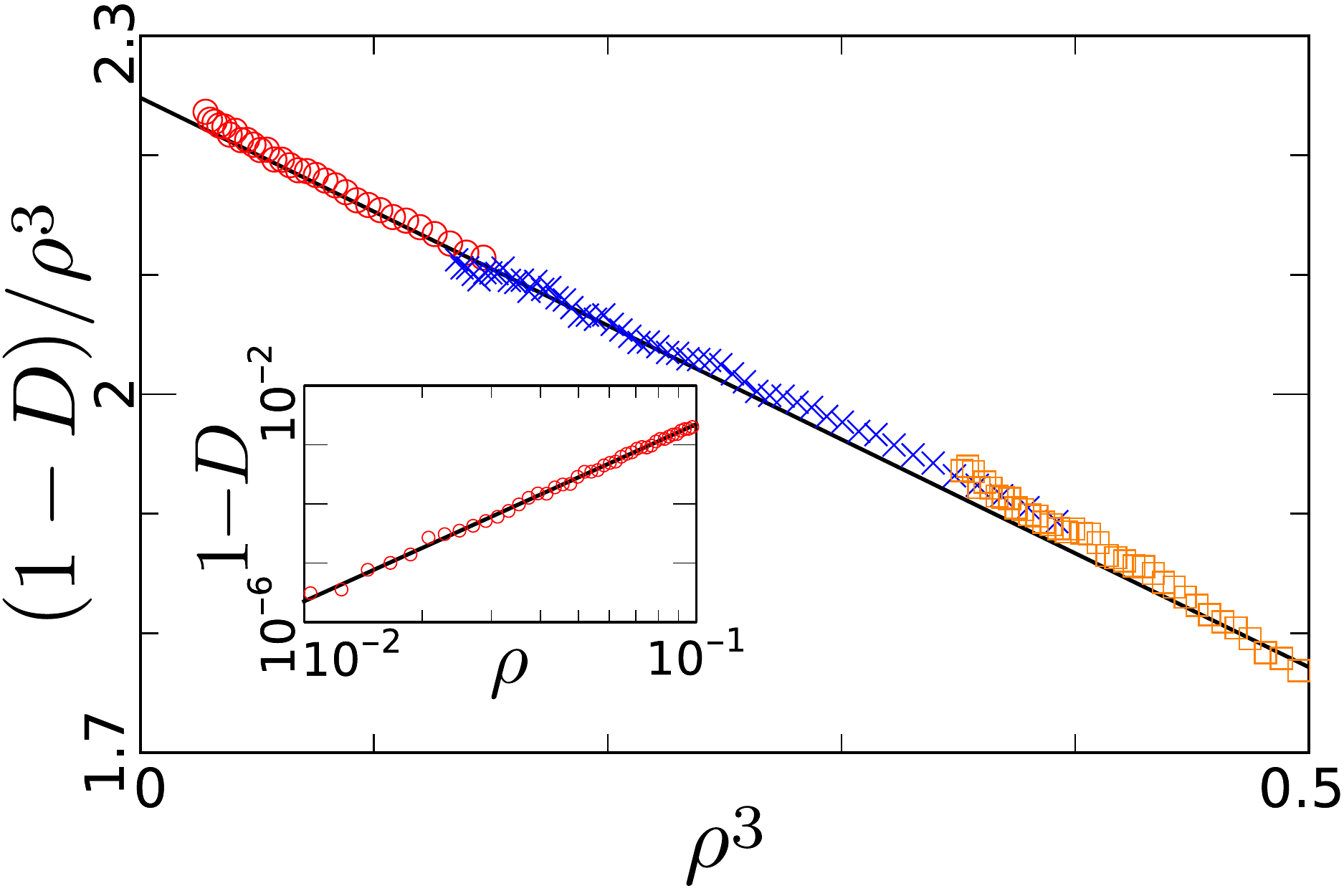}
\caption{The behavior of $ (1- D ) / \rho^3 $ as a function of $ \rho^3 $. The solid line was obtained by fitting the simulation data to the two-term expansion \eqref{D3:small}, viz. $D = 1 - a_1 \rho^3 - a_2 \rho^6 $
 with $ a_1 \approx 2.25 $ and $a_2 \approx -0.953 $.
 The inset confirms the low density asymptotic: $1-D \sim \rho^3$. }
\label{fig:fitting-entirerho}
\end{center}
\end{figure}

\begin{figure}
\begin{center}
\includegraphics[width=80mm]{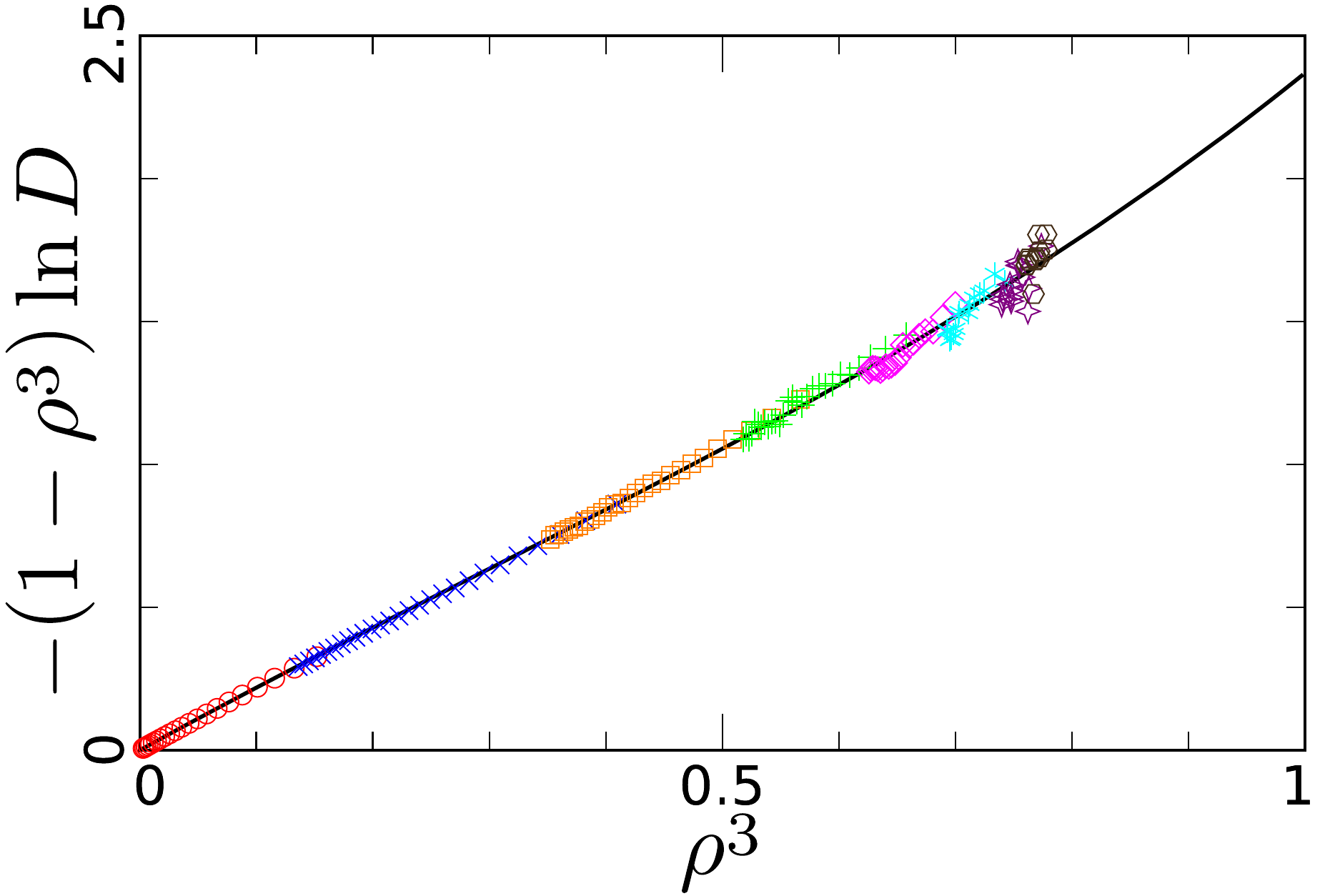}
\caption{Plots of $ - (1-\rho^3) \ln D $ versus $ \rho^3 $.
The solid line corresponds to a fitting curve 
$ - (1-\rho^3) \ln D(\rho) = a_1 \rho^3 - (a_1-a_1^2/2-a_2) \rho^6 + c \rho^9$ with $ c\approx0.796 $.}
\label{fig:nice}
\end{center}
\end{figure}

We now discuss the behavior of the diffusion coefficient $D(\rho)$ in the $\rho\to 0$ limit. The small density behavior of the upper bounds \eqref{D11}--\eqref{D5} is
\begin{equation*}
 1-D(\rho) = A\rho^3 + O(\rho^4)
\end{equation*}
with $A=2$ for the bounds \eqref{D11}--\eqref{D22} and $A=\frac{20}{9}=2.222\dots$ for the bound \eqref{D5}. Therefore one anticipates the expansion of the form
\begin{equation}
\label{D:small}
 1 - D (\rho) = \sum_{\nu\geq 3} A_\nu \rho^\nu . 
\end{equation}
In the bounds \eqref{D14} and \eqref{D22}, the sub-leading term in the expansion \eqref{D:small} does not vanish, $A_4\ne 0$, but our simulations indicate that $A_4=0$ and even $A_5=0$, see figure~\ref{fig:fitting-entirerho}. 
Furthermore, simulation data support a tantalizing conjecture that the diffusion coefficient $D(\rho)$ is an analytical function of $\rho^3$ (figures~\ref{fig:fitting-entirerho} and \ref{fig:nice}), and hence suggest the small density expansion
\begin{equation}\label{D3:small}
 1 - D (\rho) = \sum_{\mu\geq 1} a_\mu \rho^{3\mu} . 
\end{equation}

One would like to compute at least the leading term in the expansion \eqref{D3:small}. The upper bound \eqref{D5} yields the lower bound $a_1\ge \frac{20}{9}$ for the amplitude. To determine better bounds one can try to extract upper bounds for the diffusion coefficient corresponding to larger sets than that we considered before. The first new set to consider is the $3\times 3$ square and since this set includes the rhombus $\cplus$, the corresponding bound is certainly better than the bound \eqref{D5}. 

Although we have not been able to determine $D[3\times 3]$ in the entire density range, we combined our restricted minimization procedure with perturbation techniques using $\rho$ as a small parameter and extracted the amplitude
\begin{equation}
\label{a:3}
a_1[3\times 3] = \frac{65031366433372758107}{29102666440685008803} = 2.23455\dots\, . 
\end{equation}
For the $4\times 4$ square we similarly obtained 
\begin{equation}
\label{a:4}
a_1[4\times 4] = \frac{r}{s} = 2.24065\dots\, . 
\end{equation}
The numbers $r$ and $s$ are 238 digits integers (see \ref{Ap:max}).
Thus $a_1\ge 2.24065\dots$, and this lower bound is just $0.4\%$ smaller
than the estimate $a_1\approx 2.25$ obtained by fitting simulation data.

\section{Asymptotic behavior in the high-density limit}
\label{sec:high-density}

\newcommand{\ba}{\boldsymbol{a}}
\newcommand{\balpha}{\boldsymbol{\alpha}}
\newcommand{\bc}{\boldsymbol{c}}

Our formulas for the diffusion coefficient provide excellent approximations as long as the density is not too high. When $\rho\to 1$, however, the predictions based on the restricted minimization become very bad.
For $ S= \emptyset , ~\ctwotwo $ and $ \cplus $ we obtain the high-density expansions 
\begin{align*}
 D [ \emptyset ] &= 9 v^2 - 18 v^3 + 15 v^4 - 6 v^5 + v^6 , \\ 
 D \Big[\ctwotwo \Big] &= 7 v^2 + 4 v^3 - 163 v^4 + 1384 v^5 -10843 v^6 + O(v^7) , \\ 
 D \Big[\cplus \Big] &= 7 v^2 - \tfrac{ 77 }{34} v^3 - \tfrac{ 168851 }{ 2312 } v^5 
 - \tfrac{ 71610651 }{ 157216 } v^5 + \tfrac{ 23879122075 }{ 10690688 } v^6 + O(v^7) , 
\end{align*}
where $ v = 1- \rho $. In all these examples the diffusion coefficient vanishes algebraically in the $v\to 0$ limit, more precisely as $v^2$. Simulations indicate, however, that the diffusion coefficient decays much faster than $ v^2 $.

The $v\to 0$ behavior of the KA model is very interesting. Earlier simulations led to the conjecture \cite{bib:KA,Kurchan,Parisi} about the break of ergodicity at a certain $\rho_c<1$ and non-standard mechanism for the glass transition. It has been later understood that the KA model on the square lattice is ergodic \cite{TBF,BT17}. (The same is true for the KA models on hyper-cubic lattices \cite{TBF,BT17} and for other kinetically-constraint lattice gases \cite{TB}.) The KA model apparently exhibits the hydrodynamic behavior in the entire density range $0<\rho<1$. (Most rigorous analyses of the KA models were actually focused on the behavior of a tagged particle at equilibrium, rather than on the relaxation to equilibrium, and it was demonstrated \cite{TBF,BT17} that the tracer behaves diffusively.) 

The relationship between the self-diffusion coefficient $\mathcal{D}(\rho)$ and the diffusion coefficient $D(\rho)$ is generally unknown. In the symmetric simple exclusion process, the self-diffusion coefficient satisfies $\mathcal{D}(\rho)<D(\rho)=1$. It seems plausible that this inequality 
\begin{equation}
\label{ineq:DD}
\mathcal{D}(\rho)<D(\rho)
\end{equation}
holds for general exclusion processes including the KA model. 

The dependence of the self-diffusion coefficient of the KA model on the density has attracted considerable attention \cite{TBF,TB} and it was shown that the self-diffusion coefficient vanishes faster than any power of $v$ in the $v\to 0$ limit. An interesting connection between the KA model and bootstrap percolation together with exact results for the latter \cite{AL,Holroyd} lead to a more precise prediction \cite{TBF}
\begin{equation}
\label{SD:exp}
\lim_{\rho\to 1} (1-\rho) \ln\mathcal{D}(\rho)= -\frac{\pi^2}{9}
\end{equation}

One could anticipate a similar asymptotic behavior of the diffusion coefficient:
\begin{equation}
\label{D:exp}
\lim_{\rho\to 1} (1-\rho) \ln D(\rho)= -C
\end{equation}
Our simulations indeed support \eqref{D:exp}, see figure~\ref{fig:nice}. Assuming additionally the inequality \eqref{ineq:DD} one arrives at the inequality $C < \frac{\pi^2}{9} $. The fitting curve in figure~\ref{fig:nice} indicates $ C \approx 0.79 $, satisfying $C < \frac{\pi^2}{9} = 1.0966\dots $\,.
 
Our restricted minimization procedure requires solving a finite number of linear equations. The total number of equations is $2^{|S|}$. In the class of symmetric convex sets, we performed exact minimizations when $|S|=1,4,5$, but we have not succeeded for the next symmetric convex set, namely for the $3\times 3$ square. 
The difficulty in the computation is as follows: One has to calculate $ \langle Q \rangle $, the expectation value of the functional $Q$ in the equilibrium state. In other words, one calculates $ \sum \text{(weight)} \times Q $ over all configurations of a certain finite subset of $ \mathbb Z^2 $, e.g. $ 2 n ( 2 n -1 )$ sites in the case of the $n\times n$ square (with $n\geq 2$). Therefore the number of summands in $ \langle Q \rangle $, $ 2^{ 2 n ( 2 n -1 ) } $, dwarfs the number of equations, $2^{n^2}$, see \ref{Ap:22}.
For $ n=3 $, the number of terms $2^{30}$ is already huge. One gets the same problem for other large sets.

It might be possible to deal with the sum $ \langle Q \rangle $ by devising a more efficient algorithm. We leave this for the future, and here we show how to handle larger sets if we merely want to extract the behavior in the $\rho\to 1$ limit. We accomplish this by combining the variational method with perturbation techniques. We use the density $ v = 1- \rho $ of vacancies as the small parameter. The basic idea is to obtain the expansion of $ q [ S ] $ in powers of $v$. First we expand the test function 
\begin{align}
\varphi (\tau) = \varphi_0 (\tau)+ v \varphi_1 (\tau)+ v^2 \varphi_2 (\tau) + \cdots 
\end{align}
and the functional
\begin{align}
 \langle Q ( \varphi ) \rangle = Q_3 v^3 + Q_4 v^4 + Q_5 v^5 + \cdots 
\label{eq:<Q>=v3Q3...}
\end{align}
We notice that the expansion \eqref{eq:<Q>=v3Q3...} begins from the cubic term. This is due to the fact that the hopping in our model is allowed only if there are more than two vacant sites. Then we perform minimizations, term by term, starting from $ Q_3 $. The biggest advantage of the perturbation approach is that 
the effective numbers of summands in $ Q_j $ $(j\ge 3)$ are much smaller than 
the number of terms in the full average $ \langle Q ( \varphi ) \rangle $. In principle one can choose any $ S $. We argued before in favor of choosing symmetric convex sets. 
Squares constitute one simple class of such sets. In the case of the $ n\times n $ square, the numbers of non-zero terms in $ Q_3 $, $ Q_4 $, and $ Q_5 $ are $ 36 $, $\simeq 144n^2 $, and $ \simeq 288 n^4 $, respectively, 
much smaller than $2^{ |S| } = 2^{ 2n(2n-1) } $.

Performing minimizations for $ Q_3 $ and $ Q_4 $, we obtained the following asymptotic behaviors of the upper bounds (see \ref{Ap:max} for details):
\begin{align}
\begin{split}
 D [2 \times 2 ] &= 7 v^2 + 4 v^3 + O(v^4) , \\ 
 D[3\times 3] &= v^2 + \tfrac{ 14408 }{ 307 } v^3 + O(v^4), \\ 
 D [4 \times 4 ] 
 &= 0 v^2 + \tfrac{ 453068679808 }{ 66311971451 } v^3 + O(v^4), \\
 D [ 5 \times 5 ] &= 0 v^2 + \tfrac 1 2 v^3 + O(v^4), \\ 
 D [ 6 \times 6 ] & =0 v^2 + 0 v^3 + O(v^4 ) . 
\end{split}
\label{eq:D[nbyn]=...}
\end{align}
Thus the upper bound vanishes as $v^2$ when $n=1,2,3$; as $v^3$ when $n=4,5$; and at least as $v^4$ when $n\geq 6$. These observations suggest that the true diffusion coefficient vanishes faster than algebraically as $v\to 0$.

\section{Details of simulations}
\label{sec:simulations}

For the Monte Carlo simulations, we imposed cylindrical boundary conditions on a finite square lattice, as illustrated in figure~\ref{fig:cylinder}. We think that the virtual sites are in equilibrium with densities $ \rho_0 $ and $ \rho_L $. The transition rates on the left boundary are
\begin{align*}
 \tau_{ 1, j } = 0 \to 1 \qquad
 \textrm{with\ rate}\quad \rho_0 ( 1 - \rho_0^3 ) 
\end{align*} 
if at least one of the sites $ (1,j\pm1)$ and $ (2,j) $ is empty, and
\begin{align*}
 \tau_{ 1, j } = 1 \to 0 \qquad
 \textrm{with\ rate}\quad ( 1 - \rho_0 ) ( 1 - \rho_0^3 ) 
\end{align*}
if at least one of the sites $ (1,j\pm1)$ and $ (2,j) $ is empty. Similarly on the right boundary:
\begin{align*}
\tau_{ L-1, j } = 0 \to 1 \qquad \textrm{with\ rate}\quad \rho_L ( 1 - \rho_L^3 ) 
\end{align*} 
if at least one of the sites $ (L-1,j\pm1)$ and $ (L-2,j) $ is empty, and 
\begin{align*}
\tau_{ L-1, j } = 1 \to 0 \qquad
 \textrm{with\ rate}\quad ( 1 - \rho_L ) ( 1 - \rho_L^3 ) 
\end{align*} 
if at least one of the sites $ (L-1,j\pm1)$ and $ (L-2,j) $ is empty. 

In simulations, we chose the following boundary densities,
 which were used to plot the numerical data of the diffusivity in the previous sections: 
\begin{align*}
\begin{array}{cllll}
\text{marker} & \rho_0 & \rho_L & \text{time average} \\ \hline 
 \includegraphics[width=3.5mm]{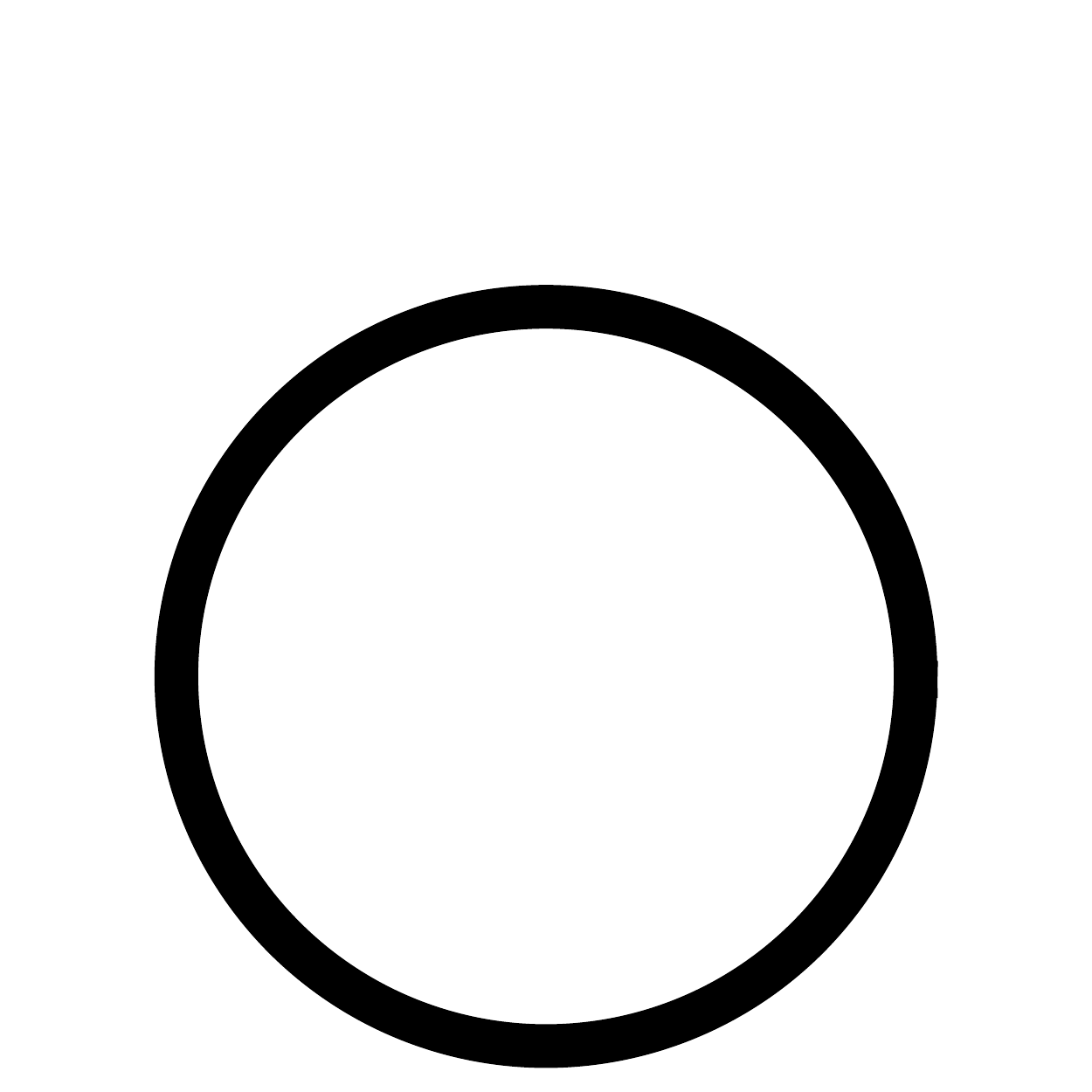} & 0.6 & 0 & t_0 \le t \le 3 t_0 \\ 
 \includegraphics[width=3.5mm]{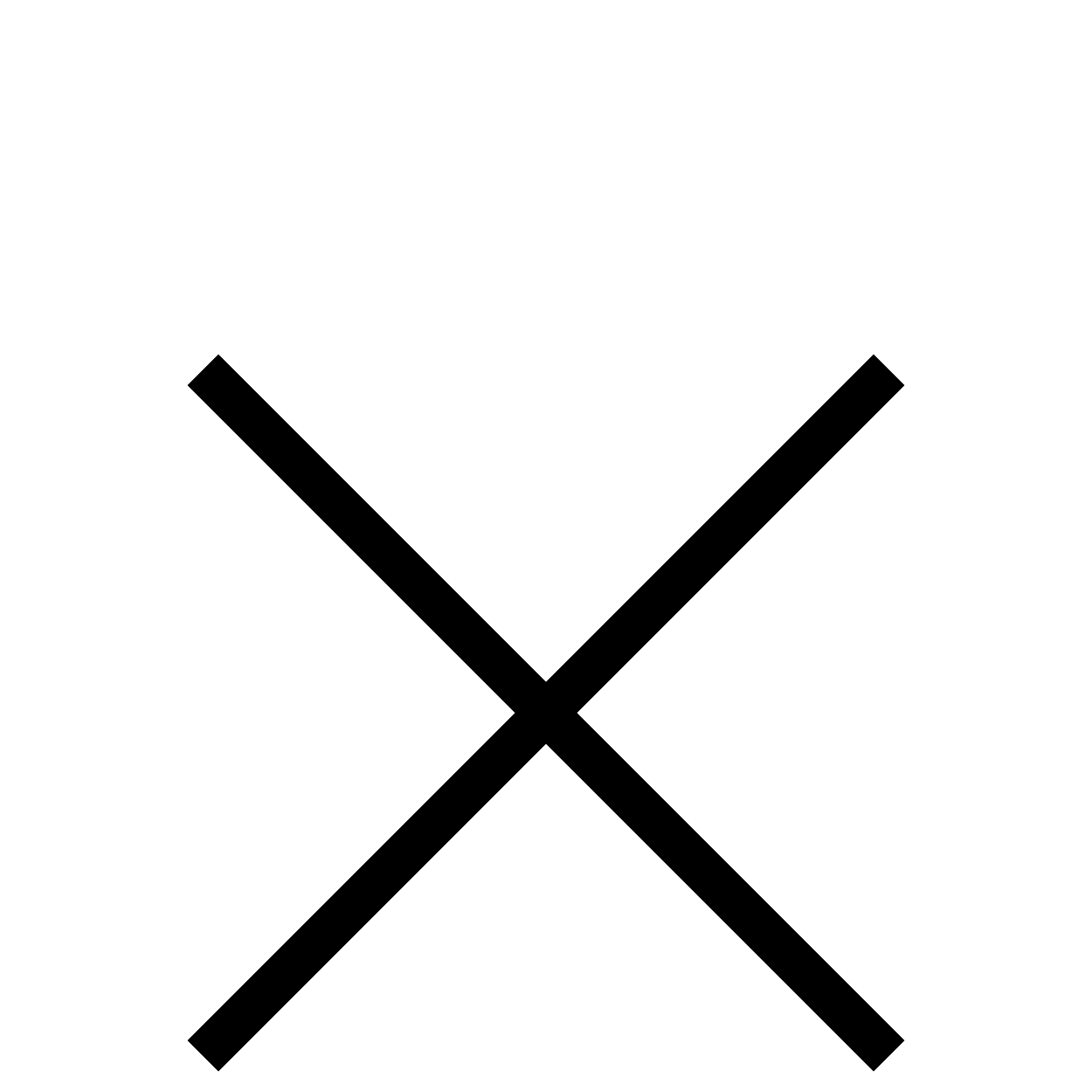} & 0.8 & 0.5 & t_0 \le t \le 5 t_0 \\ 
 \includegraphics[width=3.5mm]{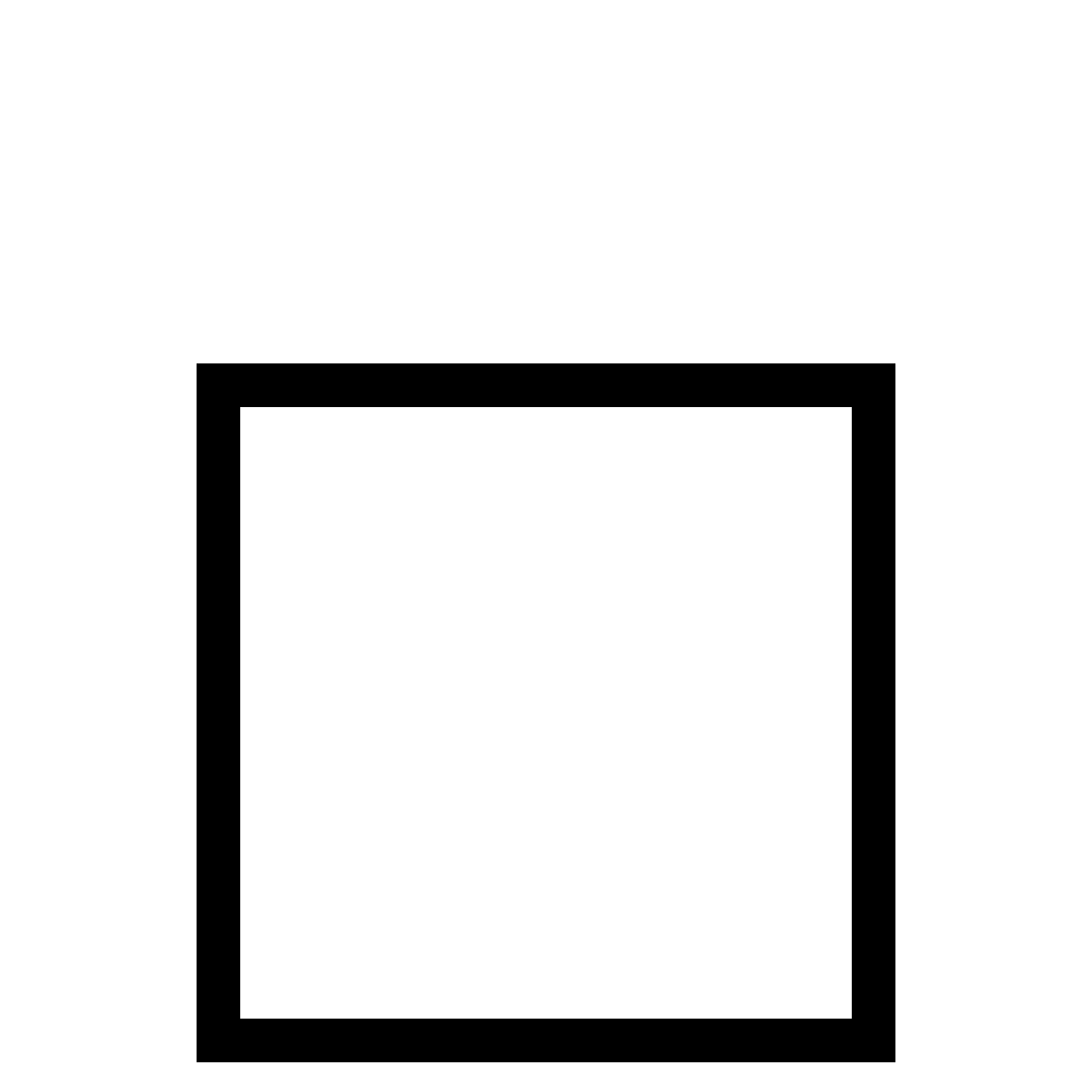} & 0.9 & 0.7 & t_0 \le t \le 7 t_0 \\ 
 \includegraphics[width=3.5mm]{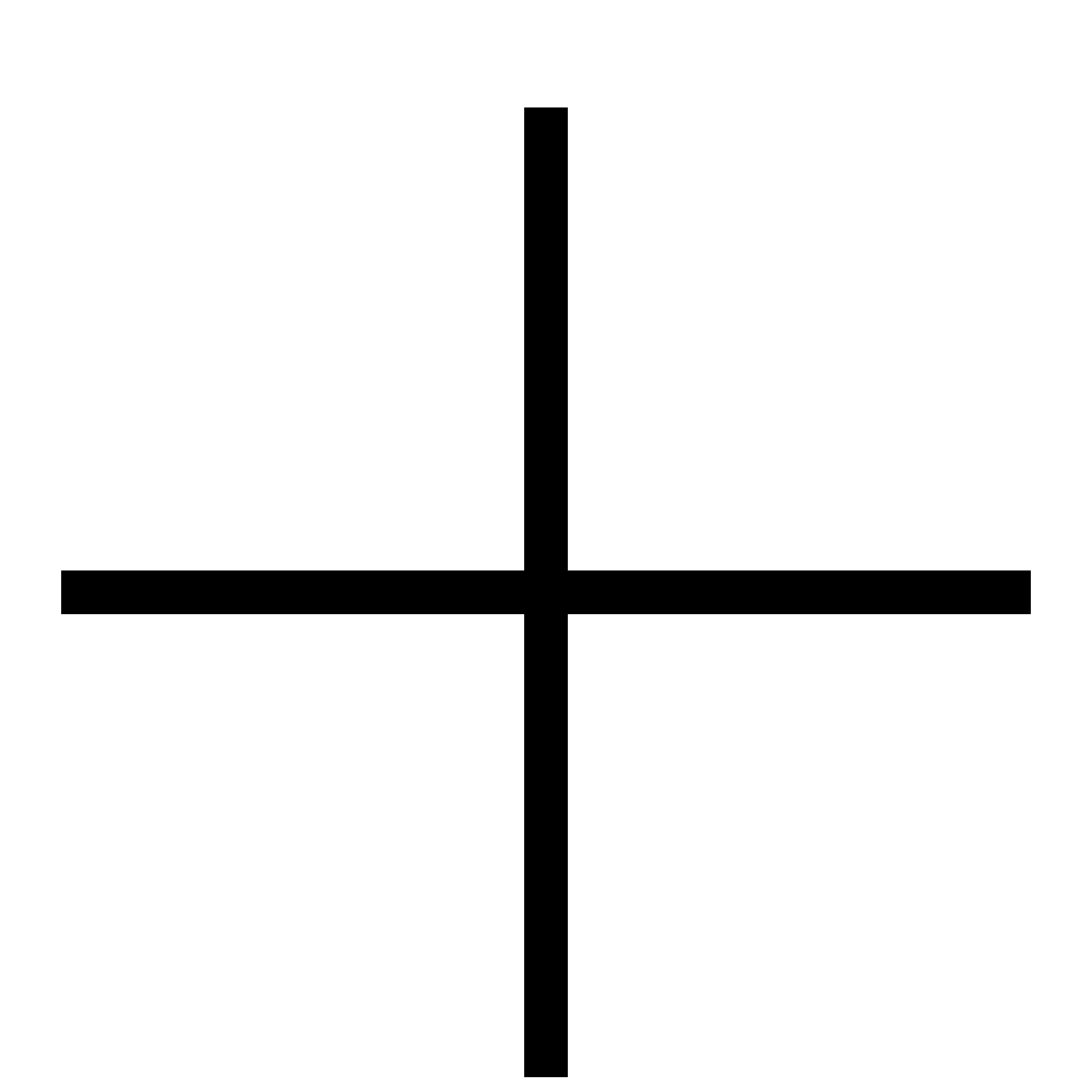} & 0.95 & 0.8 & t_0 \le t \le 9 t_0 \\ 
 \includegraphics[width=3.5mm]{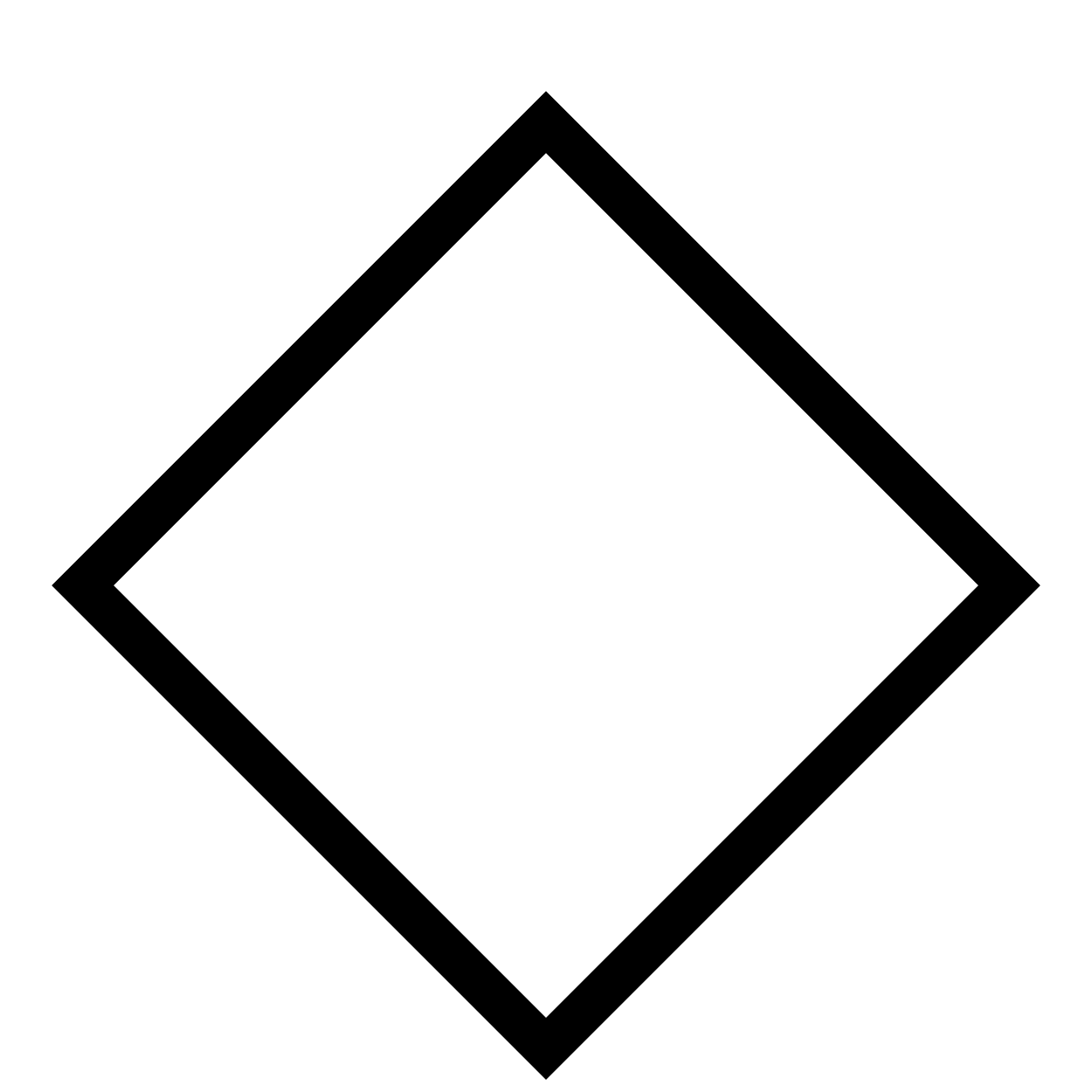} & 0.97 & 0.85 & t_0 \le t \le 11 t_0 \\ 
 \includegraphics[width=3.5mm]{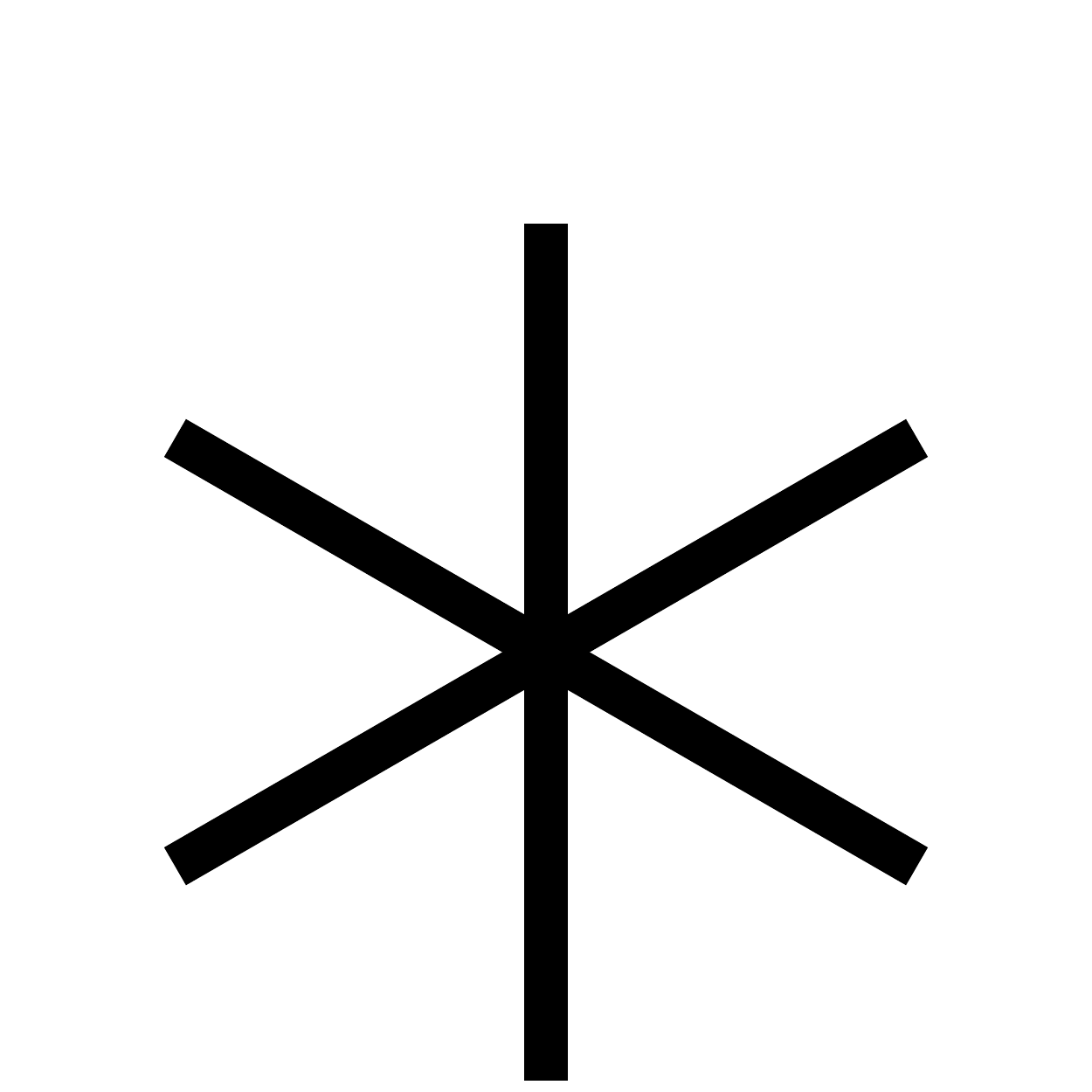} & 0.98 & 0.88 & t_0 \le t \le 13 t_0 \\ 
 \includegraphics[width=3.5mm]{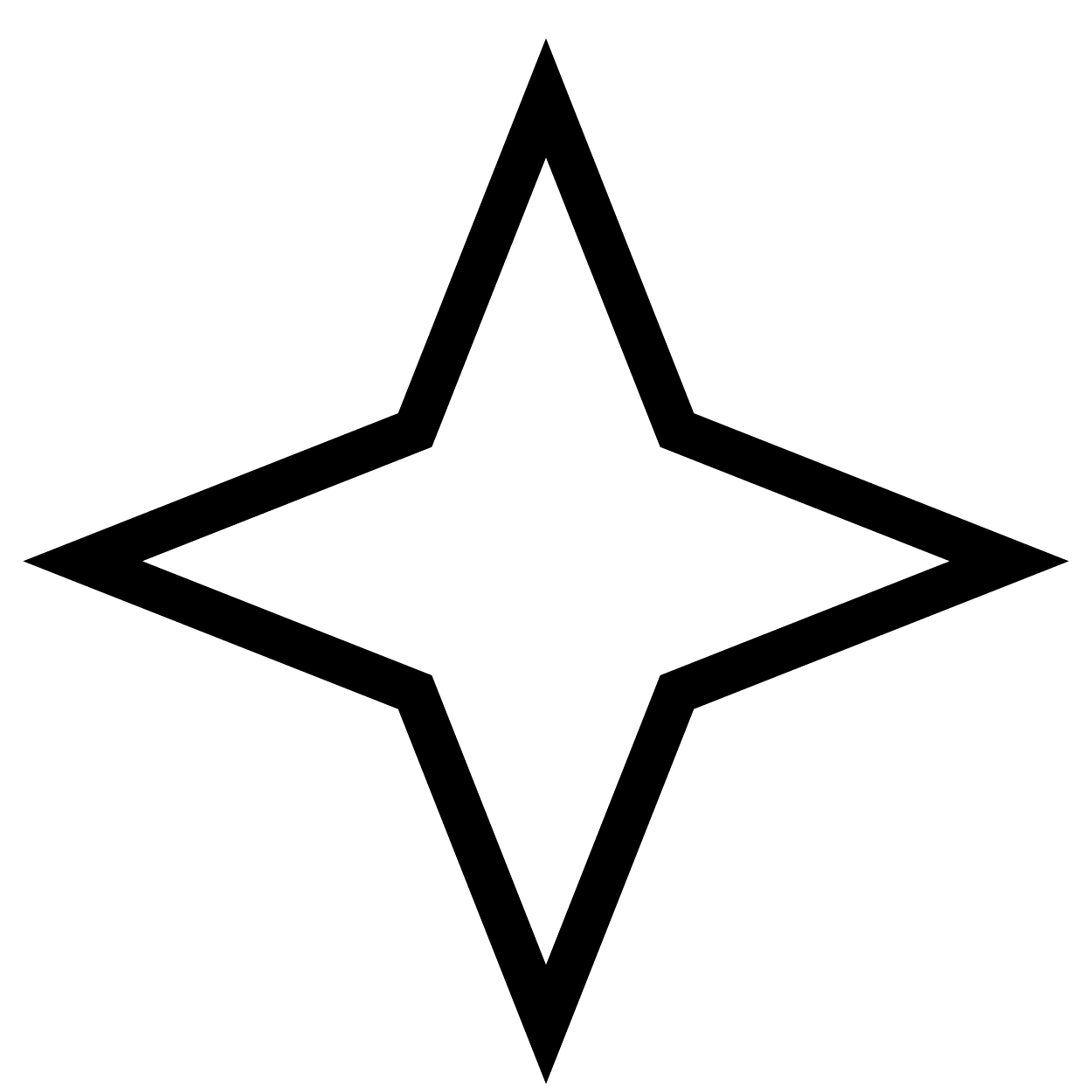} & 0.99 & 0.9 & t_0 \le t \le 15 t_0 \\
 \includegraphics[width=3.5mm]{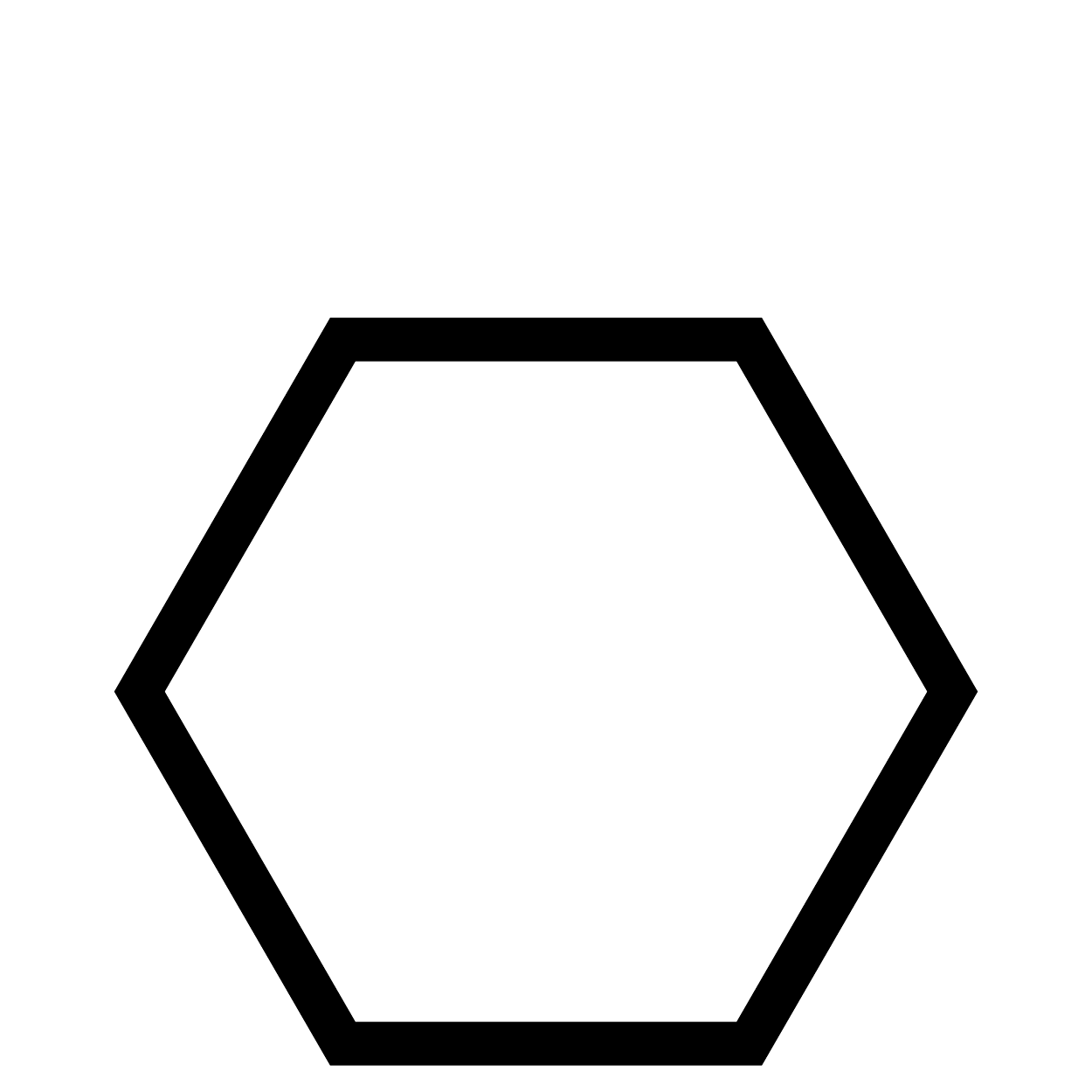} & 0.99 & 0.91 & t_0 \le t \le 16 t_0 
\end{array}
\end{align*}
Here the intervals over which we took average are also shown with $ t_0 = 5\times 10^6 $. Note that longer simulation times are needed for higher boundary densities to obtain accurate results. Indeed, the diffusion coefficient is extremely small as $ \rho \to 1 $, therefore the current is very small and relative statistical errors become large, as compared to low-density cases. Concerning the system size, we fixed the vertical length, $L_y = 200$, and varied the horizontal length: $ L=16,32,\dots , 512 $. For the plots, we used the data corresponding to $ L=256 $. 

The equilibrium stationary state in the infinite lattice or in the periodic boundary conditions (torus) is simply described by the product measure. In our cylindrical boundary conditions, however, details of the stationary state are highly non-trivial except for the special case $ \rho_0 = \rho_L $, where the same product measure with density $ \rho = \rho_0 = \rho_L $ is valid. 
When $ \rho_0 \neq \rho_L $, the horizontal current does not vanish even in the stationary state. By taking an average of the difference between the rightward and leftward instantaneous currents \eqref{eq:instant10}, \eqref{eq:instant-10}, 
we find that the current in the horizontal direction has the form 
\begin{align}
\label{eq:Jij=}
 J_{i,j} = 
 \Big\langle P^{ ( 1, 0 ) }_{i,j} (\tau) - P_{i,j}^{ (-1, 0 ) } (\tau) \Big\rangle 
 = \big\langle ( \tau_{ i,j } - \tau_{ i+1,j } ) H_{i,j} (\tau ) \big\rangle 
\end{align}
with $ H $ defined in equation~\eqref{eq:H=}. The current of vertical direction vanishes because of the symmetry.

We extract the density profile $ \rho(x) $ from simulations and then numerically evaluate $\frac{d\rho}{dx}$. The current, that is the expectation value \eqref{eq:Jij=}, is also directly observed. Using data for $ J $ and $ \frac{ d \rho }{ d x }$ we evaluate the diffusion coefficient via $ D \text{\bf(} \rho (x) \text{\bf)} = - J L \big/ \big[ \frac{d \rho }{ d x } (x) \big] $ following from the Fick law \eqref{eq:J=}. This method was used in preparing figures~\ref{fig:diffusivity}, \ref{fig:fitting-entirerho}, \ref{fig:nice}.

\section{Discussion}
\label{sec:conclusions}

In a previous study \cite{bib:AKM} we developed an approximation scheme that yields upper bounds for the diffusion coefficient of lattice gases with known equilibrium properties, specifically we used a one-dimensional 
generalized exclusion processes with maximal occupancy number 2. The scheme is based on the exact variational formula for the diffusion coefficient, so it is a variant of the Ritz method as the minimization is performed on finite-dimensional sub-spaces. 

In this article we showed how to apply this scheme to two-dimensional lattice gases. As an example, we chose the Kob-Andersen with $m=2$ on the square lattice: A hop to an empty site is allowed only if before and after the hop the particle has at least $m=2$ empty neighbors. This lattice gas does not satisfy the gradient property, and therefore the diffusion coefficient is impossible to obtain analytically. The KA model is a very useful toy model mimicking dynamics of glasses. The model exhibits extremely slow `glassy' relaxation in the $\rho\to 1$ limit, e.g. the coefficient of self-diffusion vanishes faster than any power of the density $v=1-\rho$ of vacancies, see \eqref{SD:exp}. This suggests that the diffusion coefficient can also be anomalously small when $v\ll 1$ which we indeed confirm numerically. Overall, the KA model provides a stringent test for approximation approaches since these approaches cannot detect a non-analytic behavior.

We derived upper bounds for the diffusion coefficient of the KA model with $ d=m=2 $, by using the variational approach. For some simple sets $ S\subset \mathbb Z^2 $, e.g. $ S= \emptyset, \square $,
and vertical hard rods, the upper bounds coincide with the naively derived diffusivity $ (1-\rho^3)^2 $ under the mean-field assumption. We performed calculations for symmetric convex finite sets $S$, which give improved bounds. 
The results are very accurate for moderate densities, e.g. for $\rho\leq 0.5$.

 The dimension of the sub-space of functions over which we perform the minimization is $2^{|S|}$, where $|S|$ is the cardinality of the set $S$. Performing the minimization is formally simple since the functional is quadratic, so one ends up with linear equations. The number of equations grows as $2^{|S|}$. The chief numerical obstacle is that the number of summands in $ \langle Q \rangle $ grows much faster than $2^{|S|}$. For instance, for the $n\times n$ square the dimension of the number of equations is $2^{n^2}$, while the number of summands in $\langle Q \rangle $ is $2^{2n(2n-1)}$. For the $3\times 3$ square the number of summands is $2^{30}$ which is already on the verge of what is feasible using straightforward algorithms.
 
 As expected, in the $v\to 0$ limit the relative discrepancy between the bounds and the diffusion coefficient (determined through simulations) is huge. The cause of the problem is that the dimensionality of the `relevant' sub-space diverges as $v\to 0$. To probe the behavior in the $v\to 0$ limit we combined the variational approach with perturbation techniques (using $v$ as a small parameter). This allowed us to reduce the number of summands and gave a power series expansion of $D$ for larger sets $S$. Specifically, we extracted the expansions for $n\times n$ squares with $n\leq 6$. The resulting upper bounds vanish algebraically, but with exponents growing with $n$. This provides an indication that the actual behavior might be non-analytic, and perhaps $D$ indeed vanishes according to equation~\eqref{D:exp} resembling the behavior of the coefficient of self-diffusion. 
 
The diffusion coefficient appears to admit an expansion in powers of $\rho^3$ rather than $\rho$. Although we do not have theoretical evidence in favor of this tantalizing property, it is supported by our simulation results. If this conjecture is true, the self-diffusion coefficient is probably also a function of $\rho^3$. 

In this article, we limited ourselves to the simplest non-trivial KA model, viz. the KA model on the square lattice with $m=2$. For the KA models on $\mathbb{Z}^d$, the interesting range is $2\leq m\leq d$. 
For $ S=\emptyset $, the upper bound is the same as the mean-field prediction, which can be obtained by straightforward extension of the argument presented at the end of section \ref{sec:model}. The answer is particularly simple for the KA model on $\mathbb{Z}^d$ with $m=2$:
\begin{equation*}
D = \left( 1-\rho^{2d-1} \right)^2 . 
\end{equation*}
The mean-field prediction for general $d$ and $m$ is 
\begin{equation}
\label{eq:general}
D = \left[1-\sum_{k=0}^{m-1}\binom{2d-1}{k}\rho^{2d-1-k}(1-\rho)^k\right]^2 . 
\end{equation}
On the cubic lattice, the two interesting cases are $m=2$ and $m=3$ (originally considered by Kob and Andersen \cite{bib:KA}), where the corresponding predictions are
\begin{equation*}
D = 
\begin{cases}
\left( 1-\rho^5\right)^2 & m=2\, , \\
\left( 1-5\rho^4 +4\rho^5\right)^2 & m=3\, . 
\end{cases}
\end{equation*}

We anticipate that the result of the minimization coincides with \eqref{eq:general}, 
when one chooses the class $S$ of functions depending on one site. 
To perform a minimization procedure relying on symmetric convex sets one should probably start with (non-trivial) ``hyper-rhombus'' for which $|S|=2d+1$. For instance, the hypercube of length two has cardinality $|S|=2^d$ which is larger than $2d+1$ for $d\geq 3$. We expect that the minimization over hyper-rhombus functions with $|S|=2d+1$ gives a better explicit upper bound than \eqref{eq:general}.

It would be interesting to establish connection with other variational approaches (see e.g. \cite{bib:Zaluska-Kotur_etal}) applied to the calculation of the diffusion coefficient in lattice gas models. Finally, we mention that there is a variational formula for the self-diffusion coefficient \cite{bib:Spohn1991}. This more special transport coefficient is unknown even in the simplest gradient lattice gases such as the symmetric simple exclusion process in two dimensions, so it would be interesting to obtain upper bounds for the self-diffusion coefficient using the procedure described in this paper.

\section*{Acknowledgements}
We are grateful to G. Biroli for discussions and to S. Mallick for a careful reading of the manuscript.

\appendix

 \section{$1\times 1$ square}
 \label{Ap:11}

\begin{figure}
\begin{center}
 \includegraphics[width=40mm]{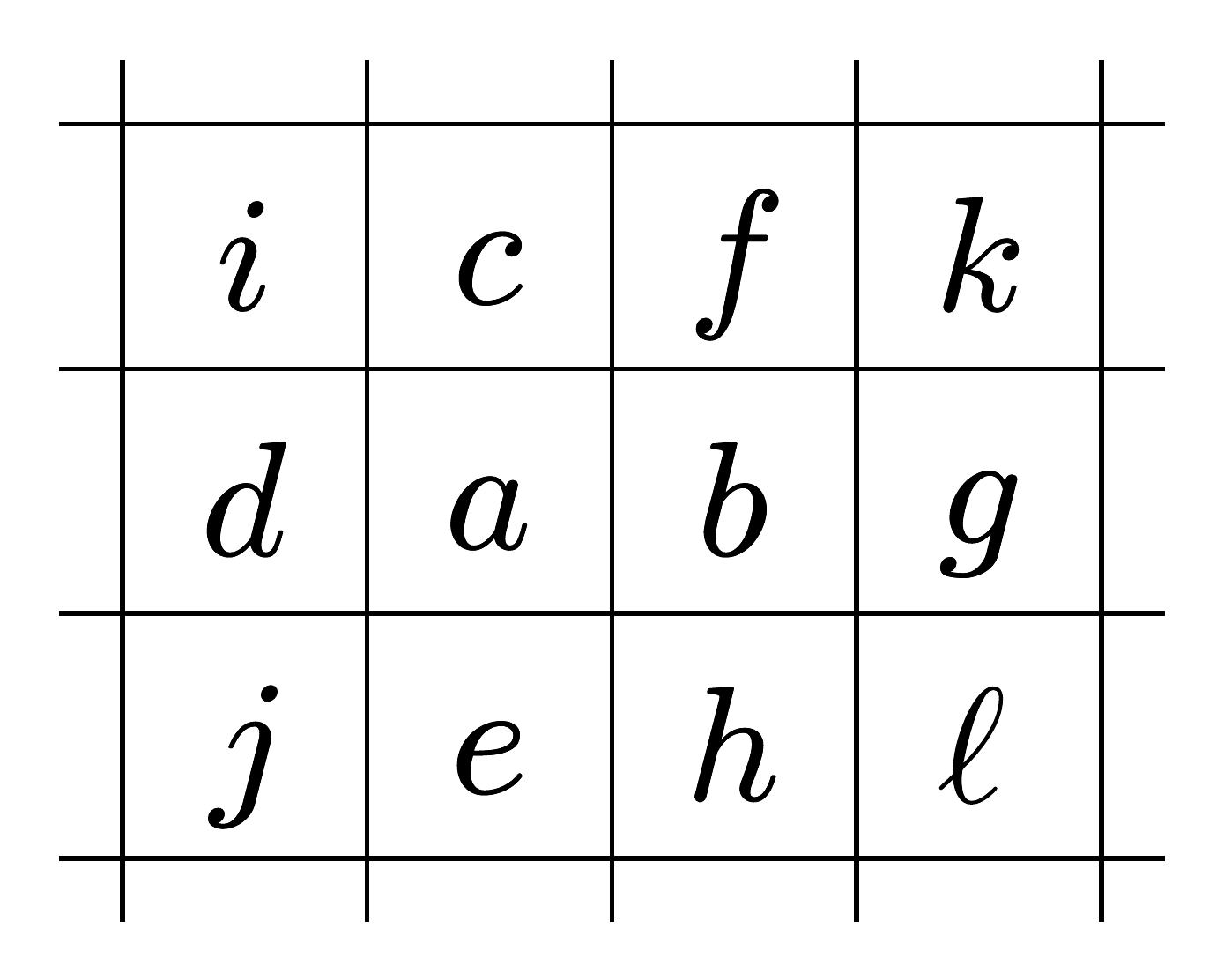}
\quad\includegraphics[width=32mm]{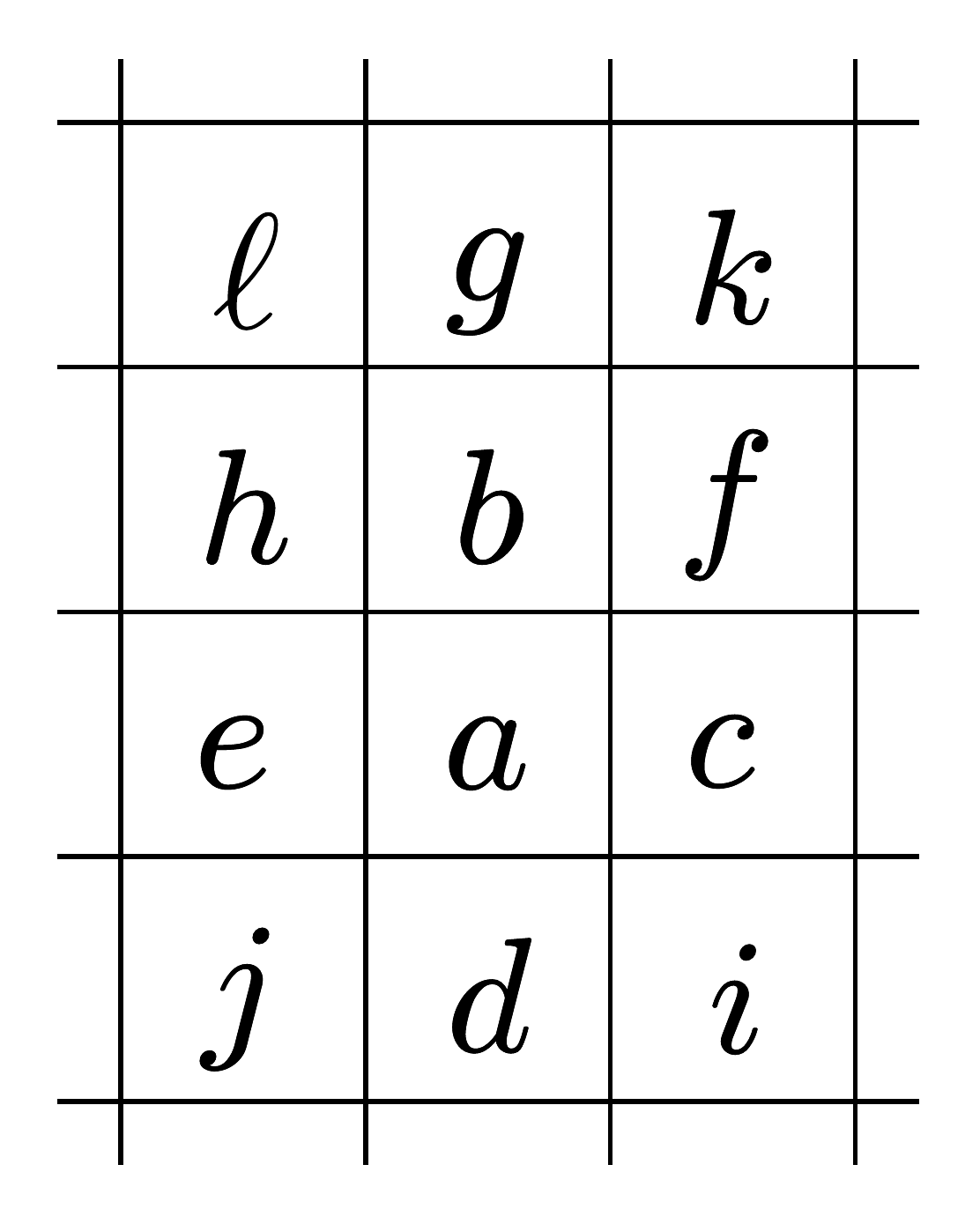}
 \caption{Notations for the formulas $\langle Q^{ (\pm1,0) }\rangle $ (left) and $\langle Q^{ (0, \pm1) } \rangle $ (right). 
We use letters from $ a $ to $ h $ for the $1\times 1 $ square [equations \eqref{eq:1by1Qpm10=} and \eqref{eq:Q0pm1=}],
and from $a$ to $ \ell $ for the $ 2\times 2 $ square [equations \eqref{eq:2times2:Qpm10=} and \eqref{eq:2times2:Q0pm1=}]. We always set $a=\tau_{0,0}$, so e.g. $\ell=\tau_{2,-1}$ on the left panel and $\ell=\tau_{-1,2}$ on the right panel. 
 }\label{fig:alphabet}
\end{center}
\end{figure}

\begin{figure}
\begin{center}
 \includegraphics[width=40mm]{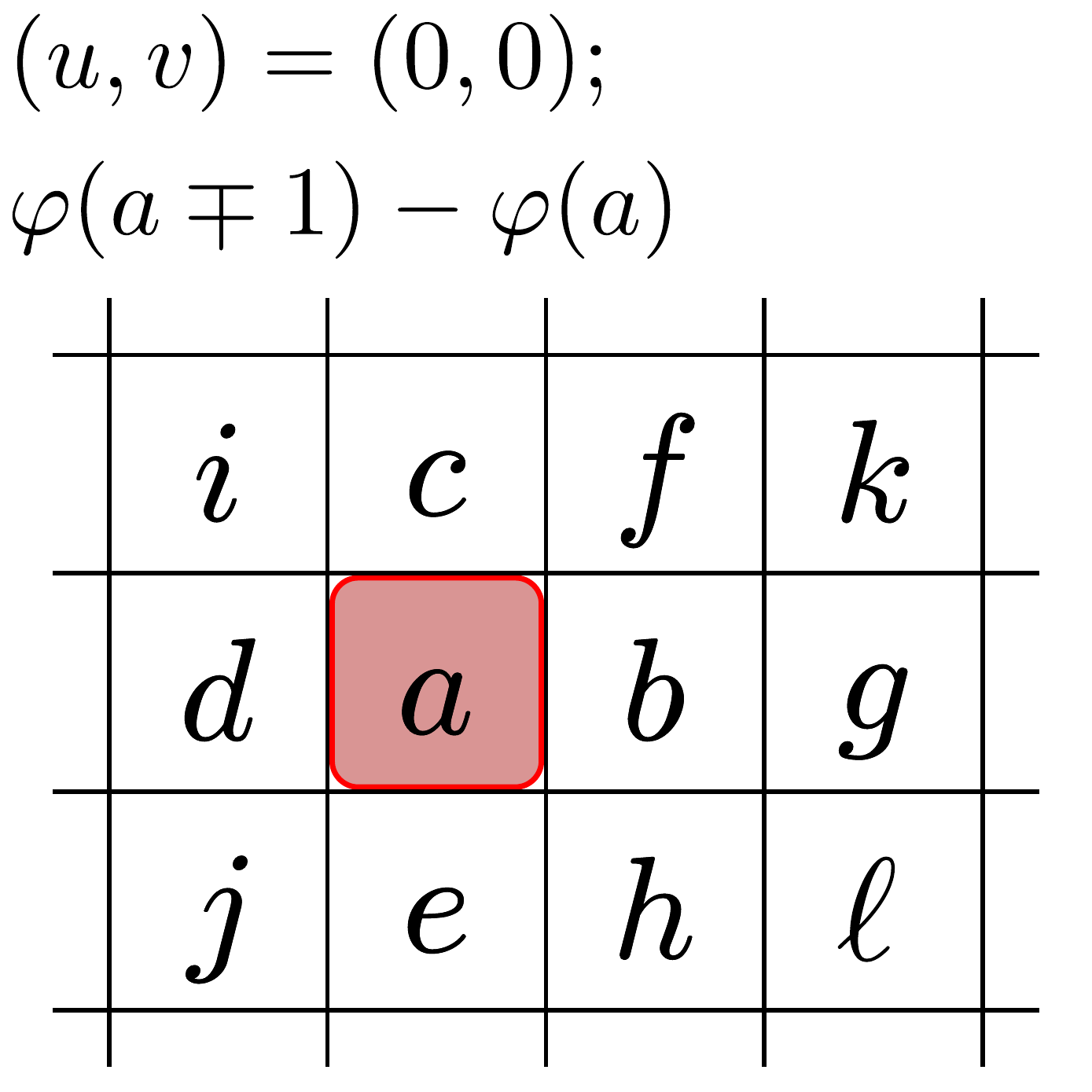}
\quad \includegraphics[width=40mm]{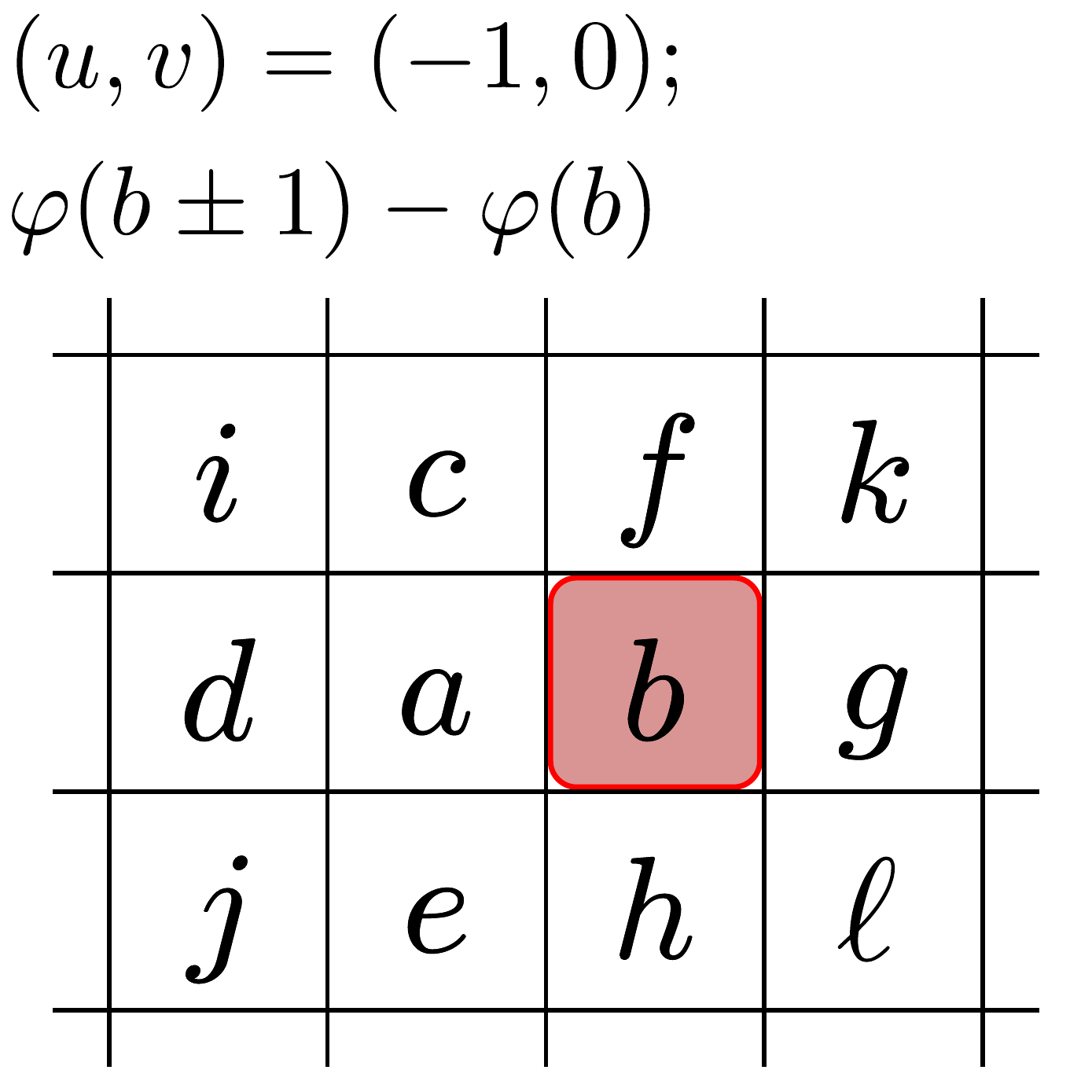}
 \caption{Illustration for the functionals $ Q^{ ( \pm 1, 0 ) } $ given by \eqref{eq:1by1Qpm10=}. The summands indexed by $ (u,v) = (0,0) $ and $ (-1,0) $ in \eqref{eq:Qalphabeta} provide non-vanishing contributions. 
The function $ \varphi \big( A_{ u,v } \, \tau^{ ( \alpha, \beta ) }\big) - \varphi ( A_{ u,v } \, \tau) $
depends only on the shaded sites. }
\label{fig:11covering}
\end{center}
\end{figure}

For the $1\times 1$ square, schematically $ S= \square$, functions $ \varphi $ depend on one site $ \tau_{ 0,0 } $
and we shortly write $ \varphi( \tau ) = \varphi( \tau_{0,0} ) $. The expectation value $ \langle Q^{ (\pm 1, 0 )} \rangle $ reads
\begin{align}
\label{eq:1by1Qpm10=}
 \langle Q^{ (\pm 1, 0 )} ( \varphi ) \rangle &= \sum_{a,\dots, h\in \{ 0,1 \} } 
 W_a \cdots W_h\, P^{ (\pm 1,0) } \left[ \pm 1 - \Phi^{\pm}(a,b)\right]^2 
\end{align}
where we replaced $ \tau_{ 0,0 } , \tau_{ 1 ,0 } , \dots $ by letters $ a,b,\dots$ as shown in figure~\ref{fig:alphabet} (left) and denoted by $W_0$ [resp. $W_1$] the probability that the site is empty [resp. occupied], that is 
\begin{align}
\label{WW}
W_0 =1-\rho, \qquad W_1=\rho.
\end{align}
We also used the shorthand notation
\begin{align*}
\Phi^{\pm}(a,b) = \varphi(a\mp 1) - \varphi(a) + \varphi(b\pm 1) -\varphi(b).
\end{align*}
The leftward and rightward hopping rates can be written as 
\begin{equation}
\label{eq:P10,P-10} 
\begin{split}
 P^{ (1,0) } &= a ( 1 - b ) ( 1 - cde) ( 1 - fgh ), \\
 P^{ (-1,0) }& = b ( 1 - a ) ( 1 - cde ) (1-fgh).
\end{split}
\end{equation}
Only two summands indexed by $ (u,v) = (0,0) $ and $ (-1,0) $ in \eqref{eq:Qalphabeta} contribute (see figure~\ref{fig:11covering}). For any other pair of $ (u,v) $, we have 
$ \varphi \big( A_{ u,v } \, \tau^{ ( \pm 1 , 0 ) }\big) = \varphi ( A_{ u,v } \, \tau) $. 
 
For the `vertical' functionals, one finds 
\begin{align}
\label{eq:Q0pm1=}
 \langle Q^{ ( 0,\pm 1 )} ( \varphi ) \rangle &= \sum_{a,\dots, h\in \{ 0,1 \} } 
 W_a \cdots W_h\, P^{ (0,\pm 1) } \left[ \Phi^{\pm}(a,b) \right]^2. 
\end{align}
Here we transposed the alphabetic notations of the sites, see figure~\ref{fig:alphabet}(right); 
the upward and downward hopping rates are 
\begin{equation}
\label{eq:P01,P0-1} 
\begin{split}
 P^{ (0,1) } &= a ( 1 - b ) ( 1 - cde) ( 1 - fgh ), \\
 P^{ (0,-1) } &= b ( 1 - a ) ( 1 - cde ) (1-fgh) . 
 \end{split}
\end{equation}
The sums in \eqref{eq:1by1Qpm10=} and \eqref{eq:Q0pm1=} are calculated to give
$ \langle Q^{ (\pm 1, 0 )} (\varphi) \rangle = \rho(1-\rho) (1-\rho^3)^2 $ and $ \langle Q^{ ( 0, \pm 1 )} (\varphi) \rangle = 0 $. Hence $ D[ \square ] = (1-\rho^3)^2 $ which is the same result as the mean-field approximation \eqref{MF}.

\section{$2\times 2$ square}
\label{Ap:22}

\begin{figure}
\begin{center}

 \includegraphics[width=42mm]{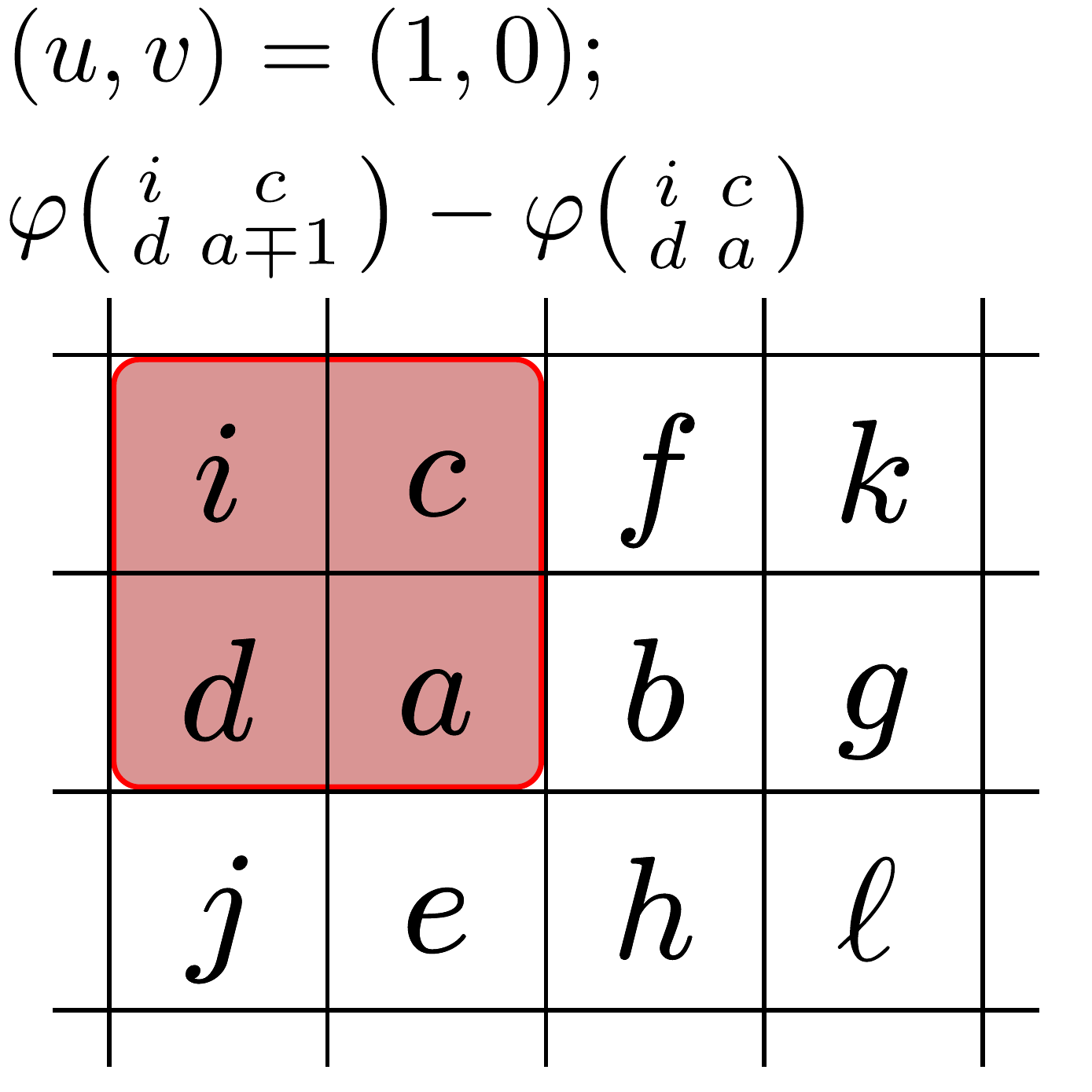} \quad 
 \includegraphics[width=42mm]{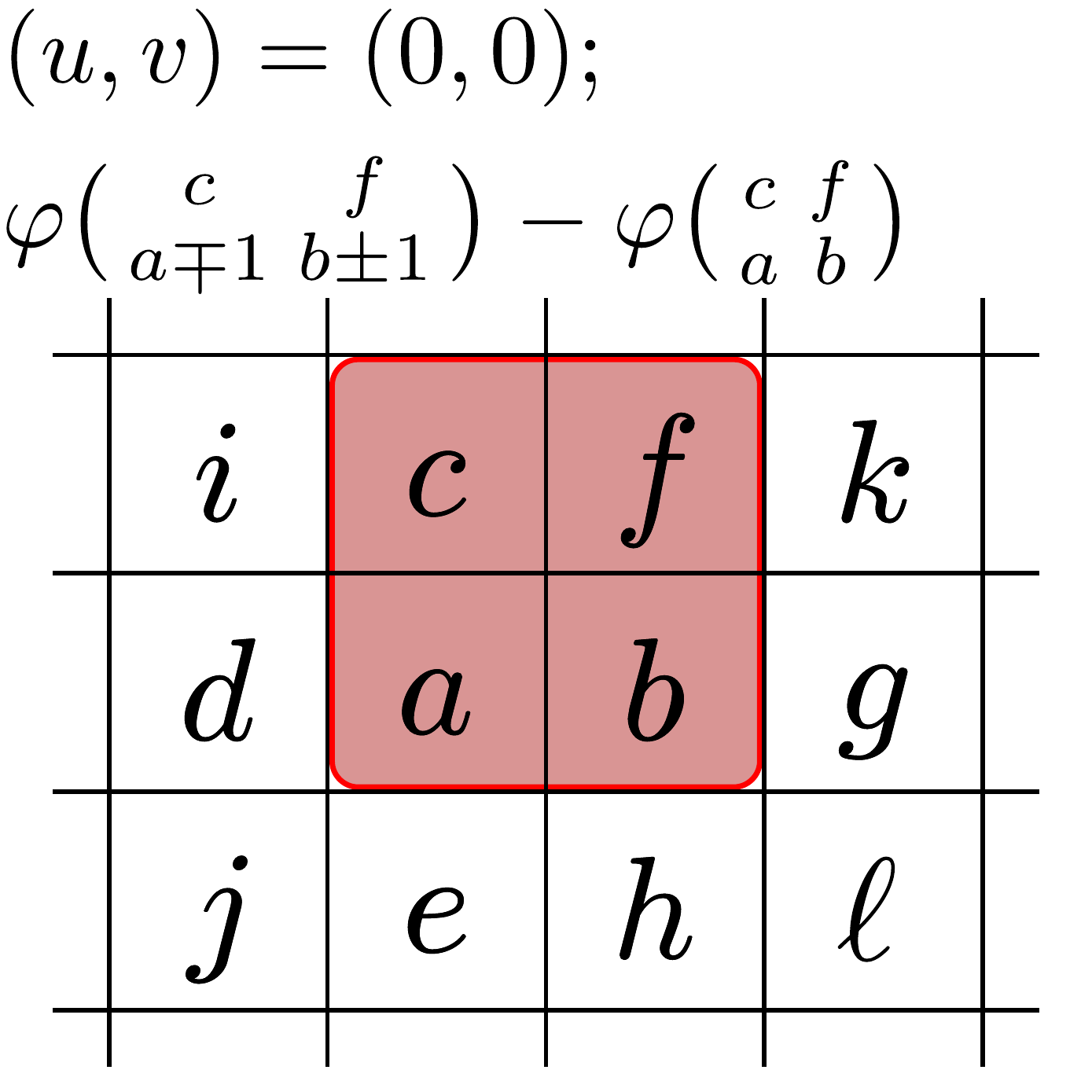} \quad 
 \includegraphics[width=42mm]{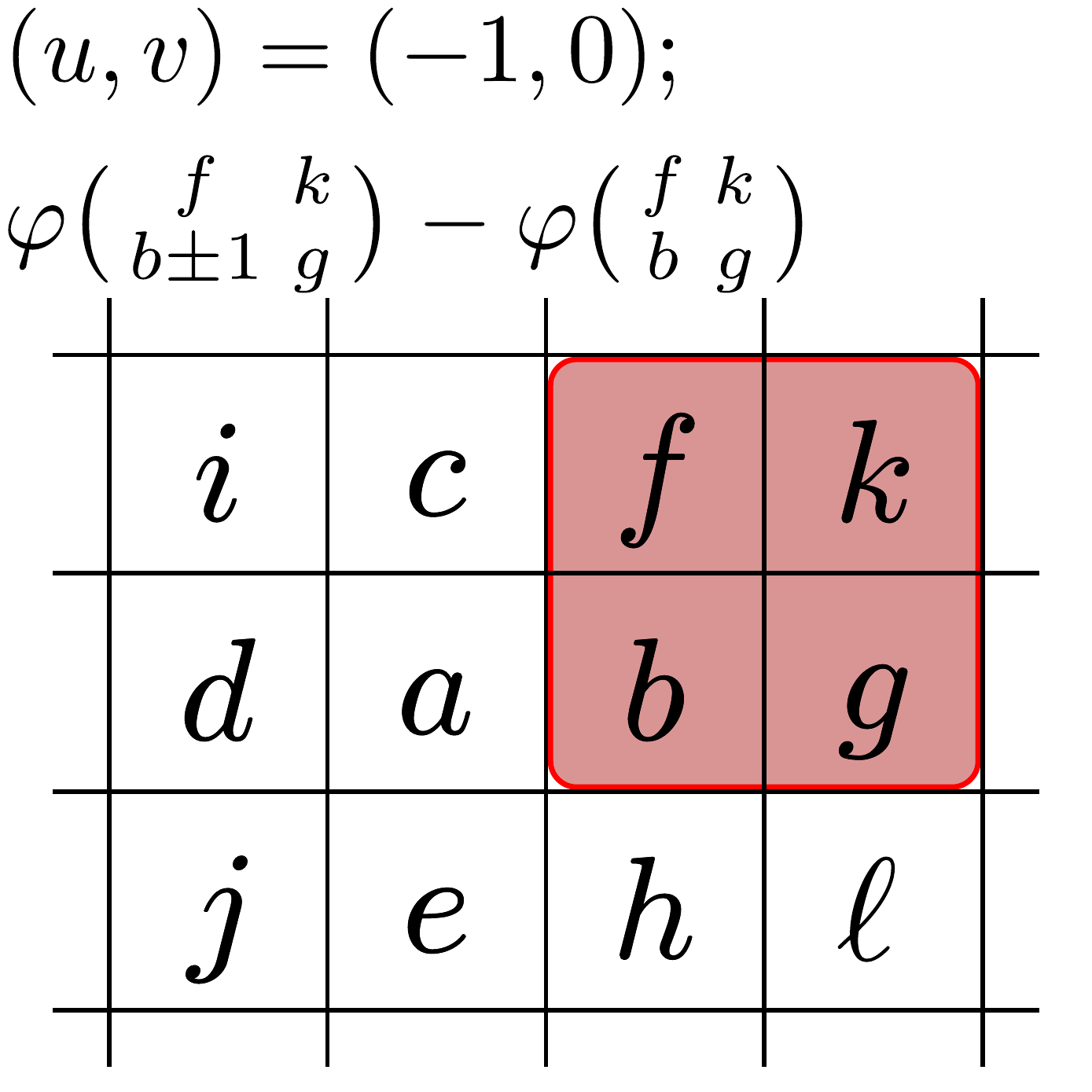}
\\[5mm]
 \includegraphics[width=42mm]{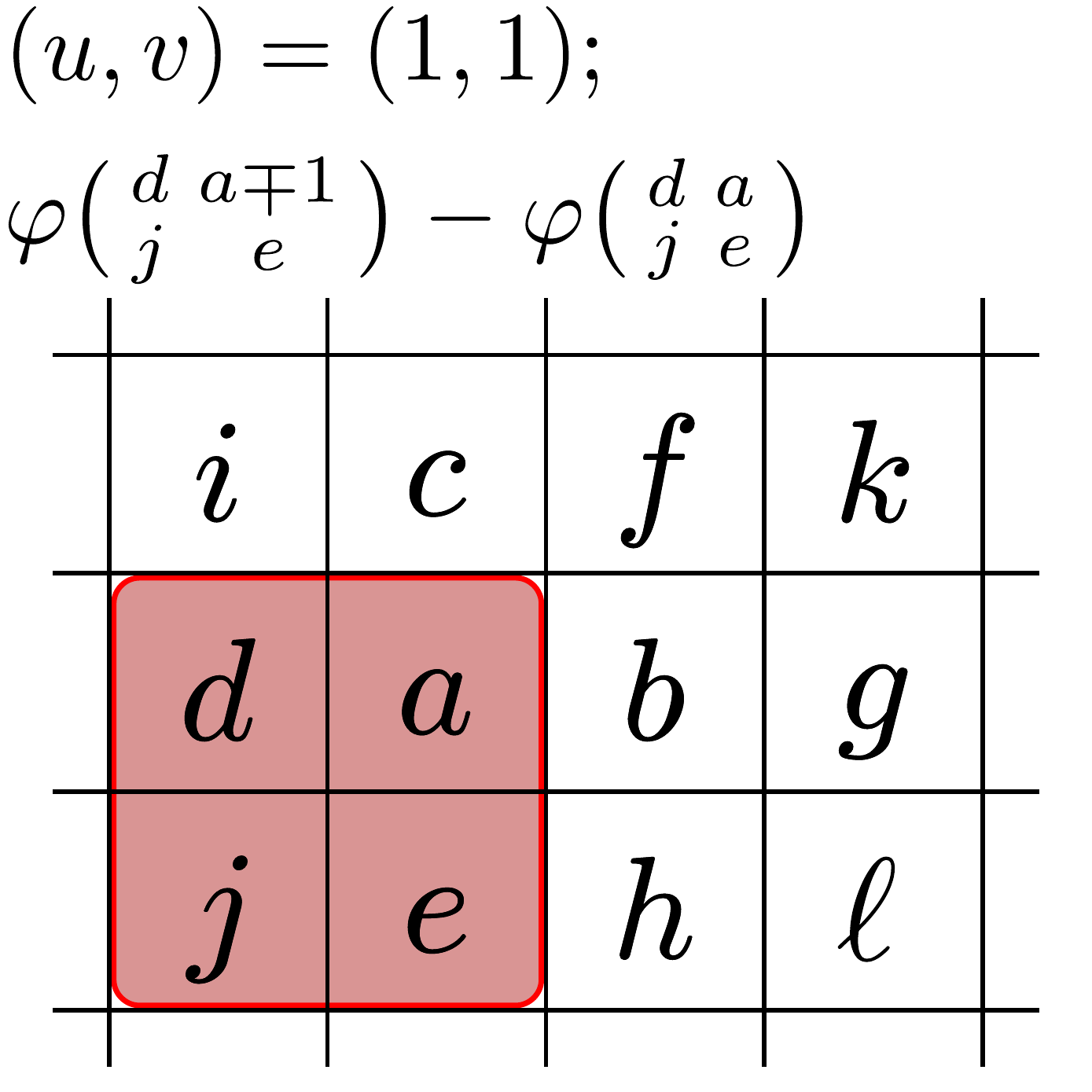} \quad 
\includegraphics[width=42mm]{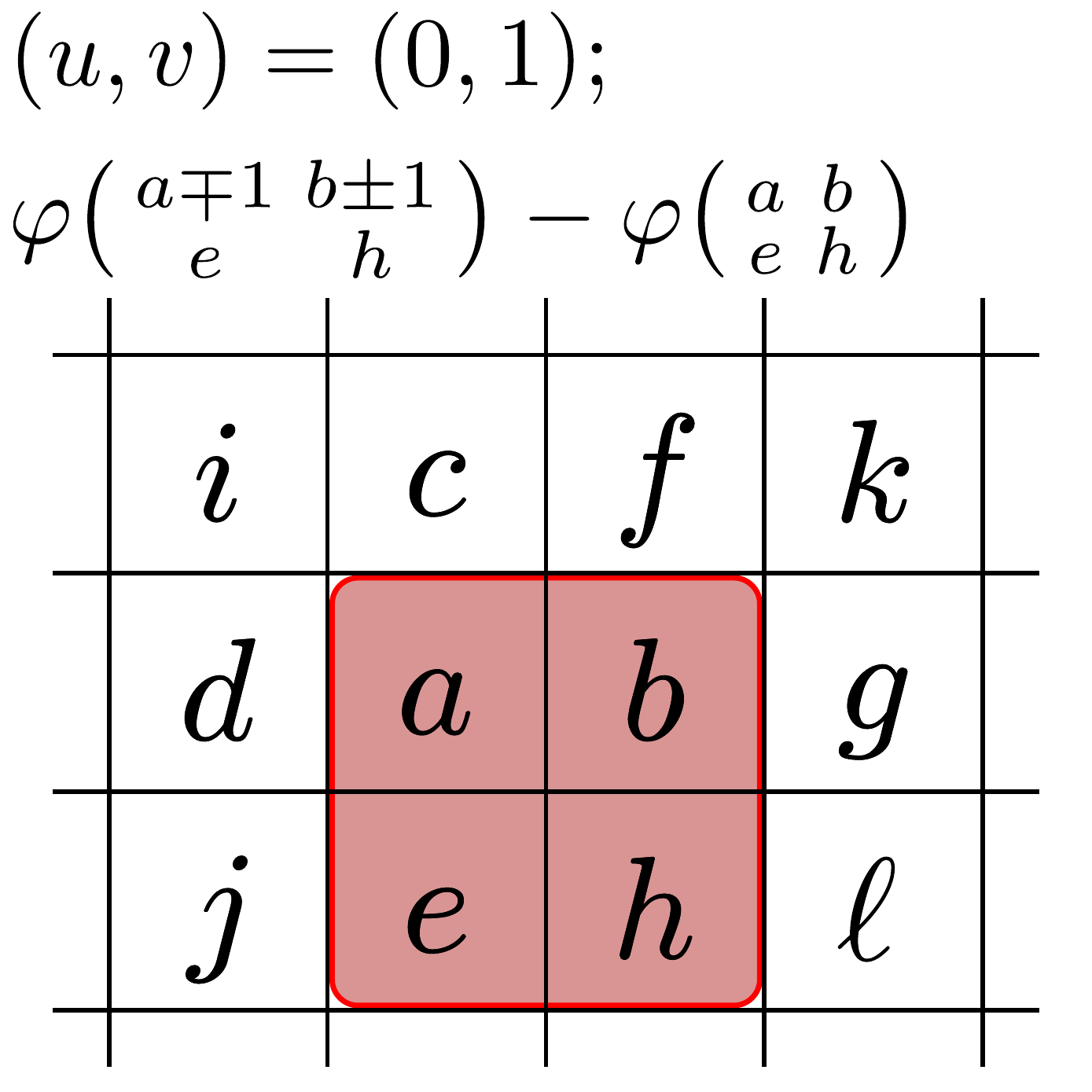} \quad 
\includegraphics[width=42mm]{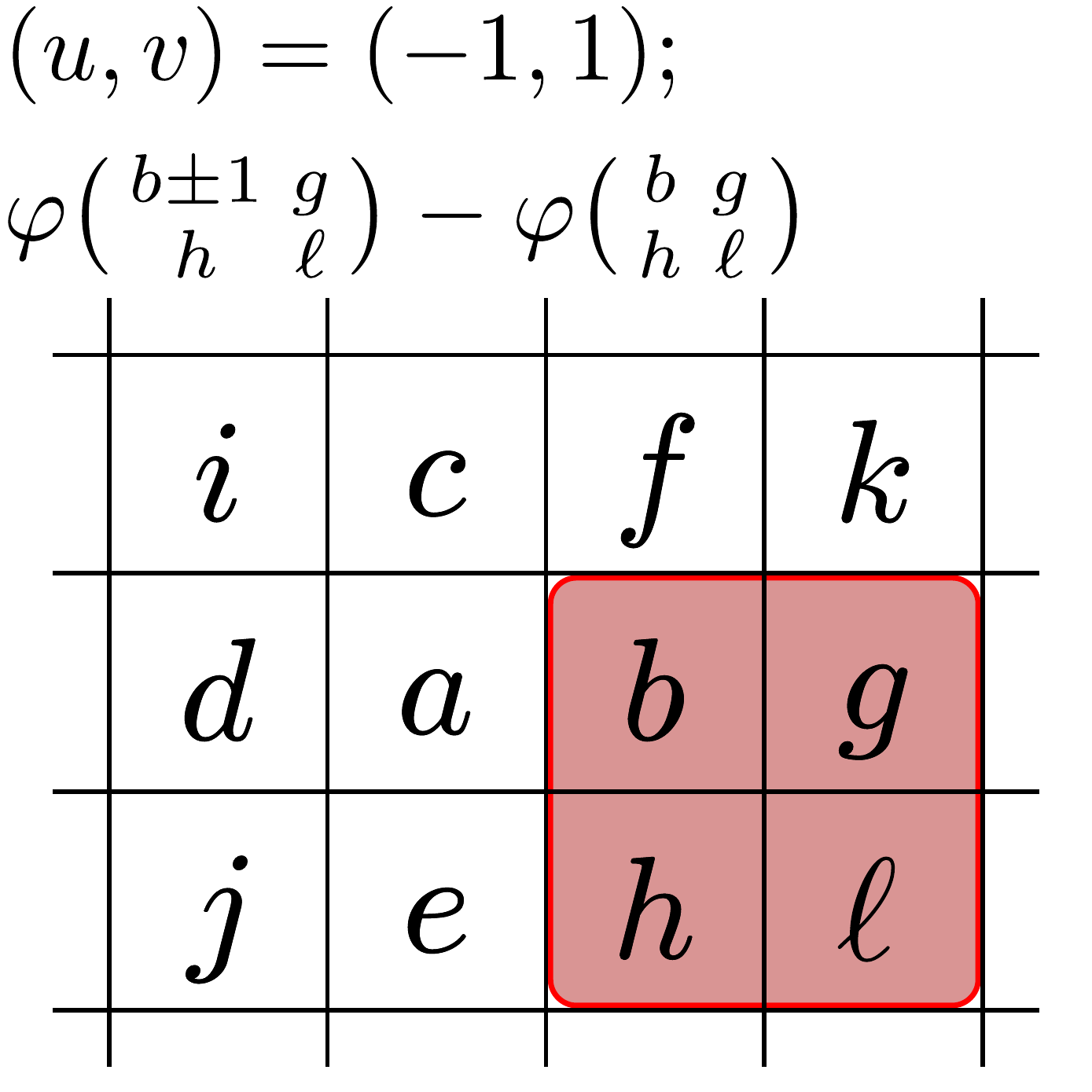}
 \caption{Illustration for the functionals $ Q^{ ( \pm 1, 0 ) } $ in the case of the $2 \times 2$ square. 
The shown six patterns correspond to non-vanishing $ \varphi \big( A_{ u,v } \, \tau^{ ( \alpha, \beta ) }\big) - \varphi ( A_{ u,v } \, \tau) $ terms in equation~\eqref{eq:2times2:Qpm10=}. 
}\label{fig:22covering}
\end{center}
\end{figure}

Here we consider the $2\times 2$ square, schematically $ S= \ctwotwo $. Thus $ \varphi (\tau ) = 
 \varphi \big( \begin{smallmatrix}
 \tau_{0,1} & \tau_{1,1} \\ 
 \tau_{0,0} & \tau_{1,0} \end{smallmatrix}\big) $. 
The expectation values of the functionals for the horizontal directions are 
\begin{align}
\label{eq:2times2:Qpm10=}
 \big\langle Q^{ ( \pm 1, 0 ) } ( \varphi ) \big\rangle
 &= \sum_{a,\dots, \ell \in \{ 0,1 \}} P^{ ( \pm 1, 0 ) } 
 W_a \cdots W_\ell 
 \big[ \pm 1 - R^{ ( \pm 1 , 0 ) } ( \varphi ) \big]^2 , \\
\begin{split}
 R^{ ( \pm 1 , 0 ) }( \varphi ) &= 
 \varphi \big( \begin{smallmatrix} i & c \\ d & a \mp 1 \end{smallmatrix}\big) 
- \varphi \big( \begin{smallmatrix} i & c \\ d & a \end{smallmatrix}\big) 
+ \varphi \big( \begin{smallmatrix} c & f \\ a \mp 1 & b \pm 1 \end{smallmatrix}\big) 
- \varphi \big( \begin{smallmatrix} c & f \\ a & b \end{smallmatrix}\big) 
+ \varphi \big( \begin{smallmatrix} f & k \\ b \pm 1 & g \end{smallmatrix}\big) 
- \varphi \big( \begin{smallmatrix} f & k \\ b & g \end{smallmatrix}\big) \\
& 
+ \varphi \big( \begin{smallmatrix} d & a \mp 1 \\ j & e \end{smallmatrix}\big) 
- \varphi \big( \begin{smallmatrix} d & a \\ j & e \end{smallmatrix}\big) 
+ \varphi \big( \begin{smallmatrix} a \mp 1 & b \pm 1 \\ e & h \end{smallmatrix}\big) 
- \varphi \big( \begin{smallmatrix} a & b \\ e & h \end{smallmatrix}\big) 
+ \varphi \big( \begin{smallmatrix} b \pm 1 & g \\ h & \ell \end{smallmatrix}\big) 
- \varphi \big( \begin{smallmatrix} b & g \\ h & \ell \end{smallmatrix}\big).
\end{split} 
\end{align} 
The notations are explained in figure~\ref{fig:alphabet} (left), and the transition rates are still given by \eqref{eq:P10,P-10}. We have a finite sum, because $ \varphi \big( A_{ u,v } \, \tau^{ ( \pm 1, 0 ) }\big) - \varphi ( A_{ u,v } \, \tau) $ vanishes unless $ u = -1,0,1 $ and $ v = 0,1 $. The remaining cases are illustrated in figure~\ref{fig:22covering}.

The vertical ones are 
\begin{align} 
\label{eq:2times2:Q0pm1=}
\big\langle Q^{ ( 0, \pm 1 ) }( \varphi ) \big\rangle
 &= \sum_{a,\dots, \ell \in \{ 0,1 \}} P^{ ( 0, \pm 1 ) }
 W_a \cdots W_\ell 
 \big[ R^{ ( 0, \pm 1 ) } ( \varphi ) \big]^2 , \\
\begin{split}
 R^{ ( 0, \pm 1 ) } ( \varphi ) &= 
 \varphi \big(\begin{smallmatrix} e & a\mp 1 \\ j & d \end{smallmatrix}\big)
- \varphi \big(\begin{smallmatrix} e & a \\ j & d \end{smallmatrix}\big)
+ \varphi \big(\begin{smallmatrix} a\mp 1 & c \\ d & i \end{smallmatrix}\big)
- \varphi \big(\begin{smallmatrix} a & c \\ d & i \end{smallmatrix}\big)
+ \varphi \big(\begin{smallmatrix} h & b\pm 1 \\ e & a\mp 1 \end{smallmatrix}\big)
- \varphi \big(\begin{smallmatrix} h & b \\ e & a \end{smallmatrix}\big) \\
& \quad
+ \varphi \big(\begin{smallmatrix} b\pm1 & f \\ a\mp 1 & c \end{smallmatrix}\big)
- \varphi \big(\begin{smallmatrix} b & f \\ a & c \end{smallmatrix}\big)
+ \varphi \big(\begin{smallmatrix} \ell & g \\ h & b\pm1 \end{smallmatrix}\big)
- \varphi \big(\begin{smallmatrix} \ell & g \\ h & b \end{smallmatrix}\big)
+ \varphi \big(\begin{smallmatrix} g & k \\ b\pm1 & f \end{smallmatrix}\big)
- \varphi \big(\begin{smallmatrix} g & k \\ b & f \end{smallmatrix}\big). 
\end{split} 
\end{align} 
The notations are explained in figure~\ref{fig:alphabet} (right), the transition rates are given by \eqref{eq:P01,P0-1}. 

One wants to solve $ \frac{ \partial }{ \partial \xi } \langle Q \rangle = 0 $, 
where $ \xi = \varphi \big(\begin{smallmatrix} \gamma & \delta \\ \alpha & \beta \end{smallmatrix}\big) $ for all $ \alpha,\beta,\gamma,\delta \in \{0,1\} $. The solution space is characterized by the following nine relations
\noindent
\begin{align} 
\label{9-square}
\begin{split} 
& \varphi \left(\begin{smallmatrix} 
 0 & 0 \\
 0 & 0 
\end{smallmatrix}\right)
+\varphi \left(\begin{smallmatrix} 
 1 & 1 \\
 1 & 1 
\end{smallmatrix}\right)
=
S_\leftrightarrow\left(\begin{smallmatrix} 
 1 & 0 \\
 1 & 0 
\end{smallmatrix}\right) =
S_\updownarrow\left(\begin{smallmatrix} 
 1 & 1 \\
 0 & 0 
\end{smallmatrix}\right) 
=S_\updownarrow\left(\begin{smallmatrix} 
 1 & 0 \\
 0 & 1 
\end{smallmatrix}\right) , \\
& \varphi \left(\begin{smallmatrix} 
 0 & 0 \\
 0 & 0 
\end{smallmatrix}\right)
+3 \varphi \left(\begin{smallmatrix} 
 1 & 1 \\
 1 & 1 
\end{smallmatrix}\right)
=S_\updownarrow\left(\begin{smallmatrix} 
 1 & 1 \\
 0 & 1 
\end{smallmatrix}\right)
+S_\updownarrow\left(\begin{smallmatrix} 
 1 & 1 \\
 1 & 0 
\end{smallmatrix}\right) , \\
 & 3 \varphi \left(\begin{smallmatrix} 
 0 & 0 \\
 0 & 0 
\end{smallmatrix}\right)
+\varphi \left(\begin{smallmatrix} 
 1 & 1 \\
 1 & 1 
\end{smallmatrix}\right)
=S_\updownarrow\left(\begin{smallmatrix} 
 0 & 1 \\
 0 & 0 
\end{smallmatrix}\right)
+S_\updownarrow\left(\begin{smallmatrix} 
 1 & 0 \\
 0 & 0 
\end{smallmatrix}\right) , \\ 
& A_\updownarrow\left(
\begin{smallmatrix} 
 1 & 0 \\
 0 & 1 
\end{smallmatrix}\right) 
=
A_\updownarrow\left(\begin{smallmatrix} 
 1 & 0 \\
 0 & 0 
\end{smallmatrix}\right)
- A_\updownarrow\left(\begin{smallmatrix} 
 0 & 1 \\
 0 & 0 
\end{smallmatrix}\right)
=A_\updownarrow\left(\begin{smallmatrix} 
 1 & 1 \\
 0 & 1 
\end{smallmatrix}\right)
-A_\updownarrow\left(\begin{smallmatrix} 
 1 & 1 \\
 1 & 0 
\end{smallmatrix}\right) , \\
& 2 A_\updownarrow\left(\begin{smallmatrix} 
 1 & 1 \\
 0 & 0 
\end{smallmatrix}\right)
=
 S_\leftrightarrow\left(\begin{smallmatrix} 
 0 & 0 \\
 1 & 0 
\end{smallmatrix}\right)
- S_\leftrightarrow\left(\begin{smallmatrix} 
 1 & 0 \\
 0 & 0 
\end{smallmatrix}\right)
+S_\leftrightarrow\left(\begin{smallmatrix} 
 1 & 0 \\
 1 & 1 
\end{smallmatrix}\right)
- S_\leftrightarrow\left(\begin{smallmatrix} 
 1 & 1 \\
 1 & 0 
\end{smallmatrix}\right) , \\ 
& 2 A_\leftrightarrow\left(\begin{smallmatrix} 
 1 & 0 \\
 1 & 0 
\end{smallmatrix}\right)
= A_\leftrightarrow\left(\begin{smallmatrix} 
 0 & 0 \\
 1 & 0 
\end{smallmatrix}\right)
+A_\leftrightarrow\left(\begin{smallmatrix} 
 1 & 0 \\
 0 & 0 
\end{smallmatrix}\right)
+A_\leftrightarrow\left(\begin{smallmatrix} 
 1 & 0 \\
 1 & 1 
\end{smallmatrix}\right)
+A_\leftrightarrow\left(\begin{smallmatrix} 
 1 & 1 \\
 1 & 0 \\
\end{smallmatrix}
\right)
 - \frac{2 \rho}{4 - \rho - 2 \rho^3}\, , 
\end{split} 
\end{align} 
where we used shorthand notations $S$ and $A$ for symmetric and asymmetric combinations of $\varphi$, namely 
\begin{align*} 
\begin{split} 
 S_\leftrightarrow\left(\begin{smallmatrix} 
 \gamma & \delta \\
 \alpha & \beta 
\end{smallmatrix}\right)
=
 \varphi\left(\begin{smallmatrix} 
 \gamma & \delta \\
 \alpha & \beta 
 \end{smallmatrix}\right)
+ \varphi \left(\begin{smallmatrix} 
 \delta & \gamma \\
 \beta & \alpha 
 \end{smallmatrix}\right), \quad 
 %%%%%%%%%%%%%%%%%%%%%%%%%%%%%%%%%
S_\updownarrow\left(\begin{smallmatrix} 
 \gamma & \delta \\
 \alpha & \beta 
\end{smallmatrix}\right)
=
 \varphi\left(\begin{smallmatrix} 
 \gamma & \delta \\
 \alpha & \beta 
\end{smallmatrix}\right)
+ \varphi \left(\begin{smallmatrix} 
 \alpha & \beta \\
 \gamma & \delta 
\end{smallmatrix}\right),
\\
%%%%%%%%%%%%%%%%%%%%%%%%%%%%%%%%%
 A_\leftrightarrow\left(\begin{smallmatrix} 
 \gamma & \delta \\
 \alpha & \beta 
 \end{smallmatrix}\right)
=
 \varphi\left(\begin{smallmatrix} 
 \gamma & \delta \\
 \alpha & \beta 
\end{smallmatrix}\right)
- \varphi \left(\begin{smallmatrix} 
 \delta & \gamma \\
 \beta & \alpha 
\end{smallmatrix}\right), \quad 
%%%%%%%%%%%%%%%%%%%%%%%%%%%%%%%%%
A_\updownarrow\left(\begin{smallmatrix} 
 \gamma & \delta \\
 \alpha & \beta 
\end{smallmatrix}\right)
=
 \varphi\left(\begin{smallmatrix} 
 \gamma & \delta \\
 \alpha & \beta 
\end{smallmatrix}\right)
- \varphi \left(\begin{smallmatrix} 
 \alpha & \beta \\
 \gamma & \delta
\end{smallmatrix}\right) .
\end{split} 
\end{align*} 
Substituting \eqref{9-square} into (\ref{eq:2times2:Qpm10=}) and (\ref{eq:2times2:Q0pm1=}) and performing straightforward calculations yield
\begin{align}
\label{q-square}
 q \Big[ \ctwotwo \Big] 
 =2 \rho (1-\rho) (1-\rho^3)^2 - \frac{4 (1-\rho)^3 \rho^5 }{ 4-\rho-2\rho^3 }
\end{align}
leading to the announced bound \eqref{D22}.

\section{The simplest non-trivial rhombus}
\label{Ap:5}

\begin{figure}
\begin{center}
 \includegraphics[width=56mm]{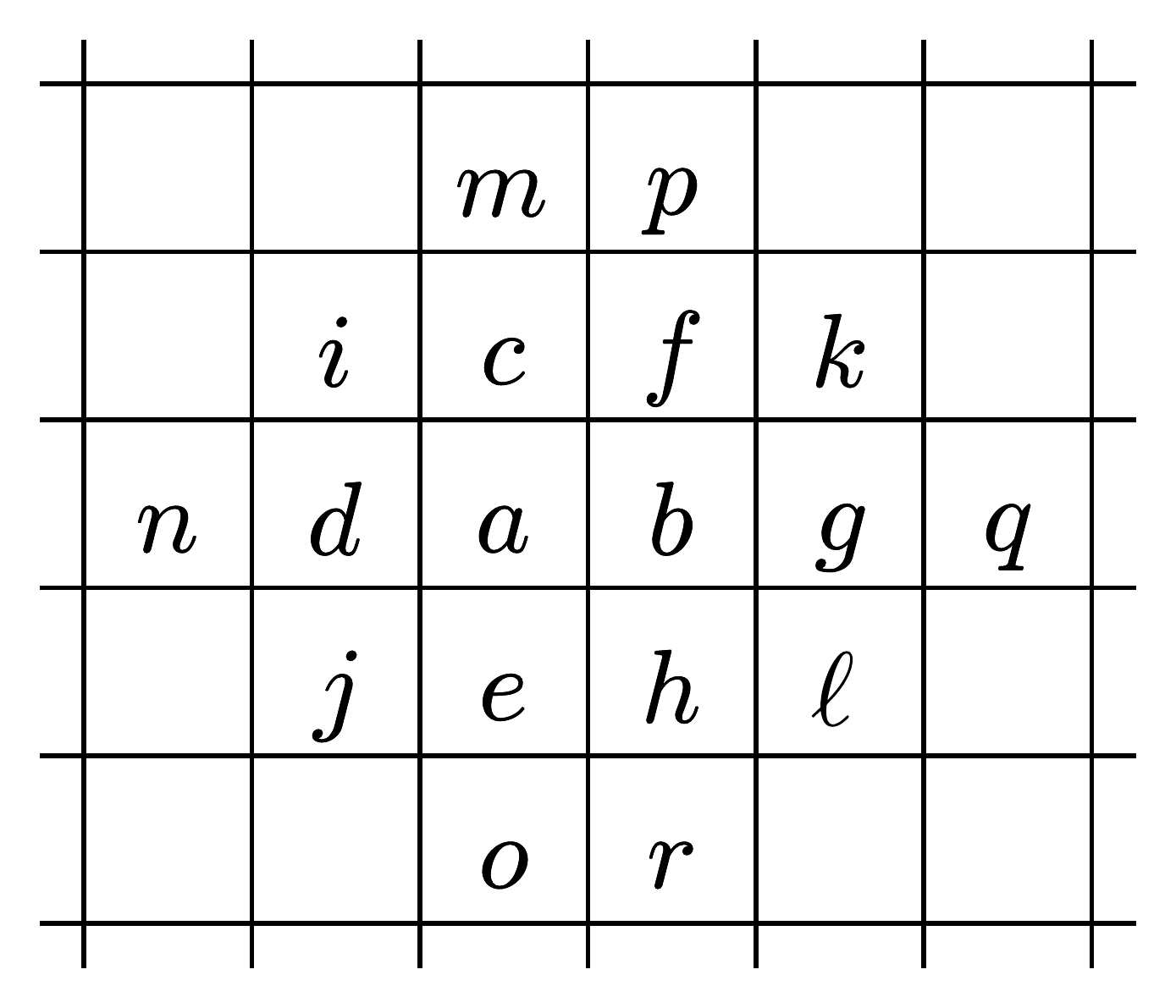}
\quad\includegraphics[width=49mm]{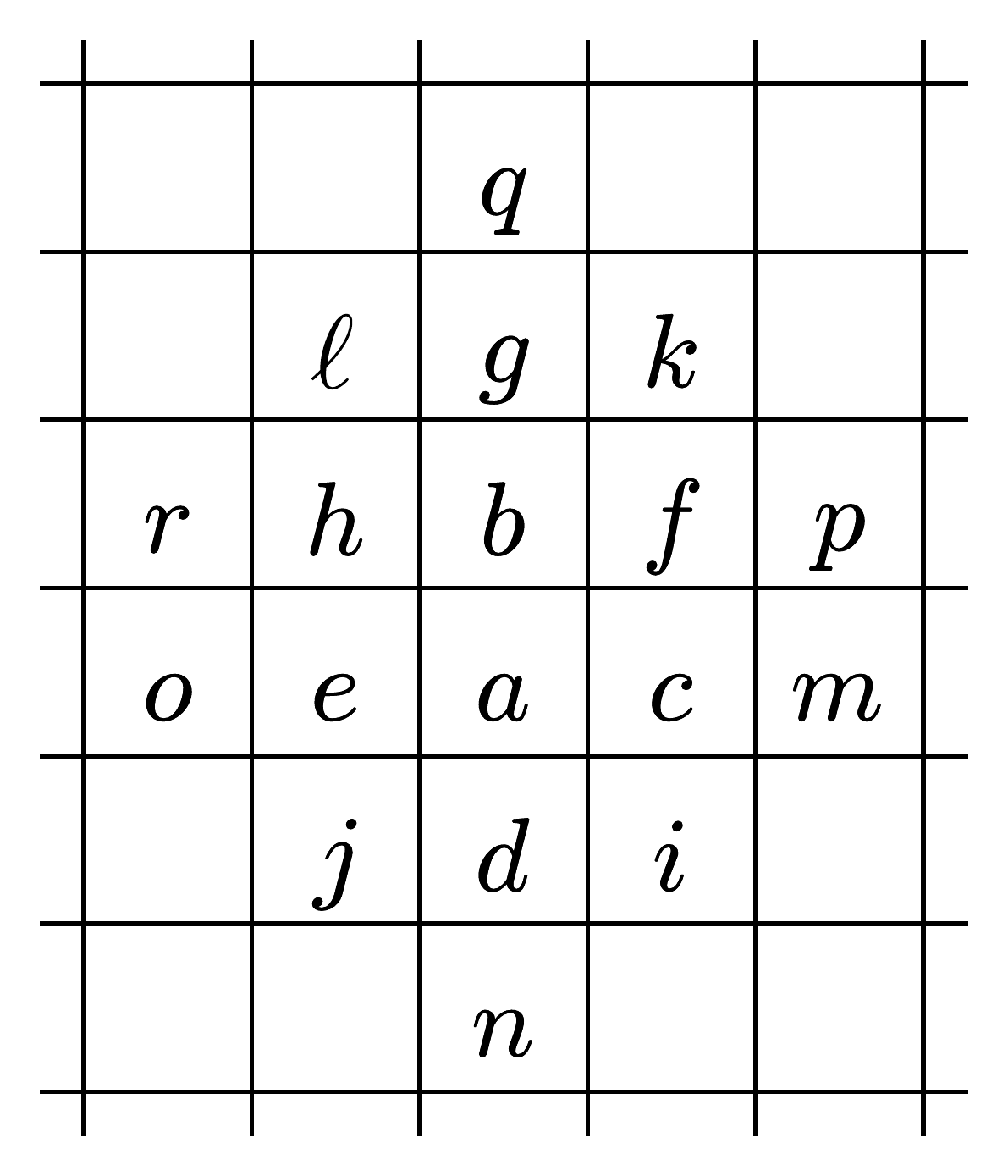}
 \caption{The lattice sites that appear in the formulas for 
 $\langle Q^{ (\pm1,0) }\rangle $ (left panel) and $\langle Q^{ (0, \pm1) } \rangle $ (right panel) in the case of the rhombus. 
 }\label{fig:alphabet-plus}
\end{center}\end{figure}

For the rhombus $S = \cplus$, functions $ \varphi $ depend on five sites: 
 $ \varphi ( \tau ) 
 = \varphi \left(\begin{smallmatrix}
 {} & \tau_{0,1 } & {} \\
\tau_{-1,0 } & \tau_{0, 0 } & \tau_{1, 0 } \\
 {} & \tau_{0,-1 } & {} 
\end{smallmatrix}\right). 
$
The expectation values of the functionals for the horizontal directions read
\begin{align}
 \big\langle Q^{ ( \pm 1, 0 ) } \big\rangle &= \sum_{a,\dots, r \in \{ 0,1 \}} P^{ ( \pm 1, 0 ) } 
 W_a \cdots W_r 
 \big[ \pm 1 - R^{ ( \pm 1 , 0 ) } \big]^2 , \\
\begin{split}
 R^{ ( \pm 1 , 0 ) } &= 
 \varphi \left(\begin{smallmatrix}
 {} & c & {} \\
 d & a\mp 1 & b\pm 1 \\
 {} & e & {} 
\end{smallmatrix}\right)
-\varphi \left(\begin{smallmatrix}
 {} & c & {} \\
 d & a & b \\
 {} & e & {} 
\end{smallmatrix}\right)
+\varphi \left(\begin{smallmatrix}
 {} & f & {} \\
 a\mp 1 & b\pm 1 & g \\
 {} & h & {} 
\end{smallmatrix}\right)
-\varphi \left(\begin{smallmatrix}
 {} & f & {} \\
 a & b & g \\
 {} & h & {} 
\end{smallmatrix}\right)\\
&\quad 
+\varphi \left(\begin{smallmatrix}
 {} & m & {} \\
 i & c & f \\
 {} & a\mp 1 & {} 
\end{smallmatrix}\right)
-\varphi \left(\begin{smallmatrix}
 {} & m & {} \\
 i & c & f \\
 {} & a & {} 
\end{smallmatrix}\right)
+ \varphi \left(\begin{smallmatrix}
 {} & i & {} \\
 n & d & a\mp 1 \\
 {} & j & {} 
\end{smallmatrix}\right)-
\varphi \left(\begin{smallmatrix}
 {} & i & {} \\
 n & d & a \\
 {} & j & {} 
\end{smallmatrix}\right)\\
&\quad 
+\varphi \left(\begin{smallmatrix}
 {} & a\mp 1 & {} \\
 j & e & h \\
 {} & o & {} 
\end{smallmatrix}\right)
-\varphi \left(\begin{smallmatrix}
 {} & a & {} \\
 j & e & h \\
 {} & o & {} 
\end{smallmatrix}\right)
+\varphi \left(\begin{smallmatrix}
 {} & p & {} \\
 c & f & k \\
 {} & b\pm 1 & {} 
\end{smallmatrix}\right)
-\varphi \left(\begin{smallmatrix}
 {} & p & {} \\
 c & f & k \\
 {} & b & {} 
\end{smallmatrix}\right)
\\
&\quad 
+\varphi \left(\begin{smallmatrix}
 {} & k & {} \\
 b\pm 1 & g & q \\
 {} & \ell & {} 
\end{smallmatrix}\right)
-\varphi \left(\begin{smallmatrix}
 {} & k & {} \\
 b & g & q \\
 {} & \ell & {} 
\end{smallmatrix}\right)
+\varphi \left(\begin{smallmatrix}
 {} & b\pm 1 & {} \\
 e & h & \ell \\
 {} & r & {} 
\end{smallmatrix}\right)
-\varphi \left(\begin{smallmatrix}
 {} & b & {} \\
 e & h & \ell \\
 {} & r & {} 
\end{smallmatrix}\right) 
\end{split} 
\end{align} 
The notations are explained in figure~\ref{fig:alphabet-plus}(left); 
the transition rates are
\begin{align*}
P^{ ( 1, 0 ) } &= a(1-b) (1-cde) ( 1- fgh ),\\
P^{ (- 1, 0 ) } &= b(1-a) (1-cde) ( 1- fgh ).
\end{align*}
Similarly for the vertical direction we have 
\begin{align}
 \big\langle Q^{ ( 0, \pm 1 ) } \big\rangle &= \sum_{a,\dots, r \in \{ 0,1 \}} P^{ ( 0, \pm 1 ) } \,
 W_a \cdots W_r 
 \big[ R^{ ( 0, \pm 1 ) } \big]^2 , \\
\begin{split}
 R^{ ( 0 , \pm 1 ) } &= 
 \varphi \left(\begin{smallmatrix}
 {} & b\pm 1 & {} \\
 e & a\mp 1 & c \\
 {} & d & {} 
\end{smallmatrix}\right)
-\varphi \left(\begin{smallmatrix}
 {} & b & {} \\
 e & a & c \\
 {} & d & {} 
\end{smallmatrix}\right)
+\varphi \left(\begin{smallmatrix}
 {} & g & {} \\
 h & b\pm 1 & f \\
 {} & a\mp 1 & {} 
\end{smallmatrix}\right)
-\varphi \left(\begin{smallmatrix}
 {} & g & {} \\
 h & b & f \\
 {} & a & {} 
\end{smallmatrix}\right)\\
&\quad 
+\varphi \left(\begin{smallmatrix}
 {} & f & {} \\
 a\mp 1 & c & m \\
 {} & i & {} 
\end{smallmatrix}\right)
-\varphi \left(\begin{smallmatrix}
 {} & f & {} \\
 a & c & m \\
 {} & i & {} 
\end{smallmatrix}\right) 
+\varphi \left(\begin{smallmatrix}
 {} & a\mp 1 & {} \\
 j & d & i \\
 {} & n & {} 
\end{smallmatrix}\right)
-\varphi \left(\begin{smallmatrix}
 {} & a & {} \\
 j & d & i \\
 {} & n & {} 
\end{smallmatrix}\right)\\
&\quad 
+\varphi \left(\begin{smallmatrix}
 {} & h & {} \\
 o & e & a\mp 1 \\
 {} & j & {} 
\end{smallmatrix}\right)
-\varphi \left(\begin{smallmatrix}
 {} & h & {} \\
 o & e & a \\
 {} & j & {} 
\end{smallmatrix}\right) 
+\varphi \left(\begin{smallmatrix}
 {} & k & {} \\
 b\pm 1 & f & p \\
 {} & c & {} 
\end{smallmatrix}\right)
-\varphi \left(\begin{smallmatrix}
 {} & k & {} \\
 b & f & p \\
 {} & c & {} 
\end{smallmatrix}\right)\\
& \quad 
+\varphi \left(\begin{smallmatrix}
 {} & q & {} \\
 \ell & g & k \\
 {} & b\pm 1 & {} 
\end{smallmatrix}\right)
-\varphi \left(\begin{smallmatrix}
 {} & q & {} \\
 \ell & g & k \\
 {} & b & {} 
\end{smallmatrix}\right)
+\varphi \left(\begin{smallmatrix}
 {} & \ell & {} \\
 r & h & b\pm 1 \\
 {} & e & {} 
\end{smallmatrix}\right)
-\varphi \left(\begin{smallmatrix}
 {} & \ell & {} \\
 r & h & b \\
 {} & e & {} 
\end{smallmatrix}\right)
\end{split} 
\end{align} 
The notations are explained in figure~\ref{fig:alphabet-plus}(right); the transition rates are
\begin{align*}
P^{ ( 0 , 1 ) } & = a(1-b) (1-cde) ( 1- fgh ), \\
P^{ ( 0,- 1 ) } & = b(1-a) (1-cde) ( 1- fgh ).
\end{align*}

We should solve 
$ \frac{ \partial }{ \partial \xi } \langle Q \rangle = 0 $, 
where $ \xi = \varphi \left(\begin{smallmatrix} 
 {} & \gamma & {} \\
 \delta & \alpha & \beta \\
 {} & \epsilon & {} \\
 \end{smallmatrix}\right) $
 for all $ \alpha,\beta,\gamma,\delta, \epsilon \in \{0,1\} $.
The solutions are given by 19 homogeneous relations 
\begin{align*} 
&
 \varphi \left(\!\!
\begin{smallmatrix}
 {} & 0 & {} \\
 0 & 0 & 0 \\
 {} & 0 & {} \\
\end{smallmatrix}\!\! 
\right) -\varphi \left(\!\!
\begin{smallmatrix}
 {} & 1 & {} \\
 0 & 0 & 0 \\
 {} & 1 & {} \\
\end{smallmatrix}\!\! 
\right) =
\varphi \left(\!\!\begin{smallmatrix}
 {} & 0 & {} \\
 0 & 1 & 0 \\
 {} & 0 & {} 
\end{smallmatrix}\!\! \right) - 
\varphi \left(\!\!\begin{smallmatrix}
 {} & 1 & {} \\
 0 & 1 & 0 \\
 {} & 1 & {} \\
\end{smallmatrix}\!\! 
\right)
=
 \varphi \left(\!\!\begin{smallmatrix}
 {} & 0 & {} \\
 1 & 0 & 1 \\
 {} & 0 & {} 
\end{smallmatrix}\!\! \right)
-\varphi \left(\!\!\begin{smallmatrix}
 {} & 1 & {} \\
 1 & 0 & 1 \\
 {} & 1 & {} 
\end{smallmatrix}\!\! \right) 
=
\varphi \left(\!\!\begin{smallmatrix}
 {} & 0 & {} \\
 1 & 1 & 1 \\
 {} & 0 & {} 
\end{smallmatrix}\!\! \right)
-\varphi \left(\!\!\begin{smallmatrix}
 {} & 1 & {} \\
 1 & 1 & 1 \\
 {} & 1 & {} 
\end{smallmatrix}\!\! \right) ,
\\ % % % % % % % % % % % % % % % % % % % % % % 
&
 A_{\updownarrow} \left(\!\!\begin{smallmatrix}
 {} & 0 & {} \\
 0 & 0 & 1 \\
 {} & 1 & {} \\
\end{smallmatrix}\!\! \right)
+A_{\updownarrow} \left(\!\!\begin{smallmatrix}
 {} & 0 & {} \\
 1 & 0 & 0 \\
 {} & 1 & {} 
\end{smallmatrix}\!\! \right)
= 
 2 A_{\updownarrow}\left(\!\!\begin{smallmatrix}
 {} & 0 & {} \\
 0 & 0 & 0 \\
 {} & 1 & {} 
\end{smallmatrix}\!\! \right)
 -A_{\updownarrow}\left(\!\!\begin{smallmatrix}
 {} & 0 & {} \\
 0 & 1 & 0 \\
 {} & 1 & {} 
\end{smallmatrix}\!\! \right) 
+A_{\updownarrow}\left(\!\!\begin{smallmatrix}
 {} & 0 & {} \\
 1 & 1 & 1 \\
 {} & 1 & {} \\
\end{smallmatrix}\!\! 
\right) \\
& =
 A_{\updownarrow} \left(\!\!\begin{smallmatrix}
 {} & 0 & {} \\
 0 & 0 & 0 \\
 {} & 1 & {} \\
\end{smallmatrix}\!\! 
\right) +A_{\updownarrow} \left(\!\!
\begin{smallmatrix}
 {} & 0 & {} \\
 1 & 0 & 1 \\
 {} & 1 & {} \\
\end{smallmatrix}\!\! 
\right) +S_{\leftrightarrow} \left(\!\!
\begin{smallmatrix}
 {} & 0 & {} \\
 1 & 1 & 0 \\
 {} & 0 & {} \\
\end{smallmatrix}\!\! \right) 
+ 2 \varphi \left(\!\!
\begin{smallmatrix}
 {} & 0 & {} \\
 0 & 0 & 0 \\
 {} & 0 & {} \\
\end{smallmatrix}\!\! \right) 
-3 \varphi \left(\!\!
\begin{smallmatrix}
 {} & 0 & {} \\
 0 & 1 & 0 \\
 {} & 0 & {} \\
\end{smallmatrix}\!\! 
\right) -2 \varphi \left(\!\!
\begin{smallmatrix}
 {} & 0 & {} \\
 1 & 0 & 1 \\
 {} & 0 & {} 
\end{smallmatrix}\!\! \right)
 +\varphi \left(\!\!\begin{smallmatrix}
 {} & 0 & {} \\
 1 & 1 & 1 \\
 {} & 0 & {} \\
\end{smallmatrix}\!\! \right ) , \\ % % % % % % % % % % % % % % % % % % % % % % 
&
 A_{\updownarrow}\left(\!\!\begin{smallmatrix}
 {} & 0 & {} \\
 0 & 1 & 1 \\
 {} & 1 & {} 
\end{smallmatrix}\!\! \right) 
+A_{\updownarrow}\left(\!\!\begin{smallmatrix}
 {} & 0 & {} \\
 1 & 0 & 1 \\
 {} & 1 & {} 
\end{smallmatrix}\!\! \right)
 +A_{\updownarrow}\left(\!\!\begin{smallmatrix}
 {} & 0 & {} \\
 1 & 1 & 0 \\
 {} & 1 & {} 
\end{smallmatrix}\!\! \right) 
 -A_{\updownarrow}\left(\!\!\begin{smallmatrix}
 {} & 0 & {} \\
 0 & 0 & 0 \\
 {} & 1 & {} 
\end{smallmatrix}\!\! \right) 
-2 A_{\updownarrow}\left(\!\! \begin{smallmatrix}
 {} & 0 & {} \\
 1 & 1 & 1 \\
 {} & 1 & {} \\
\end{smallmatrix}\!\! \right) 
\\ 
&
= 3 \varphi \left(\!\!\begin{smallmatrix}
 {} & 0 & {} \\
 0 & 0 & 0 \\
 {} & 0 & {} 
\end{smallmatrix}\!\! \right)
 -3 \varphi \left(\!\!\begin{smallmatrix}
 {} & 0 & {} \\
 1 & 0 & 1 \\
 {} & 0 & {} 
\end{smallmatrix}\!\! \right)
 -3 \varphi \left(\!\!\begin{smallmatrix}
 {} & 1 & {} \\
 0 & 0 & 0 \\
 {} & 1 & {} 
\end{smallmatrix}\!\! \right) 
+3 \varphi \left(\!\!\begin{smallmatrix}
 {} & 1 & {} \\
 1 & 0 & 1 \\
 {} & 1 & {} \\
\end{smallmatrix}\!\! 
\right), \\ % % % % % % % % % % % % % % % % % % % % % % 
&
 A_{\updownarrow}\left(\!\!
\begin{smallmatrix}
 {} & 0 & {} \\
 0 & 0 & 1 \\
 {} & 1 & {} \\
\end{smallmatrix}\!\! 
\right) =A_{\updownarrow}\left(\!\!
\begin{smallmatrix}
 {} & 0 & {} \\
 1 & 0 & 0 \\
 {} & 1 & {} 
\end{smallmatrix}\!\! \right), \quad 
 A_{\updownarrow}\left(\!\!
\begin{smallmatrix}
 {} & 0 & {} \\
 0 & 1 & 1 \\
 {} & 1 & {} \\
\end{smallmatrix}\!\! \right) 
=
A_{\updownarrow}\left(\!\!\begin{smallmatrix}
 {} & 0 & {} \\
 1 & 1 & 0 \\
 {} & 1 & {} 
\end{smallmatrix}\!\! \right) , \\ 
& \varphi \left(\!\!\begin{smallmatrix}
 {} & 0 & {} \\
 1 & 0 & 1 \\
 {} & 0 & {} 
\end{smallmatrix}\!\! \right) 
= S_{\updownarrow}\left(\!\!\begin{smallmatrix}
 {} & 0 & {} \\
 1 & 1 & 1 \\
 {} & 1 & {} 
\end{smallmatrix}\!\! \right) +
\varphi \left(\!\!\begin{smallmatrix}
 {} & 1 & {} \\
 0 & 0 & 0 \\
 {} & 1 & {} 
\end{smallmatrix}\!\! \right)
 -S_\leftrightarrow \left(\!\!
\begin{smallmatrix}
 {} & 1 & {} \\
 1 & 1 & 0 \\
 {} & 1 & {} \\
\end{smallmatrix}\!\! \right) , 
\\
& \varphi \left(\!\!\begin{smallmatrix}
 {} & 0 & {} \\
 1 & 0 & 1 \\
 {} & 0 & {} 
\end{smallmatrix}\!\! \right) 
 = 
 S_\leftrightarrow\left(\!\!
\begin{smallmatrix}
 {} & 0 & {} \\
 1 & 0 & 0 \\
 {} & 0 & {} \\
\end{smallmatrix}\!\! 
\right) -\varphi \left(\!\!
\begin{smallmatrix}
 {} & 0 & {} \\
 0 & 0 & 0 \\
 {} & 0 & {} \\
\end{smallmatrix}\!\! 
\right)
= S_{\updownarrow}\left(\!\!\begin{smallmatrix}
 {} & 0 & {} \\
 1 & 0 & 1 \\
 {} & 1 & {} 
\end{smallmatrix}\!\! \right)
 - \varphi \left(\!\!\begin{smallmatrix}
 {} & 0 & {} \\
 1 & 0 & 1 \\
 {} & 0 & {} 
\end{smallmatrix}\!\! \right), \\
% % % % % % % % % % % % % % % % % % % % % % % % % %
& \varphi \left(\!\!\begin{smallmatrix}
 {} & 0 & {} \\
 0 & 1 & 0 \\
 {} & 0 & {} 
\end{smallmatrix}\!\! \right)
= 
 S_\leftrightarrow\left(\!\!\begin{smallmatrix}
 {} & 0 & {} \\
 1 & 1 & 0 \\
 {} & 0 & {} 
\end{smallmatrix}\!\! \right) 
 -\varphi \left(\!\!\begin{smallmatrix}
 {} & 0 & {} \\
 1 & 1 & 1 \\
 {} & 0 & {} \\
\end{smallmatrix}\!\! 
\right) 
 = S_{\updownarrow}\left(\!\!\begin{smallmatrix}
 {} & 0 & {} \\
 0 & 1 & 0 \\
 {} & 1 & {} 
\end{smallmatrix}\!\! \right)
-\varphi \left(\!\!\begin{smallmatrix}
 {} & 1 & {} \\
 0 & 1 & 0 \\
 {} & 1 & {} 
\end{smallmatrix}\!\! \right) , \\
% % % % % % % % % % % % % % % % % % % % % % % % % %
&
 \varphi \left(\!\!\begin{smallmatrix}
 {} & 0 & {} \\
 1 & 0 & 1 \\
 {} & 0 & {} 
\end{smallmatrix}\!\! \right)
 + \varphi \left(\!\!\begin{smallmatrix}
 {} & 1 & {} \\
 0 & 0 & 0 \\
 {} & 1 & {} 
\end{smallmatrix}\!\! \right)
= \tfrac 1 2 S_{\updownarrow}\left(\!\!
\begin{smallmatrix}
 {} & 0 & {} \\
 0 & 0 & 1 \\
 {} & 1 & {} 
\end{smallmatrix}\!\! \right) 
+ \tfrac 1 2 S_{\updownarrow}\left(\!\!\begin{smallmatrix}
 {} & 0 & {} \\
 1 & 0 & 0 \\
 {} & 1 & {} 
\end{smallmatrix}\!\! \right) 
=\varphi \left(\!\!
\begin{smallmatrix}
 {} & 0 & {} \\
 1 & 1 & 1 \\
 {} & 0 & {} 
\end{smallmatrix}\!\! \right)
 + S_{\leftrightarrow} \left(\!\!\begin{smallmatrix}
 {} & 1 & {} \\
 1 & 0 & 0 \\
 {} & 1 & {} 
\end{smallmatrix}\!\! \right) 
-\varphi \left(\!\!\begin{smallmatrix}
 {} & 1 & {} \\
 1 & 1 & 1 \\
 {} & 1 & {} 
\end{smallmatrix}\!\! \right) , \\
% % % % % % % % % % % % % % % % % % % % % % % % % %
& 
\varphi \left(\!\!\begin{smallmatrix}
 {} & 0 & {} \\
 0 & 0 & 0 \\
 {} & 0 & {} 
\end{smallmatrix}\!\! \right) 
- 2 \varphi \left(\!\!\begin{smallmatrix}
 {} & 0 & {} \\
 0 & 1 & 0 \\
 {} & 0 & {} 
\end{smallmatrix}\!\! \right) = 
 \varphi \left(\!\!\begin{smallmatrix}
 {} & 0 & {} \\
 1 & 0 & 1 \\
 {} & 0 & {} 
\end{smallmatrix}\!\! \right) 
- S_\leftrightarrow\left(\!\!\begin{smallmatrix}
 {} & 0 & {} \\
 1 & 1 & 0 \\
 {} & 0 & {} 
\end{smallmatrix}\!\! \right) 
 = \varphi \left(\!\!\begin{smallmatrix}
 {} & 1 & {} \\
 1 & 0 & 1 \\
 {} & 1 & {} 
\end{smallmatrix}\!\! \right) 
-\tfrac 1 2 S_{\updownarrow}\left(\!\!\begin{smallmatrix}
 {} & 0 & {} \\
 0 & 1 & 1 \\
 {} & 1 & {} 
\end{smallmatrix}\!\! \right)
 - \tfrac 1 2 S_{\updownarrow}\left(\!\!\begin{smallmatrix}
 {} & 0 & {} \\
 1 & 1 & 0 \\
 {} & 1 & {} 
\end{smallmatrix}\!\! \right) , \\
% % % % % % % % % % % % % % % % % % % % % % % % % %
&
\varphi \left(\!\!
\begin{smallmatrix}
 {} & 1 & {} \\
 1 & 1 & 1 \\
 {} & 1 & {} 
\end{smallmatrix}\!\! \right) 
= S_{\updownarrow}\left(\!\! \begin{smallmatrix}
 {} & 0 & {} \\
 0 & 0 & 0 \\
 {} & 1 & {} \\
\end{smallmatrix}\!\! 
\right) -2 \varphi \left(\!\!
\begin{smallmatrix}
 {} & 0 & {} \\
 0 & 0 & 0 \\
 {} & 0 & {} 
\end{smallmatrix}\!\! \right)
 +\varphi \left(\!\!
\begin{smallmatrix}
 {} & 0 & {} \\
 1 & 1 & 1 \\
 {} & 0 & {} 
\end{smallmatrix}\!\! \right) 
= S_{\updownarrow}\left(\!\!
\begin{smallmatrix}
 {} & 0 & {} \\
 1 & 1 & 1 \\
 {} & 1 & {} 
\end{smallmatrix}\!\! \right) 
-S_\leftrightarrow\left(\!\!\begin{smallmatrix}
 {} & 0 & {} \\
 1 & 1 & 0 \\
 {} & 0 & {} 
\end{smallmatrix}\!\! \right)
 +\varphi \left(\!\!\begin{smallmatrix}
 {} & 0 & {} \\
 0 & 1 & 0 \\
 {} & 0 & {} 
\end{smallmatrix}\!\! \right) , 
 \end{align*}
 and 3 inhomogeneous relations 
\begin{align*} 
& A_{\leftrightarrow} \left(\!\!\begin{smallmatrix}
 {} & 0 & {} \\
 1 & 0 & 0 \\
 {} & 0 & {} 
\end{smallmatrix}\!\! \right)
+ \varphi \left(\!\!\begin{smallmatrix}
 {} & 1 & {} \\
 1 & 0 & 1 \\
 {} & 1 & {} 
\end{smallmatrix}\!\! \right) \\
& =A_{\leftrightarrow} \left(\!\!\begin{smallmatrix}
 {} & 0 & {} \\
 1 & 1 & 0 \\
 {} & 0 & {} 
\end{smallmatrix}\!\! \right) 
+A_{\leftrightarrow} \left(\!\!\begin{smallmatrix}
 {} & 1 & {} \\
 1 & 0 & 0 \\
 {} & 1 & {} 
\end{smallmatrix}\!\! \right) 
-A_{\leftrightarrow} \left(\!\!\begin{smallmatrix}
 {} & 1 & {} \\
 1 & 1 & 0 \\
 {} & 1 & {} 
\end{smallmatrix}\!\! \right) -\varphi 
 \left(\!\!\begin{smallmatrix}
 {} & 0 & {} \\
 0 & 1 & 0 \\
 {} & 0 & {} 
\end{smallmatrix}\!\! \right) 
+\varphi \left(\!\!\begin{smallmatrix}
 {} & 0 & {} \\
 1 & 1 & 1 \\
 {} & 0 & {} 
\end{smallmatrix}\!\! \right) 
+\varphi \left(\!\!\begin{smallmatrix}
 {} & 1 & {} \\
 0 & 0 & 0 \\
 {} & 1 & {} 
\end{smallmatrix}\!\! \right) 
- 2r_1 \\
& 
= 
 -A_{\leftrightarrow} \left(\!\! \begin{smallmatrix}
 {} & 1 & {} \\
 1 & 0 & 0 \\
 {} & 1 & {} \\
 \end{smallmatrix}\!\! \right) 
 -S_{\updownarrow} \left(\!\!\begin{smallmatrix}
 {} & 0 & {} \\
 0 & 0 & 1 \\
 {} & 1 & {} \\
 \end{smallmatrix}\!\! \right) 
 +S_{\updownarrow} \left(\!\! \begin{smallmatrix}
 {} & 0 & {} \\
 1 & 0 & 0 \\
 {} & 1 & {} \\
 \end{smallmatrix}\!\! \right) 
 +\varphi \left(\!\! \begin{smallmatrix}
 {} & 0 & {} \\
 1 & 0 & 1 \\
 {} & 0 & {} \\
 \end{smallmatrix}\!\! \right)
 -\varphi \left(\!\!\begin{smallmatrix}
 {} & 0 & {} \\
 1 & 1 & 1 \\
 {} & 0 & {} \\
 \end{smallmatrix}\!\! \right) 
 +\varphi \left(\!\! \begin{smallmatrix}
 {} & 1 & {} \\
 1 & 1 & 1 \\
 {} & 1 & {} \\
 \end{smallmatrix}\!\! \right) 
-2r_2 , \\
&
 A_{\leftrightarrow} \left(\!\!\begin{smallmatrix}
 {} & 0 & {} \\
 1 & 1 & 0 \\
 {} & 0 & {} 
\end{smallmatrix}\!\! \right) 
+A_{\leftrightarrow} \left(\!\!\begin{smallmatrix}
 {} & 1 & {} \\
 1 & 1 & 0 \\
 {} & 1 & {} 
\end{smallmatrix}\!\! \right)
+S_{\updownarrow} \left(\!\!
\begin{smallmatrix}
 {} & 0 & {} \\
 0 & 1 & 1 \\
 {} & 1 & {} \\
\end{smallmatrix}\!\! \right) 
-S_{\updownarrow} \left(\!\!\begin{smallmatrix}
 {} & 0 & {} \\
 1 & 1 & 0 \\
 {} & 1 & {} 
\end{smallmatrix}\!\! \right) 
-2 S_{\updownarrow} \left(\!\!\begin{smallmatrix}
 {} & 0 & {} \\
 1 & 1 & 1 \\
 {} & 1 & {} 
\end{smallmatrix}\!\! \right) \\
& = \varphi \left(\!\!
\begin{smallmatrix}
 {} & 0 & {} \\
 0 & 1 & 0 \\
 {} & 0 & {} \\
\end{smallmatrix}\!\! 
\right) -3 \varphi \left(\!\!
\begin{smallmatrix}
 {} & 0 & {} \\
 1 & 0 & 1 \\
 {} & 0 & {} \\
\end{smallmatrix}\!\! 
\right) 
+3 \varphi \left(\!\!\begin{smallmatrix}
 {} & 1 & {} \\
 0 & 0 & 0 \\
 {} & 1 & {} 
\end{smallmatrix}\!\! \right) 
-4 \varphi \left(\!\!\begin{smallmatrix}
 {} & 1 & {} \\
 0 & 1 & 0 \\
 {} & 1 & {} 
\end{smallmatrix}\!\! \right) 
-\varphi \left(\!\!\begin{smallmatrix}
 {} & 1 & {} \\
 1 & 1 & 1 \\
 {} & 1 & {} 
\end{smallmatrix}\!\! \right) 
 - 2r_3 . 
 \end{align*}
To make above equations more compact we used shorthand notations $S$ and $A$ for symmetric and asymmetric combinations of $\varphi$, namely 
 \begin{align*}
&S_{\leftrightarrow} \left(\begin{smallmatrix}
 {} & \delta & {} \\
 \alpha & \beta & \gamma\\
 {} & \epsilon & {} \\
 \end{smallmatrix}\right) 
= 
 \varphi \left(\begin{smallmatrix}
 {} & \delta & {} \\
 \alpha & \beta & \gamma\\
 {} & \epsilon & {} \\
 \end{smallmatrix}\right) 
+ \varphi \left(\begin{smallmatrix}
 {} & \delta & {} \\
 \gamma& \beta & \alpha \\
 {} & \epsilon & {} \\
 \end{smallmatrix}\right) , \quad 
%%%%%%%%%%%%%%%%%%%%%%%%%%%%%%%%%
S_{\updownarrow} \left(\begin{smallmatrix}
 {} & \delta & {} \\
 \alpha & \beta & \gamma\\
 {} & \epsilon & {} \\
 \end{smallmatrix}\right) 
= 
 \varphi \left(\begin{smallmatrix}
 {} & \delta & {} \\
 \alpha & \beta & \gamma\\
 {} & \epsilon & {} \\
 \end{smallmatrix}\right) 
+ \varphi \left(\begin{smallmatrix}
 {} & \epsilon & {} \\
 \alpha & \beta & \gamma\\
 {} & \delta & {} \\
 \end{smallmatrix}\right) , \\
& A_{\leftrightarrow} \left(\begin{smallmatrix}
 {} & \delta & {} \\
 \alpha & \beta & \gamma\\
 {} & \epsilon & {} \\
 \end{smallmatrix}\right) 
= 
 \varphi \left(\begin{smallmatrix}
 {} & \delta & {} \\
 \alpha & \beta & \gamma\\
 {} & \epsilon & {} \\
 \end{smallmatrix}\right) 
- \varphi \left(\begin{smallmatrix}
 {} & \delta & {} \\
 \gamma& \beta & \alpha \\
 {} & \epsilon & {} \\
 \end{smallmatrix}\right) , \quad 
 A_{\updownarrow} \left(\begin{smallmatrix}
 {} & \delta & {} \\
 \alpha & \beta & \gamma\\
 {} & \epsilon & {} \\
 \end{smallmatrix}\right) 
= 
 \varphi \left(\begin{smallmatrix}
 {} & \delta & {} \\
 \alpha & \beta & \gamma\\
 {} & \epsilon & {} \\
 \end{smallmatrix}\right) 
- \varphi \left(\begin{smallmatrix}
 {} & \epsilon & {} \\
 \alpha & \beta & \gamma\\
 {} & \delta & {} \\
 \end{smallmatrix}\right).
\end{align*}
 The inhomogeneous terms are given by 
 \begin{align*} 
r_1 V &= 47 + 83 \rho + 43 \rho ^2 -116 \rho ^3 -63 \rho ^4 -9 \rho ^5 +30 \rho ^6 +10\rho ^7 -2 \rho ^9 , \\
r_2 V &= 3\rho +3 \rho^2 -5 \rho^3-15 \rho^4-15 \rho^5+6 \rho^6+4 \rho^7+2 \rho^8, \\
r_3 V&= 47+50 \rho +10 \rho ^2-49 \rho ^3-40 \rho ^4-18 \rho ^5+14 \rho ^6+10 \rho ^7+4 \rho ^8 
\end{align*}
with $ V $ determined by equation~\eqref{V:def}.

\section{Perturbative calculations near the maximal density} 
\label{Ap:max}

Here we describe in more detail the perturbation approach which we use to probe the behavior of the diffusion coefficient in the $1-\rho = v \to 0$ limit. For concreteness, we choose squares as the basic symmetric sets used in our minimization procedure. 

The expectation values of the functionals \eqref{eq:Qalphabeta} read 
\begin{equation}
\label{eq:Qab}
\langle Q^{ ( \alpha,\beta ) } (\varphi ) \rangle = \sum_{ a_1,\dots ,a_N \in \{ 0 , 1 \} } 
 P^{ ( \alpha,\beta ) }_{ 0,0 } W_{ a_1 } \cdots W_{ a_N } [ \alpha - R ( \varphi) ]^2
\end{equation}
with
\begin{align}
 R ( \varphi) = \sum_{ ( u,v ) } \big\{ \varphi \big( A_{ u,v } \ba^{ ( \alpha, \beta ) }\big) 
 - \varphi ( A_{ u,v } \ba) \big\}.
\label{eq:R=}
\end{align}
In the case of the $n\times n$ square, $ \ba= ( a_1,\dots, a_N ) $ is the shorthand notation of the relevant part of the configuration $\tau$, namely $ ( \tau_{ i,j } ) $ with $ - n\le i \le n$ and $-n \le j \le n-1 $ for $ ( \alpha,\beta ) = ( \pm 1 , 0 ) $, while when $ ( \alpha,\beta ) = ( 0, \pm 1 ) $, the indexes vary in the range $ - n\le i \le n-1$ and $-n \le j \le n $. We also shortly write $ N = 2n(2n-1) $. For example, for $n=2 $ we used the letters $ a,\dots, \ell $ in \ref{Ap:22}; here we replace them by $ a_1 ,\dots, a_{12} $. The indexes $ ( u, v ) $ in the sum in equation~\eqref{eq:R=} also run over $ - n\le u \le n $ and $-n \le v \le n-1 $ for $ ( \alpha,\beta ) = ( \pm 1 , 0 ) $, and over $ - n\le u \le n-1$ and $-n \le v \le n $ for $ ( \alpha,\beta ) = ( 0, \pm 1 ) $.

We split the sum on the right-hand side of \eqref{eq:Qab} into $ N-3 $ sums according to the number of empty sites: 
\begin{align}
\label{QSSS}
 \langle Q^{ ( \alpha,\beta ) } (\varphi ) \rangle = \sum_{\ell=3}^{N-1} \Sigma^{ ( \alpha,\beta ) }_\ell (\varphi)
\end{align}
where 
\begin{align}
\label{S:L}
 \Sigma^{ ( \alpha,\beta ) }_\ell (\varphi) = \sum_{ a_1,\dots ,a_N \in \{ 0 , 1 \} \atop \# \{ a_k = 0 \} = \ell } 
 P^{ ( \alpha,\beta ) }_{ 0,0 } W_{ a_1 } \cdots W_{ a_N } [ \alpha - R ( \varphi) ]^2 . 
\end{align}
At least three empty sites and one occupied site are needed in order to have $ P^{ ( \alpha,\beta ) }_{ 0,0 } = 1 $; otherwise $ P^{ ( \alpha,\beta ) }_{ 0,0 } = 0 $. Therefore terms with $ \ell = 0,1,2, N $ vanish explaining the right-hand side of \eqref{QSSS}.

Recalling the definition \eqref{WW} we see that $ W_{ a_1 } \cdots W_{ a_N }= v^\ell ( 1-v )^{ N- \ell } $ in the sum in \eqref{S:L}. Using this observation and inserting \eqref{QSSS} into \eqref{eq:Q=Q+Q+Q+Q} we obtain
\begin{equation}
\label{eq:Q=sum}
 \langle Q (\varphi ) \rangle = \sum_{\ell=3}^{N-1} v^\ell ( 1-v )^{ N-\ell }\, \Sigma_\ell( \varphi ) 
\end{equation}
where $\Sigma_\ell( \varphi )$ do not depend explicitly on $v$. We thus arrive at the small $v$ expansion
\begin{equation}
\label{eq:<Q>=T3+T4+T5}
\langle Q (\varphi ) \rangle = \Theta_3 ( \varphi ) v^3 + \Theta_4 ( \varphi ) v^4 + \Theta_5 ( \varphi ) v^5 + \cdots
\end{equation}
with 
\begin{eqnarray*}
\Theta_3 ( \varphi ) &=& \Sigma_3 ( \varphi ), \\
\Theta_4 ( \varphi ) &=& \Sigma_4 ( \varphi ) - (N-3)\, \Sigma_3 ( \varphi ), \\
\Theta_5 ( \varphi ) &=& \Sigma_5 ( \varphi ) - (N-4)\, \Sigma_4 ( \varphi ) + 
\tfrac{1}{2} (N-3) (N-4)\, \Sigma_3 ( \varphi ),
\end{eqnarray*}
etc. To obtain the true power series expansion of $ \langle Q (\varphi ) \rangle $ we have to take into account the expansion 
\begin{align}
\label{phi:exp}
\varphi ( \bc ) = \varphi_0 ( \bc ) + v \varphi_1 ( \bc ) + v^2 \varphi_2 ( \bc ) + \cdots . 
\end{align}
The notation $\bc$ in equation~\eqref{phi:exp} we denotes the relevant part of the configuration $\tau$, so it is similar to $\ba$ in equation~\eqref{eq:R=}. Using the expansion \eqref{phi:exp} we deduce the expansion of $ \Theta_\ell( \varphi ) $:
\begin{align}
\label{eq:Qell-expansion}
\begin{split}
\Theta_\ell ( \varphi ) &= \Theta_\ell ( \varphi_ 0 ) 
 + \sum_{\bc} \big[ v \varphi_1 (\bc) + v^2 \varphi_2 ( \bc ) + \cdots \big] 
 \frac{ \partial \Theta_\ell }{ \partial \varphi (\bc) } ( \varphi_0 ) \\
& 
 + \frac{1}{2}\sum_{\bc, \bc'} \big[ v \varphi_1 (\bc) + v^2 \varphi_2 (\bc) + \cdots \big] 
 \big[ v \varphi_1 (\bc') + v^2 \varphi_2 ( \bc' ) + \cdots \big] 
 \frac{ \partial^2 \Theta_\ell }{ \partial \varphi (\bc) \partial \varphi (\bc') } 
\end{split}
\end{align}
There are no terms with higher derivatives since $ \Theta_\ell ( \varphi ) $ is quadratic. 
Substituting \eqref{eq:Qell-expansion} into \eqref{eq:<Q>=T3+T4+T5} one obtains an expansion 
\begin{equation}
\label{eq:Qexpansion-final}
\langle Q ( \varphi ) \rangle = \sum_{ \ell \geq 3} Q_\ell \, v^\ell 
\end{equation}
with
\begin{align}
Q_3 & = \Theta_3 ( \varphi_0 ) , \\
Q_4 &= \Theta_4 ( \varphi_0 ) 
+ \sum_{ \bc } \varphi_1 (\bc) \frac{ \partial \Theta_3 }{ \partial \varphi (\bc) } ( \varphi_0 ) , 
\label{eq:Q4}
 \\ 
\begin{split}
Q_5 &= \Theta_5 ( \varphi_0 )
 + \sum_{ \bc } \varphi_1 (\bc) \frac{ \partial \Theta_4 }{ \partial \varphi (\bc) } ( \varphi_0 )
 + \sum_{ \bc } \varphi_2 (\bc) \frac{ \partial \Theta_3 }{ \partial \varphi (\bc) } ( \varphi_0 )\\
& + \frac{1}{2}\sum_{ \bc , \bc' } \varphi_1 (\bc) \varphi_1 (\bc') 
 \frac{ \partial^2 \Theta_3 }{ \partial \varphi (\bc) \partial \varphi (\bc') } \, ,
\end{split}
\label{eq:Q5}
\end{align}
etc. We perform minimization at each power $ v^\ell $. The number of summands in $\Theta_\ell $ is of order $ N^{\ell-3} $, whereas that of the original $\langle Q (\varphi ) \rangle $ is of order $ 2^N $. Therefore it is helpful to use $ \Theta_\ell $ instead of $ \langle Q (\varphi ) \rangle $, in order to minimize computations. We start with $ v^3 $. We want to minimize $ \Theta_3 ( \varphi ) $. One has to solve $ \tfrac{ \partial }{ \partial \varphi (\bc) } \Theta_3 = 0 $, which is a set of linear equations. Let us denote by $ \varphi^*_0 $ the function which minimizes $ \Theta_3 $. Note that $ \varphi^*_0 $ is not unique. It is convenient to write $ \varphi^*_0 (\bc) $ as polynomials of degree one in terms of a set of parameters $ ( d_1, \dots, d_{n'} ) $
with some $ n' \le 2^{n^2} $. For $ n=2 $, in particular, $ \varphi^*_0 $ can be expressed through $ n' = 10 $ parameters, e.g. 
\begin{align*}
& \varphi_0^* \left(\begin{smallmatrix} 
 1 & 1 \\
 1 & 1 
 \end{smallmatrix}\right) = d_1 , \ 
 \varphi_0^* \left(\begin{smallmatrix} 
 1 & 1 \\
 1 & 0 
 \end{smallmatrix}\right) = d_2 , \ 
 \varphi_0^* \left(\begin{smallmatrix} 
 0 & 1 \\
 1 & 1 
 \end{smallmatrix}\right) = d_3 , \\
 &\varphi_0^* \left(\begin{smallmatrix} 
 1 & 1 \\
 0 & 1 
 \end{smallmatrix}\right) = d_4 , \ 
 \varphi_0^* \left(\begin{smallmatrix} 
 1 & 0 \\
 1 & 1 
 \end{smallmatrix}\right) = d_5 , \ 
 \varphi_0^* \left(\begin{smallmatrix} 
 1 & 1 \\
 0 & 0 
 \end{smallmatrix}\right) = d_6 , \ 
 \varphi_0^* \left(\begin{smallmatrix} 
 0 & 0 \\
 1 & 0 
 \end{smallmatrix}\right) = d_7 , \\ 
& \varphi_0^* \left(\begin{smallmatrix} 
 0 & 0 \\
 0 & 0 
 \end{smallmatrix}\right) = d_8 , \ 
 \varphi_0^* \left(\begin{smallmatrix} 
 1 & 0 \\
 1 & 0 
 \end{smallmatrix}\right) = d_9 , \ 
 \varphi_0^* \left(\begin{smallmatrix} 
 0 & 1 \\
 0 & 1 
 \end{smallmatrix}\right) = d_{10} , \ 
 \varphi_0^* \left(\begin{smallmatrix} 
 0 & 1 \\
 1 & 0 
 \end{smallmatrix}\right) = - d_1 + d_2 + d_3 , \\
& \varphi_0^* \left(\begin{smallmatrix} 
 1 & 0 \\
 0 & 1 
 \end{smallmatrix}\right) = - d_1 + d_4 + d_5 , ~~\quad 
 \varphi_0^* \left(\begin{smallmatrix} 
 0 & 0 \\
 1 & 1 
 \end{smallmatrix}\right) = - d_6 + d_9 + d_{10} , \\
& \varphi_0^* \left(\begin{smallmatrix} 
 1 & 0 \\
 0 & 0 
 \end{smallmatrix}\right) = - d_2 + d_5 + 2 d_6 + d_7 - d_9 - d_{10} , \\
& \varphi_0^* \left(\begin{smallmatrix} 
 0 & 0 \\
 0 & 1 
 \end{smallmatrix}\right) = -1 - d_3 + d_5+ d_7 - d_9 + d_{10} , \\ 
& \varphi_0^* \left(\begin{smallmatrix} 
 0 & 1 \\
 0 & 0 
 \end{smallmatrix}\right) = -1 -d_4 + d_5 + 2 d_6 + d_7- 2 d_9 .
\end{align*}
Minimizing $ \Theta_3 ( \varphi ) $ one gets $ \Theta_3 ( \varphi_0^* ) = 14 $.

Next we examine the coefficient of $ v^4 $. Since we choose $ \varphi_0 = \varphi^*_0 $, the sum in (\ref{eq:Q4}) vanishes, and the minimization should be performed for $ \Theta_4 ( \varphi_0^* )$. The variables are now $ ( d_1, \dots, d_{n'} ) $ and the minimum is reached on a certain $ \varphi_0^{**} =\varphi_0^{*} |_{d\to d^*} $ which is convenient to express through parameters $ ( e_1 , \dots, e_{n''} ) $. 
For example, for $ n=2 $, we use the parametrization 
\begin{align*}
 & d^*_k = e_k \ (1\le k \le 8) \\ 
 & d^*_9 = - \tfrac 1 2 + 2 e_1 - e_2 - e_3 - e_4 + e_6 + e_7 \\ 
 & d^*_{10} = \tfrac 1 2 - 4 e_1 + 2 e_2 + 2 e_3 + 2 e_4 + e_5 - e_6 - e_7
\end{align*}
and find $ \Theta_4( \varphi^{ ** }_0 ) = -6 $. 
In this way, one can obtain the high-density expansion of $ q[S] $ up to the order $v^4$. 
Dividing it by $ 2\chi $ one gets $ Q[S] $ up to the order $v^3$. 
Here we summarize the results for $2 \le n\le 6 $: 
\begin{align*}
 q [2 \times 2 ] &= 14 v^3 - 6 v^4 + O(v^5) , \\ 
 q [3 \times 3] &= 2 v^3 + \tfrac{ 28202 }{ 307 } v^4 + O(v^5) , \\ 
 q [4 \times 4 ] 
 &= 0 v^3 + \tfrac{ 906137359616 }{ 66311971451 } v^4 + O(v^5), \\ 
 q [ 5 \times 5 ] &= 0 v^3 + v^4 + O(v^5), \\ 
 q [ 6 \times 6 ] & =0 v^3 + 0 v^4 + O(v^5). 
\end{align*}
Dividing them by $ 2v(1-v) $ leads to the announced results \eqref{eq:D[nbyn]=...}.

We now shortly comment about the minimization of the coefficient of $v^5 $ in equation~\eqref{eq:Qexpansion-final}. 
The first sum in (\ref{eq:Q5}) vanishes due to the same reason as for $v^4$. Thus we need to minimize 
\begin{align*}
\frac{1}{2} \sum_{ \bc, \bc'} \varphi_1 (\bc) \varphi_1 (\bc') 
 \frac{ \partial^2 \Theta_3 }{ \partial \varphi (\bc) \partial \varphi (\bc') } 
 + \sum_{ \bc } \varphi_1 (\bc) \frac{ \partial \Theta_4 }{ \partial \varphi (\bc) } ( \varphi_0^{ ** } )
 \text{\quad and\quad } \Theta_5 ( \varphi_0^{ ** } ) 
\end{align*}
by varying $ \varphi_1( \bc ) $'s and $ ( e_1 , \dots, e_{N''} ) $, respectively. 
For example, for $ n=3 $ one finds 
\begin{align*}
 q [3 \times 3] 
 &= 2 v^3 + \tfrac{ 28202 }{ 307 } v^4 - \tfrac{ 37402727 }{ 94249 } v^5 + O(v^6)
 \end{align*}
and then 
\begin{align*}
 D[3\times 3] 
 &= v^2 + \tfrac{ 14408 }{ 307 } v^3 - \tfrac{ 28556215 }{ 188498 } v^4 + O(v^5).
 \end{align*}

Similar to perturbative calculations near the maximal density, one can perform perturbative calculations near the minimal density using $\rho$ as the small parameter. 
Conceptually, the method is the same; in terms of $ \rho $, equation~(\ref{eq:Q=sum}) reads 
\begin{equation}
 \langle Q (\varphi ) \rangle = \sum_{\ell=1}^{N-3} \rho^\ell ( 1- \rho )^{ N-\ell }\, \Sigma_{ N- \ell} ( \varphi ) ,
\end{equation}
and we use the test function in the form 
\begin{align}
\varphi ( \bc ) = \varphi_0 ( \bc ) + \rho \varphi_1 ( \bc ) + \rho^2 \varphi_2 ( \bc ) + \cdots 
\end{align}
instead of (\ref{phi:exp}). Then one arrives at the low-density expansion 
\begin{equation}
\langle Q ( \varphi ) \rangle = \sum_{\ell \geq 1} \widehat Q_\ell \rho^\ell , 
\end{equation}
which is similar to the high-density version (\ref{eq:Qexpansion-final}). 
It is easy to see that $ \widehat Q_1 = 2 $, so one has to perform minimizations for $ \widehat Q_\ell $ with $ \ell \ge 2$. 
On the technical level the calculations are much more demanding. For instance, for the $4\times 4$ square, the first non-trivial term in the high density limit reads
\begin{equation*}
 D [4 \times 4 ] \simeq \frac{ 453068679808 }{ 66311971451 }\, v^3 , 
\end{equation*}
while in the low density limit the leading behavior is
\begin{equation*}
1- D [4 \times 4 ] \simeq \frac{r}{s}\, \rho^3 
\end{equation*}
with 238 digits integers 
\begin{eqnarray*}
r &=&\text{\footnotesize{248318756234182766354371739342317168383056003090313982110925}}\\
 &&\text{\footnotesize{074833032524417860592986034694101352022606656385591399542945}}\\
 &&\text{\footnotesize{656491548986924927203322299708534738000477618692292211046227}}\\
 &&\text{\footnotesize{8603484915853762583795345039758918839643321845429398375204}},\\
s &=&\text{\footnotesize{110824179824578186621743868088271852384672761259627636733420}}\\
 &&\text{\footnotesize{515459361262628851512749048604531828435124815266426435799166}}\\
 &&\text{\footnotesize{615188400510629974454741744199733440547604353419033690245066}}\\
 &&\text{\footnotesize{5151170565082388955124113412600620784668852539083095494859}}.
\end{eqnarray*}

\end{document}